\documentclass[a4paper,fleqn,usenatbib]{mnras}

\usepackage[T1]{fontenc}
\usepackage{ae,aecompl}
\usepackage{threeparttable}
\usepackage{caption}

\usepackage{graphicx}	
\usepackage{amsmath}	
\usepackage{amssymb}	
\usepackage{float}
\usepackage[caption = false]{subfig}
\usepackage{pdflscape}
\usepackage{tabularx}
\usepackage{newtxtext,newtxmath}

\newcommand\T{\rule{0pt}{2.6ex}}       
\newcommand\B{\rule[-1.6ex]{0pt}{0pt}} 

\title[Progenitors and Explosion Properties of SNRs Hosting CCOs] 
{Progenitors and Explosion Properties of Supernova Remnants Hosting Central Compact Objects: II. A Global Systematic Study with a Comparison to Nucleosynthesis Models}

\author[C. Braun et al.]{
C. Braun,$^{1}$\thanks{E-mail: umbrau59@myumanitoba.ca}
S. Safi-Harb,$^{1}$\thanks{E-mail: samar.safi-harb@umanitoba.ca}
C. L. Fryer,$^{2,3,4,5}$\thanks{E-mail: fryer@lanl.gov} 
P. Zhou$^{6}$\thanks{E-mail:pingzhou@nju.edu.cn} 
\\
$^{1}$Department of Physics and Astronomy, University of Manitoba, Winnipeg, MB R3T 2N2, Canada \\
$^2$ Center for Theoretical Astrophysics, Los Alamos National Laboratory, Los Alamos,
NM 87545\\
$^3$ Department of Astronomy, The University of Arizona, Tucson,
AZ 85721\\
$^4$ Department of Physics and Astronomy, The University of
New Mexico, Albuquerque, NM 87131\\
$^5$ Department of Physics, The George Washington University, Washington, DC 20052\\
$^6$ School of Astronomy \& Space Science, Nanjing University, 163 Xianlin Avenue, Nanjing 210023, People's Republic of China
}

\date{\textit{This article has been accepted for publication in MNRAS Published by Oxford University Press on behalf of the Royal Astronomical Society. }}
\pubyear{2023}

\begin{document}
\label{firstpage}
\pagerange{\pageref{firstpage}--\pageref{lastpage}}
\maketitle

\begin{abstract}

    Core-collapse explosions of massive stars leave behind neutron stars, with a known diversity that includes the ``Central Compact Objects'' (CCOs). Typified by the neutron star discovered near the centre of the Cas~A supernova remnant (SNR), CCOs have been observed to shine only in X-rays. To address their supernova progenitors, we perform a systematic study of SNRs that contain a CCO and display X-ray emission from their shock-heated ejecta. We make use of X-ray data primarily using the \textit{Chandra} X-ray observatory, complemented with \textit{XMM-Newton}. This study uses a systematic approach to the analysis of each SNR aimed at addressing the supernova progenitor as well as the explosion properties (energy and ambient density). After fitting for the ejecta abundances estimated from a spatially resolved spectroscopic study, we compare the data to six nucleosynthesis models making predictions on supernova ejecta yields in core-collapse explosions. We find that the explosion models commonly used by the astrophysics community do not match the ejecta yields for any of the SNRs, suggesting additional physics, e.g. multi-dimensional explosion models or updated progenitor structures, are required. Overall we find low-mass ($\leq$25 solar masses) progenitors among the massive stars population and low-energy explosions ($<$10$^{51}$~ergs). We discuss degeneracies in our model fitting, particularly how altering the explosion energy affects the estimate of the progenitor mass. Our systematic study highlights the need for improving on the theoretical models for nucleosynthesis predictions as well as for sensitive, high-resolution spectroscopy observations to be acquired with next-generation X-ray missions.

\end{abstract}




\section{Introduction}
   
    Massive stars, $\gtrsim$8M$_{\odot}$, are believed to end their lives in core-collapse (CC) supernova (SN) explosions. The expanding nebula left behind the explosion is known as a supernova remnant (SNR), which is characterized by a forward expanding shock-wave that sweeps up the surrounding interstellar/circumstellar medium, a reverse shock running back into and heating the ejecta, and a compact object (often a neutron star or a pulsar); see \cite{2012A&ARv..20...49V} for a review. One such type of compact object forged in a CC-SN is the ``Central Compact Object'' (CCO), the youngest type of X-ray emitting neutron stars typified by the CCO discovered with the first light \textit{Chandra} observation of Cas~A (see \cite{2013ApJ...765...58G, neutronstarsSS, 2017JPhCS.932a2006D} for reviews). To date we know of approximately 15 such objects; the nature of their progenitors remains a puzzle. To address this puzzle, we study their hosting SNRs using a systematic spatially resolved spectroscopic study of their X-ray emission. In particular, we aim to determine the intrinsic properties of their SN explosion and the mass of their progenitor star. To do so, we select the SNRs whose emission is dominated by thermal X-ray emission arising from enhanced ejecta yields. It is in the X-ray band where the nucleosynthesis products from the explosion have prominent emission lines in the 0.3--10~keV range (O, Ne, Mg, Si, S, Ar, Ca, Fe, and Ni).
   
    The goals of this study are to: (1) perform a spatially resolved spectroscopy on each SNR to determine the plasma temperatures, ionization timescales, and chemical abundances, (2) infer the explosion properties (explosion energy and progenitor mass) by comparing the abundance yields to a suite of nucleosynthesis model yields, and (3) draw conclusions on the population of SNRs hosting CCOs. The paper is organized as follows: in $\S$2 we summarize the observations, data preparation, and region selection. In $\S$3 we describe the global \textit{Chandra} and \textit{XMM-Newton} spectroscopic study. In $\S$4 we derive and discuss the explosion properties and the progenitor mass for each SNR. In $\S$5, we probe deeper into the supernova explosion and the yields studied by remnants to better understand the difficulties in the model fits.  Finally, in $\S$6 we summarize our conclusions. The appendices contain detailed studies of each of the SNRs in our sample.

\section{Observations and Data Preparation}

    \begin{table*}
	\begin{center}
		\caption{ Observational data used in the study. All SNRs use                          \textit{Chandra} data except for \textit{Puppis A} which primarily uses          \textit{XMM-Newton} data. Distance measurements are from the listed              references: (1) \protect\cite{G15_CCO,distG15}; (2)                              \protect\cite{distKes79_2,kes79XMM}; (3) \protect\cite{distCasA}; (4)            \protect\cite{distPuppisA}; (5) \protect\cite{distG349&G350}; (6)                \protect\cite{g350_gaensler,distG349&G350}. 
		}
		\label{tbl:ExpTime}
    	\begin{tabular}{c c c c c c} 

				\hline
			
				Source   &
				ObsID    & 
				Detector & 
				\begin{tabular}[c]{@{}c@{}} Exp. Time \\ (ks) \end{tabular}  &
				\begin{tabular}[c]{@{}c@{}} Distance \\ (kpc) \B \end{tabular}  &
			References \T \\
		
		        \hline
				
				G15.9+0.2 & 5530, 6288, 6289, 16766 & ACIS--S & 120.9 & 8.5 & 1 \T \\
                Kes~79 & 1982, 17453, 17659 & ACIS & 49.15 & 7.1 & 2 \\ 
				Cas~A & 4634-4639, 5196, 5319-5320  & ACIS--S & 980.37 & 3.4 & 3 \\ 
 			Puppis A  & \begin{tabular}[c]{@{}c@{}c@{}c@{}c@{}c@{}} \textit{XMM-         Newton} \\ 0150150101, 0150150201, 0150150301 \\ 0690970201, 0690970101,
                0113020301  \\ 0606280101, 0113020101, 0303530101 \\ \textit{Chandra} \\ 6371 \end{tabular} & \begin{tabular}[c]{@{}c@{}c@{}c@{}c@{}c@{}}  \\ MOS, PN \\ \\ \\ \\ ACIS-I  \end{tabular} &  \begin{tabular}[c]{@{}c@{}c@{}c@{}c@{}c@{}}  \\ 28.3 \\ \\ \\ \\ 28.28  \end{tabular} & \begin{tabular}[c]{@{}c@{}c@{}c@{}c@{}c@{}}  \\ 1.3 \\ \\ \\ \\ \ldots \end{tabular} & \begin{tabular}[c]{@{}c@{}c@{}c@{}c@{}c@{}}  \\ 4 \\ \\ \\ \\ \ldots \end{tabular} \\
				G349.7+0.2 & 2785 & ACIS--S & 55.74 & 12 & 5 \\
				G350.1--0.3 & 10102 & ACIS--S & 82.97 & 9 & 6 \B \\
				\hline
				
		\end{tabular}
	\end{center}
    \end{table*}

    For the majority of the SNRs in this study, we used archival \textit{Chandra} data. \textit{Chandra}'s Advanced CCD Imaging Spectrometer (ACIS) has two different imaging CCD configurations, a 2x2 arrangement called ACIS-I and a 1x6 arrangement called ACIS-S. ACIS-S has 2 back-illuminated CCDs, which have better spectral resolution than the other front-illuminated CCDs. To process the data for analysis we used the \textit{Chandra} Interactive Analysis of Observations\footnote{https://cxc.cfa.harvard.edu/ciao/} (CIAO) Version 4.11 software. We first used the CIAO command \textit{chandra\_repro} to reprocess the level 2 data, filtered to energies 0.3--10~keV, and then removed periods of high background rates when applicable. The \textit{Chandra} observation IDs (obsIDs) used for each SNR and their effective exposure times are listed in Table~\ref{tbl:ExpTime}. Due to the large size of Puppis~A, we primarily used \textit{XMM-Newton}. The \textit{XMM-Newton} telescope has three detectors that make up the European Photon Imaging Camera (EPIC), the MOS--1, MOS--2, and PN camera. The data were reprocessed using the Science Analysis Software\footnote{https://www.cosmos.esa.int/web/xmm-newton/sas} (SAS) version 16.1.0 and filtered using the recommended patterns for spectroscopy. Periods of high background rates were removed using the task \textit{espfilt}. We used all 3 detectors where possible, except for the bright eastern and northern knots, which suffered from pile-up for the PN camera. The Eastern and Northern data sets (0150150101, 0150150201, 0150150301) were also in large window mode and so a portion of the SNR was cut off for all 3 cameras. The central data sets (0113020301, 0113020101, 0606280101) covering the CCO were in small window mode for the PN camera and therefore removed from the study. Furthermore, obsID 0606280101 was taken after the loss of the MOS--1 CCD chip 6, and so only the MOS--2 was used from this data set. This obsID was chosen for the central region as it had the longest exposure.
    
    Region selection was chosen using the algorithm \textit{contbin}\footnote{https://www-xray.ast.cam.ac.uk/papers/contbin/}, an adaptively smoothed binning program that selects regions based on surface brightness \citep{contBinCasA}. \textit{Contbin} allows for selection of regions that maintains the morphology of the SNR, selects the entire SNR for study leaving no gaps, and allows the same amount of counts per region. We use region selections based on surface brightness because there is evidence to suggest a correlation between surface brightness and physical parameters like density, temperatures, and abundances \citep{contBinCasA}. The \textit{contbin} algorithm is run on a binned image, allowing for smoothed or exposure-corrected images. If there are multiple obsIDs for an SNR, the data were first merged, a mask was created to cut out around the edges of the SNR, and then binned to an appropriate bin size. The \textit{contbin} algorithm adaptively smooths the image and then selects regions based on user input criteria following a geometrical constraint (how circular a region is) and signal-to-noise parameter chosen to have smoother region edges which may miss fine structure details. The final number of regions is chosen to allow for adequate statistics for each region during the spectral fitting process. 
    
    For Puppis~A's \textit{XMM-Newton} data, the process is not as straightforward. The entire SNR would need approximately 5 different observations to observe the entire SNR and so it was broken up into different obsIDs. The \textit{contbin} algorithm was run on individual obsIDs rather than a fully mosaic-ed image, so as to remove the complication of extracting spectra from regions that spanned multiple data sets. For the obsIDs where the cameras were not in full frame mode (0150150101, 0150150201, 0150150301), the individual instrument data were first corrected for vignetting using the task \textit{evigweigth}, exposure-corrected, and then merged. The \textit{contbin} algorithm was applied to this final merged image. For the rest of the obsIDs, we used the Extended Source Analysis Software\footnote{https://heasarc.gsfc.nasa.gov/docs/xmm/esas/cookbook/xmm-esas.html} (ESAS) pipeline to combine the camera data where we corrected for vignetting, exposure, and quiescent particle background before merging. The merged images were masked, and then the \textit{contbin} algorithm was run on the exposure-corrected images. For obsIDs not in full frame mode, a region was extracted from a camera as long as it covers at least 80\% of the region's area. Although we had 56 regions covering Puppis~A, almost none of the regions show evidence of enhanced ejecta. This is likely due to the diffuse nature of the SNR, where the algorithm would group large regions in order to have the same number of counts per region. As a consequence, this ends up grouping bright knots with less bright, diffuse regions, which tended to infer the spectra's global, averaged-out parameters rather than the parameters of the smaller, enhanced ejecta knots. Subsequently, we searched manually for enhanced ejecta regions using previously detected enhanced emission: the small knot and filaments in the eastern portion of the SNR \citep{puppAEastKnots}; the ``omega filament'' near the centre of the SNR \citep{puppAOmegaKnot}; knots in the north as hinted at by a \textit{Suzaku} study \citep{puppASuzaku}, and a western knot for a total of 4 regions (2 regions had poor statistics when allowing oxygen to vary and so were excluded from the study; see $\S$A4 for details). 
    
\begin{figure*}
		\begin{center}
		\subfloat[(a) G15.9+0.2 (5 Regions)]{\includegraphics[angle=0,width=0.3\textwidth]{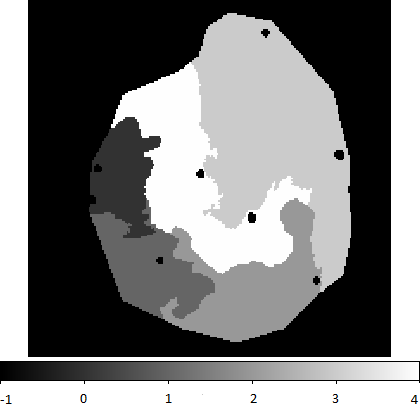}}
		\subfloat[(b) Kes~79 (12 Regions)]{\includegraphics[angle=0,width=0.3\textwidth]{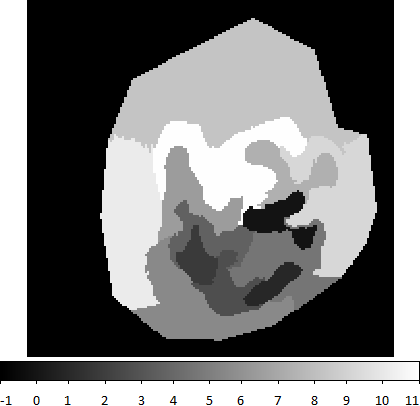}}
		\subfloat[(c) Cas~A (6251 Regions)]{\includegraphics[angle=0,width=0.3\textwidth]{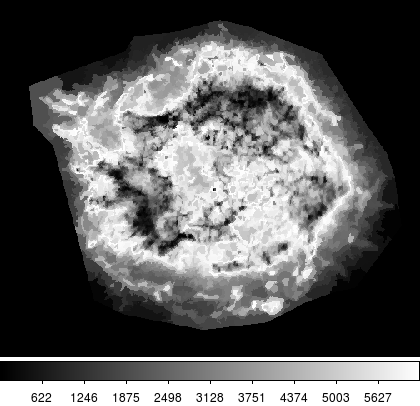}}
		\\
		\subfloat[(d) Puppis~A (56 Regions; 6 Ejecta Knots)]{\includegraphics[angle=0,width=0.28\textwidth]{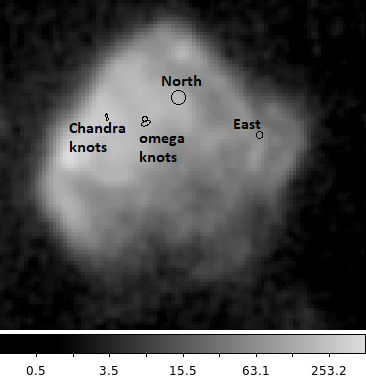}} \
		\subfloat[(e) G349.7+0.2 (6 Regions)]{\includegraphics[angle=0,width=0.3\textwidth]{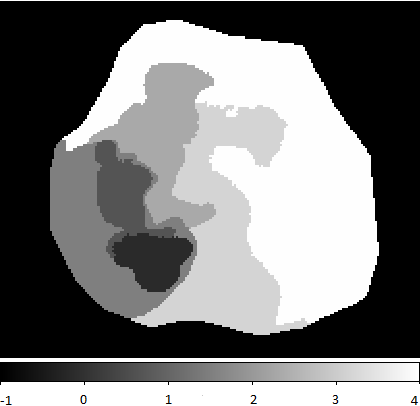}}
		\subfloat[(f) G350.1--0.3 (12 Regions)]{\includegraphics[angle=0,width=0.3\textwidth]{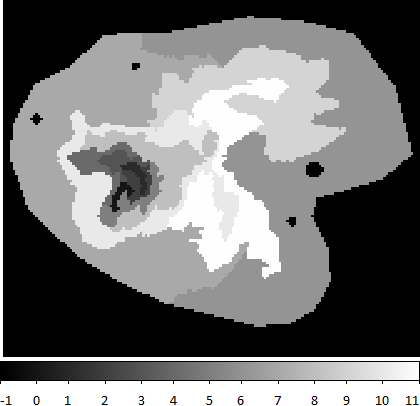}}

		\caption{Selected regions for spectral analysis. The greyscale colour bar value refers to the region number distribution, except for Puppis~A which has the regions labeled. Black regions (-1 on the colour bar) were masked out and not included in the spectral analysis. The regions containing the CCO had the CCO removed from the spectral extraction regions. For Puppis~A, the regions displayed are the enhanced ejecta knots manually selected (using previous studies), overlaid on a ROSAT image given the large size of this remnant.  }
		\label{fig:regions}	
		\end{center}
	\end{figure*}

    The final selected regions can be found in Figure~\ref{fig:regions}, where the greyscale represents the different regions for spectral extraction. For G15.9+0.2 and G350.1--0.3 we masked out particularly bright point sources as to not skew the region selection algorithm. Background regions were selected from source-free regions from the same CCD and nearby the source region. If that is not possible, then a background is selected from a region on the same obsID. For Puppis~A, we used local background regions as concentric rings surrounding the ejecta knots.

\subsection{A note on the regions selection methodology}
    Traditionally, region selection is done visually. In this study, as explained above, we use an algorithm to select our regions in order to remove bias from manually selecting regions and to ensure all SNRs are consistently analyzed. A potential concern is that the algorithm may focus on ISM interaction and not necessarily intrinsic physical properties. However, in our previous paper of this series \citep{Mine}, we selected the regions in two ways: 1) visually, as typically done; and 2) from line/equivalent width maps which are not necessarily ISM-dominated. In both cases we obtain results consistent to what we find here, where the progenitor explosion models struggle to replicate the abundance ratios compared to observations (\S4). Additionally, we also examined regions selected via \textit{contbin} for RCW~103 to check if we get consistent results with the more traditional methods. The data were well fit by two-component plasma models with similar results for temperatures, ionization age, and abundances as described in our previous paper. While the manually selected region results will not be identical to the \textit{contbin} selected region results, we still cannot find a single explosion model to account for all the abundance ratios.

\section{Spectral Analysis}

    We performed a spatially resolved spectroscopic study from the regions selected as described in $\S$2. We here elaborate on our global spectral analysis methodology, and in Appendix~A, we detail the analysis of each individual SNR. The data are modelled using the software XSPEC version 12.10.1f \citep{XSPEC}. The abundance tables were set using the XSPEC command \textit{abund wilm} \citep{Wilms}. The model TBABS is used to account for soft X-ray absorption and is characterized by the hydrogen column density, N$_{\rm H}$. We used two different non-equilibrium ionization models, VPSHOCK and VNEI, which are appropriate for modelling young SNRs whose plasma has not yet reached ionization equilibrium. They are characterized by a temperature, $kT$, an ionization timescales, $n_\text{e}~t$, where $n_\text{e}$ is the post-shock electron density and $t$ is the time since the passage of the shock, variable abundances (O, Ne, Mg, Si, S, Ar, Ca, Fe, and Ni), and a normalization constant $K=\frac{10^{-14}}{4\pi D^2}\int n_en_HdV$. These models differ in the treatment of the ionization timescale, where the VPSHOCK model considers a superposition of different ionization timescales and the VNEI model considers only a single ionization timescale. The APEC model, a collisionally-ionized plasma model, is characterized by a constant electron temperature, $kT$, a single abundance variable for various elements, the normalization constant $K$, but with no ionization timescale since the plasma is assumed to have reached collisional ionization equilibrium (CIE). The statistical quality of the data was based on reduced chi-squared statistics, $\chi^2_\nu$, where a good fit is considered $\chi^2_\nu<2$ with the exception of Cas~A where we have stricter requirements (given the quality of data), and the error for individual parameters was calculated using the \textit{Markov Chain Monte Carlo} (MCMC) method in {\tt XSPEC}\footnote{\url{https://heasarc.gsfc.nasa.gov/xanadu/xspec/manual/node43.html}}, using the Goodman-Weare algorithm with 10 walkers, a total chain of 10,000 steps, and a burn of 5,000.

    Firstly, for regions containing the CCOs, they were excluded from the extraction regions since we are focusing on the hosting SNR's emission. Each SNR region is fit by either a one- or two-component thermal plasma model in order to separate the forward-shocked ISM from the reverse-shocked ejecta. Single component modelled regions are either categorized as forward-shocked ISM if they contain solar or sub-solar abundances or as reverse-shocked ejecta if they contain super solar ejecta. For regions that are not dominated by either the forward shock or the enhanced ejecta, we then fit the data with two-component plasma models to account for any mixing between the two components. Cas~A, due to the quality and wealth of the data, and Puppis~A, due to the physical size of the SNR, had a more complicated analysis that is described in more detail, along with the rest of the SNRs in the sample, in Appendix~A.

\section{Results}
    We have performed a spatially resolved spectroscopy on each of the SNRs in our study. The emissions from these SNRs are diverse, but an effort was made to separate the forward shock from the shock-heated ejecta, in each case, in order to interpret the explosion properties and progenitor mass as described in the following section. Enhanced ejecta are characterized as having above-solar abundance values corresponding to \cite{Wilms}.

\subsection{Derived X-ray Properties}    

    We can derive the explosion properties and progenitor mass from the X-ray properties using the distance to the SNR and the corresponding physical size, as indicated in Table~\ref{tbl:ExpTime}. The X-ray emitting regions from the previous section take up a volume of space $V$, which will be defined as the area of the region as selected by \textit{contbin} multiplied by the radius of the SNR $R_s$ to account for the depth along the line of sight. Cas~A is a special case with much smaller regions, and so the volume of the regions is defined as $V=A^{3/2}$ where $A$ is the area of the region. The physical radius is highly dependent on the distance to the SNR and so we subsequently introduce a scaling factor into our calculations where, for example, $D_{3.1} = D/3.1$~kpc, such that each calculation can be easily rescaled to a different distance. The emission measure (EM) is the amount of plasma available to produce the observed flux from the volume V, and is given as $\text{EM} = \int n_en_HdV \sim fn_en_HV$ where $n_e$ is the post-shock electron density, $n_H$ is the proton density, and $f$ is the filling factor as a number between 0 and 1 where 1 corresponds to the entire volume contributing to the observed flux and 0 corresponds to none. The filling factor will be different for the hard and soft components of the SNR models, and can be distinguished as $f_h$ and $f_s$, respectively. Additionally, the different elemental contributions to the total gas flux is accounted for in the abundance parameter of the model fits rather than the filling factor. We do not calculate the filling factor in our analysis but represent it in our final results such that the values can easily be recalculated with a given filling factor. If we assume cosmic abundances, we can relate the electron density to the proton density as $n_e=1.2n_H$. For strong shocks, as is the case for SNRs, they follow the Rankine-Hugoniot jump conditions that relate the pre- and post-shock parameters. Assuming cosmic abundances, we can estimate the ambient density $n_0$ from the electron density as $n_e=4.8n_0$, where $n_0$ includes only hydrogen \citep{Borkowski2,Samar&Borkowski}. The normalization factor, $K=\frac{10^{-14}}{4\pi D^2}\int n_en_HdV$, is a parameter directly fit by the plasma models and can be used to derive the EM. We can also derive the time since the passage of the shock t$_\text{shock}$ from the ionization timescale $n_et$.
    In the following, we consider two cases for the supernova explosion properties estimates: expansion of the SNR in a uniform medium (\S4.1.1) and expansion into a wind-blown bubble (\S4.1.2).

\subsubsection{Uniform Ambient Medium}
    
    The evolutionary phase of the SNR is important for estimating the age of the remnant. If we assume the SNR is still in the early free expansion phase of its evolution, we can estimate a lower age limit. We initially assume an expansion velocity of $V_0\sim5000$~km~s$^{-1}$ which is consistent for young SNRs (see \cite{Reynoso}), and given the size of the SNR, we can estimate the age from $t=V_0/R_s$. When the swept-up mass (M$_{\text{sw}}$) becomes comparable to the mass of the ejecta (M$_{\text{ej}}$), the shock wave can no longer propagate freely into the surrounding medium and the SNR enters the Sedov-Taylor phase. If we assume the SNR is expanding into a uniform ambient medium, we can estimate the swept-up mass as $\text{M}_{\text{sw}}=1.4m_pn_0\times(4/3\pi $R$^3_sf)$, where $m_p$ is the mass of a proton. The upper limit on the age of the SNR is given as t$_{\text{SNR}}=\eta R_s/V_s$, where $\eta=0.4$ for a shock in the Sedov phase \citep{Sedov}. In the Sedov-Taylor phase, the assumed expansion velocity of $V_0\sim5000$~km~s$^{-1}$ no longer holds, as the surrounding medium will have slowed down the shock wave. However, we can derive the shock speed V$_S$ from the shock temperature T$_S$ from the Rankine-Hugoniot jump conditions such that, assuming the ions and electrons are coupled, $\text{V}_S=(16k_B\text{T}_s/3\mu m_h)^{1/2}$, where $\mu=0.604$ is the mean mass per free particle of a fully ionized plasma, $k_B=1.38\times 10^{-16}$~erg~K$^{-1}$ is the Boltzmann's constant, and $m_h$ is the mass of a hydrogen atom. The model component with solar abundances is considered the forward shock component that has swept-up and heated the ISM, and so we would associate the shock temperature with the blast wave component temperature. The Sedov blast wave model in which a supernova with explosion energy E$_*$ expands into a uniform ambient medium with density $n_0$ is given by E$_*=(1/2.02)R^5_Sm_nn_0t^{-2}_{\text{SNR}}$ where $m_n=1.4m_p$ is the mean mass of the nuclei, $m_p$ is the mass of a proton, and t$_{\text{SNR}}$ is the age of the SNR.  We calculate the X-ray properties for each SNR expanding into a uniform medium and tabulated them in Table~\ref{tbl:xrayProp}, with derived X-ray properties for the individual regions of each SNR summarized in Appendix~B. We make note here on the forward shocked ISM for Kes~79. There appears to be a range of temperatures for the forward shock component, with lower temperatures ($\approx0.2$~keV) from the two-component model regions (1, 4, 7, 8) and higher temperatures ($\approx0.8$~keV) from the single component regions (0, 9). In Table~\ref{tbl:xrayProp} we calculated the explosion properties of the forward shock using the soft component associated with the forward shock from the region~8 fits. However, if we consider the higher temperature from the single component regions in our estimates, this would result in a higher shock velocity of V$_S=800$~km~s$^{-1}$, a lower Sedov age of t$_{\text{SNR}}=5900$~D$_{7.1}$~yr, and a higher explosion energy of E$_*=2.3\times 10^{51}~f_s^{-1/2}$~D$_{7.1}^{5/2}$~erg. 
    
    We tend to find low explosion energies for most of the SNRs in Table~\ref{tbl:xrayProp}. The canonical value for the explosion energy is $10^{51}$~erg, however, lower explosion energies ($10^{49}$--$10^{50}$~erg) have been also estimated for other SNRs (see e.g., \cite{HarshaG292,Ben,Mine}). The study by \cite{Ghavamian} for young SNRs, where the plasma is generally not in ionization equilibrium, has shown that the electron temperature can be much smaller than the ion temperature, which indicates that the general assumption that the ion temperature is equal to the electron temperature may not hold. This would lead to an underestimate of the shock speed and a subsequent underestimate of the explosion energy. As well, in some cases, the assumption that the ambient density is uniform may also not hold, and so shock speed may be variable, leading to more uncertainty in the explosion energy.

\subsubsection{SNR Expanding in a Wind-Blown Bubble}

    We can subsequently consider the model of a SNR expanding into the wind-blown bubble of its own massive progenitor star, which follows an $r^{-2}$ profile as described by \cite{Chevalier} and considered for other SNRs (see e.g., \cite{Castro,HarshaKes73}). We similarly investigated this scenario for RCW~103 \citep{Mine}. If we assume the SNR is in the adiabatic phase when inferring its expansion velocity and age, the Sedov profiles can be analytically solved for a blast wave expanding in a $r^{-2}$ wind density distribution \citep{Cox}. The circumstellar wind density is described as $\rho_{cs}=\dot{M}/4\pi r^2 v_w=Dr^{-2}$ with a dimensionless parameter $D_*=D/D_{ch}$, where $D_{ch}=1\times 10^{14}$~g~cm$^{-1}$ is the coefficient of the density profile, $\dot{M}$ is the mass loss, and $v_w$ is the RSG wind velocity \citep{Chevalier}. Assuming an adiabatic approximation, the mass swept-up by the SNR shock to a radius \textit{R} is given by M$_{\text{sw}}=9.8D_*(R/\text{ 5 pc})$\,M$_\odot$ with the shell expanding to a radius $R=(3E_*/2\pi D)^{1/3}t^{2/3}$ at a velocity $V_s=2R/3t$, where $E_*$ is the explosion energy \citep{Chevalier1982,Chevalier,Castro}. From \cite{Cox}, the ratio between the average temperature and the shock temperature (weighted by $n^2$) in the RSG wind scenario is $\langle T\rangle/T_s=5/7$ whereas in the uniform density case is $\langle T\rangle/T_s=1.276$. If we assume that the X-ray spectrum mostly depends on the average temperature, we can determine a relation between the uniform and $r^{-2}$ density profiles as $T_{\text{s,wind}}/T_{\text{s,uniform}}=1.786$ and $V_{\text{s,wind}}/V_{\text{s,uniform}}=\sqrt{1.786}$. The expansion speed is then $V_s=\eta R/t$ where for the adiabatic wind phase, $\eta=2/5$, leading to an estimate for the age ratio of $t_{\text{s,wind}}/t_{\text{s,Sedov}}=(5/3)/\sqrt{1.786}=1.247$. 
    
    We can estimate the swept-up mass for each SNR using their EM and volume, assuming the shocked wind fills the full volume of the SNR (a reasonable assumption in the adiabatic phase). Using the swept-up mass allows us to derive the dimensionless parameter $D_*$ and subsequently derive the explosion energy. We use the ratios between the different density profiles to derive the age and shock speed and tabulate the results for each SNR in Table~\ref{tbl:xrayPropBubble}. The overall effect is an increase in the explosion energy by a factor of a few, as expected; however, except for Kes~79 and Cas~A, the energies remain below the canonical supernova explosion energy of 10$^{51}$~ergs.

\subsection{Explosion Models and Progenitor Mass}   
    
    The nucleosynthesis yields can be used to constrain the progenitor mass. We limit ourselves to core-collapse nucleosynthesis models since our selected SNRs are associated with neutron stars. We compare the nucleosynthesis yields from a range of progenitor mass and explosion energy models to the abundances of the ejecta component obtained from the fitted X-ray data. The abundance ratios are ratios with respects to Si given by ($X$/Si)/($X$/Si)$_{\odot}$, where $X$ is the measured ejecta mass of either O, Ne, Mg, Si, S, Ar, Ca, or Fe with respect to Si and with respect to their solar values from \cite{Wilms}. Here we use 5 sets of models for comparison: one of the first nucleosynthesis spherical explosion models by which many other models are based upon \citep{W95} hereafter labelled WW95; bipolar explosion models \citep{M03} hereafter labelled as M03; hypernova models \citep{N06} hereafter labelled as N06; a suite of spherical explosions using a range of progenitor masses \citep{S16} hereafter labelled S16; a recent set of explosion models using 3 progenitors but with a broad range of explosion energies including a mix of explosion energies per mass in order to simulate asymmetrical explosions \citep{fryer18}, hereafter labelled F18; and finally a recent electron capture supernova model by \cite{J19}, hereafter labelled J19. 
    In addition to the abundance ratios, we estimate the ejecta mass for each element using the method described in \cite{Zhou_2016}, as this would be another good proxy for comparing the data to model predictions. The mass for each element (Z) is estimated from the abundance $A_Z$ and the mass of the shock heated ejecta component of the model $M_{ej}=1.4n_\text{H}m_\text{H}V$ such that $M_Z=A_ZM_{ej}(M_{Z\odot}/M_\odot)$, where $(M_{Z\odot}/M_\odot)$ is the solar mass fraction of the element Z from the solar abundances from \cite{Wilms}. Since the ejecta masses are functions of the filling factor, these estimate act as upper limits on ejecta yields. Table~\ref{tbl:massYields} summarizes the elemental mass yields for our sample, and in the following subsections, we discuss the results for each of the SNRs in our sample.
    
\begin{landscape}
  
    \begin{table}
		\begin{center}
		  \caption{Derived explosion properties assuming each SNR is expanding into a          uniform ambient medium. The subscript $X$ is a substitute variable that refers to 
            the specific SNR's distance $D$ and filling factor $f$. Distances used in the calculations are listed in Table~\ref{tbl:ExpTime}.}
		  \label{tbl:xrayProp}
            \resizebox{\linewidth}{!}{
    		\begin{tabular}{ccccccccc} 
				\hline
				SNR & Emission Measure (EM) & Electron Density (n$_e$) & Ambient Density (n$_0$) & Swept-up Mass (M$_{sw}$) & Shock Velocity (V$_S$) & Free Expansion Age (t$_\text{SNR}$) & Sedov Age (t$_\text{SNR}$) & Explosion Energy (E$_*$) \T \\
				& $f_X$~D$_{X}^{-2}$cm$^{-3}$ & $f_X^{-1/2}$D$_{X}^{-1/2}$cm$^{-3}$ & $f_X^{-1/2}$D$_{X}^{-1/2}$cm$^{-3}$ & $f_X^{1/2}$D$_{X}^{5/2}$M$_\odot$ & km~s$^{-1}$ & D$_{X}$~yr & D$_{X}$~yr &  $\times 10^{50}~f_X^{-1/2}$~D$_{X}^{5/2}$~erg \B \\ 
		        \hline
				
                G15.9+0.2 & $2.9^{+0.3}_{-0.4}\times 10^{58}$ & $1.04^{+0.05}_{-0.07}$ & $0.22^{+0.03}_{-0.01}$ & $8.2^{+0.4}_{-0.6}$ & $1000^{+300}_{-200}$ & 1250 & $2420^{+700}_{-500}$ & $1.29^{+0.04}_{-0.06}$ \T\B \\
                Kes~79 & $7.8\pm{0.1}\times 10^{58}$ & $4.8\pm{0.3}$ & $1.00\pm{0.06}$ & 280$\pm{17}$ & $430^{+40}_{-20}$ & 2430 & $11.8^{+0.5}_{-1.0}$ & $6.9\pm{0.5}$ \B \\
                Cas~A & $4.4\times 10^{55}$ & 240 & 50 & 37 & 1400 & 336 & 470 & 11 \B \\
                Puppis~A & $5.47^{+1.4}_{-0.03}\times 10^{56}$ & $1.07^{+0.54}_{-0.08}$ & $0.22^{+0.11}_{-0.02}$ & $36^{+89}_{-14}$ & $700\pm{100}$ & 2035 & $5500^{+1000}_{-900}$ & $3.0^{+1.5}_{-0.3}$ \B \\
                G349.8+0.2 & $2.9_{-0.2}^{+0.1} \times 10^{59}$ & $6.1\pm{0.2}$ &  $1.27_{-0.04}^{+0.03}$ & $14.2\pm{0.4}$ & $850_{-170}^{+340}$ & 830 & $1960^{+780}_{-390}$ & $1.52_{-0.05}^{+0.04}$ \B \\
                G350.1$-$0.3 & $2.0^{+0.6}_{-0.3}\times 10^{58}$ & $30^{+4}_{-3}$ & $6.3^{+0.9}_{-0.5}$ & $25^{+4}_{-2}$ & $520^{+20}_{-13}$ & 670 & $2570^{+60}_{-110}$ & $1.1^{+0.2}_{-0.1}$ \B \\           
        	\hline
			\end{tabular}}
            \scriptsize \raggedright Note - The free expansion age is a lower age estimate whereas the Sedov age is an upper age estimate. G15.9+0.2: We use the full SNR fit to derive the explosion properties where the soft component was associated with the forward shock. Kes 79: We use region 8 to derive the explosion properties where the soft component is associated with the forward shock. Cas~A: We use a forward shocked CSM region to derive the explosion properties. Puppis~A: We use a centrally located region from the \textit{contbin} chosen regions to derive the explosion properties. G349.8+0.2: We use the full SNR fit to derive the explosion properties where the soft component was associated with the forward shock. G350.1$-$0.3: We use region 4 to derive the explosion properties where it is a one-component model associated with the forward shocked ISM.
		\end{center}
	\end{table}
    \begin{table}
		\begin{center}
		  \caption{Derived explosion properties assuming each SNR is expanding into a wind-    blown bubble formed from its own massive progenitor in a $r^{-2}$ density profile.   The subscript $X$ is a substitute variable that refers to the specific SNR's         distance $D$ and filling factor $f$. Distances used in the calculations are listed   in Table~\ref{tbl:ExpTime}.}
		  \label{tbl:xrayPropBubble}
    		\begin{tabular}{ccccc} 
				\hline
				SNR & Swept-up Mass (M$_{sw}$) & Shock Velocity (V$_S$) & Age (t$_\text{s,wind}$) & Explosion Energy (E$_*$) \T \\
				& $f_X^{1/2}$D$_{X}^{5/2}$M$_\odot$ & km~s$^{-1}$ & D$_{X}$~yr &  $\times 10^{51}~f_X^{-1/2}$~D$_{X}^{5/2}$~erg \B \\ 
		        \hline
				
                G15.9+0.2 & $23^{+1}_{-2}$ & $1400^{+400}_{-300}$ & $3000^{+900}_{-600}$ & $0.31\pm{0.02}$ \B\T \\
                Kes~79    & $790\pm{50}$ & $550^{+50}_{-30}$ & $14700^{+1300}_{-700}$ & $1.6\pm{0.1}$ \B \\
                Cas~A     & 125 & 1900 & 590 & 3.1 \B \\
                Puppis~A  & $104^{+13}_{-3}$ & $1000^{+200}_{-100}$ & $7000\pm{1000}$ & $0.70^{+0.1}_{0.02}$ \B \\
                G349.8+0.2 & $40\pm{1}$ & $1100^{+500}_{-200}$ & $2400^{+1000}_{-500}$ & $0.36^{+.11}_{-0.09}$ \B \\
                G350.1$-$0.3 & $87^{+13}_{-8}$ & $690^{+30}_{-20}$ & $3210^{+130}_{-80}$ & $0.29^{+0.05}_{-0.03}$ \B \\
        	\hline
			\end{tabular}
		\end{center}
	\end{table}
    \begin{table}
		\begin{center}
		  \caption{Elemental mass yields for each SNR. The subscript $X$ is a substitute       variable that refers to the specific SNR's distance $D$ and filling factor $f$. 
            Puppis~A yields were not calculated since we only found small knots with enhanced ejecta. }
		  \label{tbl:massYields}
            \resizebox{\textwidth}{!}{
    		\begin{tabular}{cccccccc} 
				\hline
				SNR & M$_\text{Ne}$ & M$_\text{Mg}$ & M$_\text{Si}$ & M$_\text{S}$ & M$_\text{Ar}$ & M$_\text{Ca}$ & M$_\text{Fe}$ \T \\
                & ($f_X^{1/2}$D$_{X}^{5/2}$~M$_\odot$) & ($f_X^{1/2}$D$_{X}^{5/2}$~M$_\odot$) & ($f_X^{1/2}$D$_{X}^{5/2}$~M$_\odot$) & ($f_X^{1/2}$D$_{X}^{5/2}$~M$_\odot$) & ($f_X^{1/2}$D$_{X}^{5/2}$~M$_\odot$) & ($f_X^{1/2}$D$_{X}^{5/2}$~M$_\odot$) & ($f_X^{1/2}$D$_{X}^{5/2}$~M$_\odot$) \B \\
		      \hline
				
                G15.9+0.2 & ... & $0.039_{-0.004}^{+0.010}$ & $0.033_{-0.002}^{+0.006}$ & $0.038_{-0.003}^{+0.006}$ & $0.018_{-0.003}^{+0.007}$ & $0.084_{-0.003}^{+0.002}$ & $0.14_{-0.02}^{+0.11}$ \T\B \\
                Kes~79 & $0.06_{-0.01}^{+0.02}$ & $0.026_{-0.002}^{+0.005}$ & $0.025_{-0.001}^{+0.004}$ & $0.013_{-0.001}^{+0.002}$ & $0.0009_{-0.0003}^{+0.0008}$ & ... & $0.059_{-0.007}^{+0.026}$ \B \\
                Cas~A & $0.046$ & $0.011$ & $0.038$ & $0.019$ & $0.0047$ & $0.0041$ & $0.062$ \B \\
                G349.8+0.2 & ... & $0.0394_{-0.009}^{+0.042}$ & $0.027_{-0.002}^{+0.008}$ & $0.0097_{-0.0005}^{+0.0034}$ & $0.0016_{-0.0002}^{+0.0045}$ & $0.0010\pm{0.0004}$ & $0.08_{-0.02}^{+0.03}$ \B \\
                G350.1$-$0.3 & ... & $0.020_{-0.001}^{+0.002}$ & $0.0235_{-0.0008}^{+0.0015}$ & $0.0085_{-0.0003}^{+0.0005}$ & $0.0026_{-0.0002}^{+0.0002}$ & $0.0032_{-0.0003}^{+0.0004}$ & $0.029_{-0.004}^{+0.009}$ \B \\
        	  \hline
			\end{tabular}}
		\end{center}
	\end{table} 
\end{landscape}  
    
\subsubsection{G15.9+0.2}
    
\begin{figure*}
		\begin{center}
    		\subfloat[(a) M03 Model]{\includegraphics[angle=0,width=0.5\textwidth,scale=0.5]{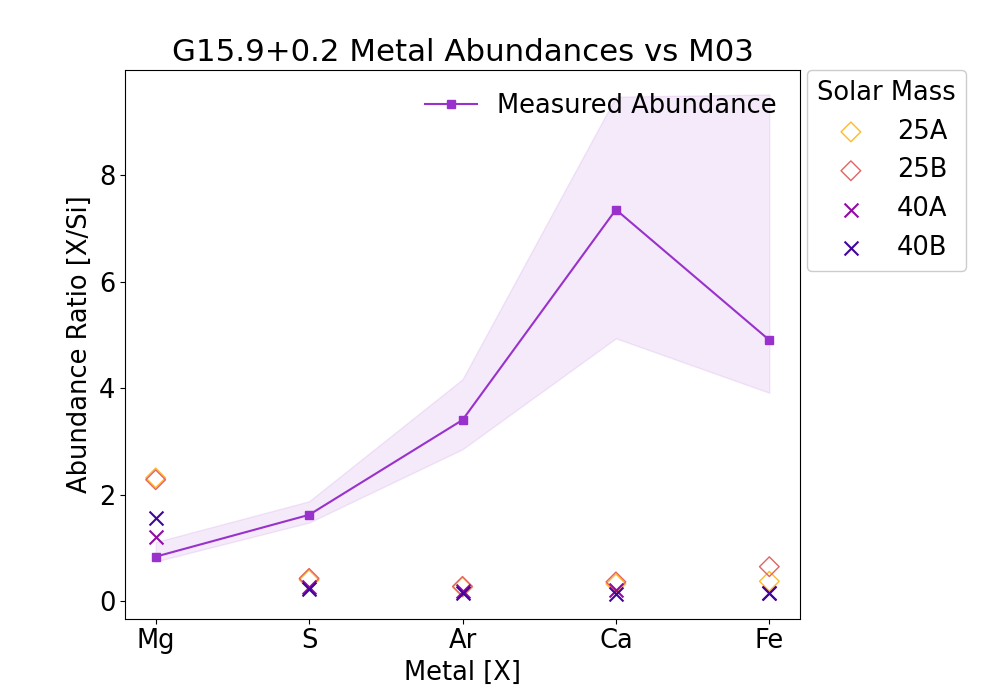}}
    		\subfloat[(b) N06 Model]{\includegraphics[angle=0,width=0.5\textwidth]{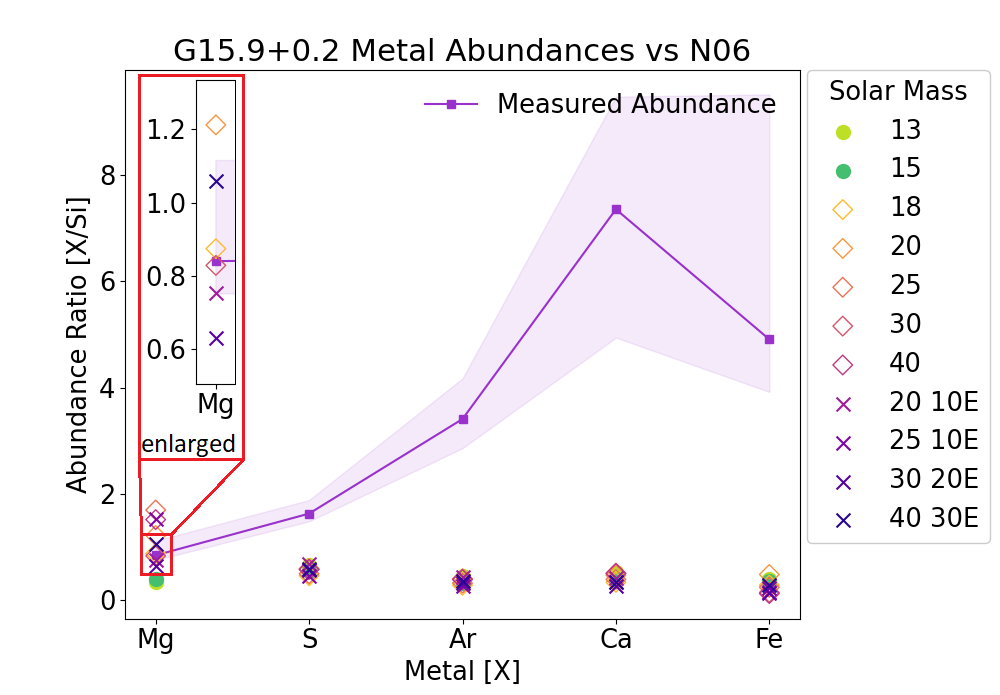}} \\
    		\subfloat[(c) WW95 Model]{\includegraphics[angle=0,width=0.5\textwidth]{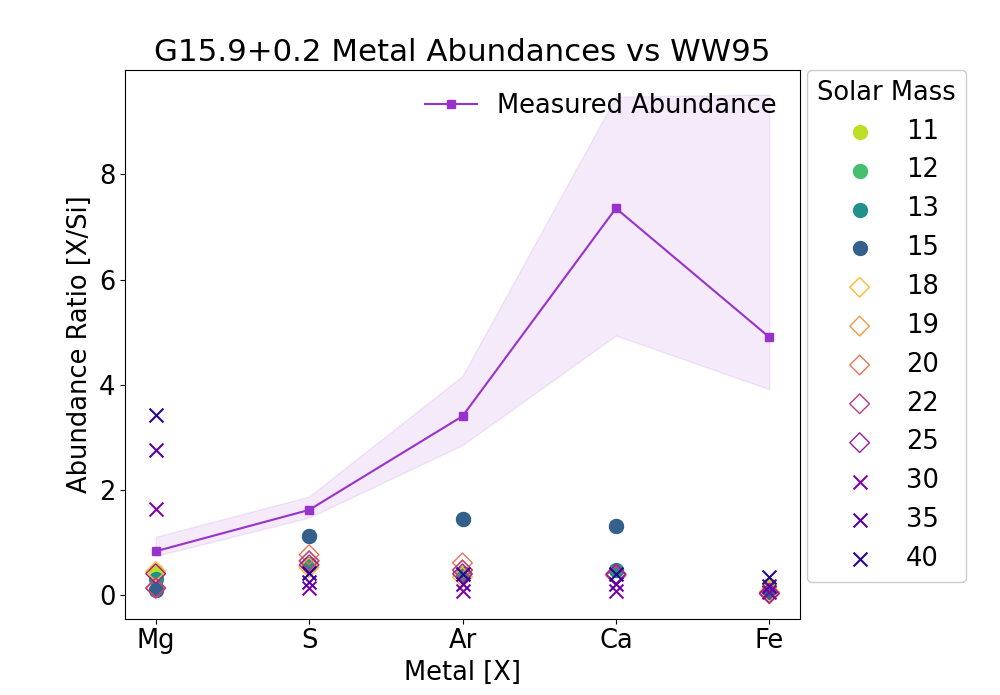}}
    		\subfloat[(d) S16 Model]{\includegraphics[angle=0,width=0.5\textwidth]{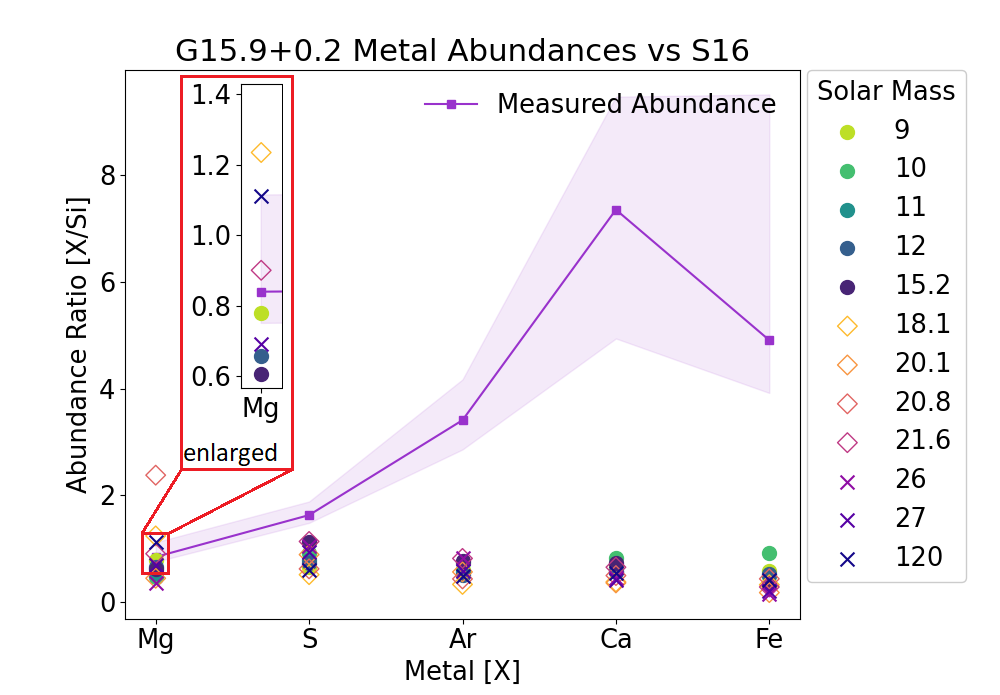}} \\
    		\subfloat[(e) F18 Model]{\includegraphics[angle=0,width=0.5\textwidth]{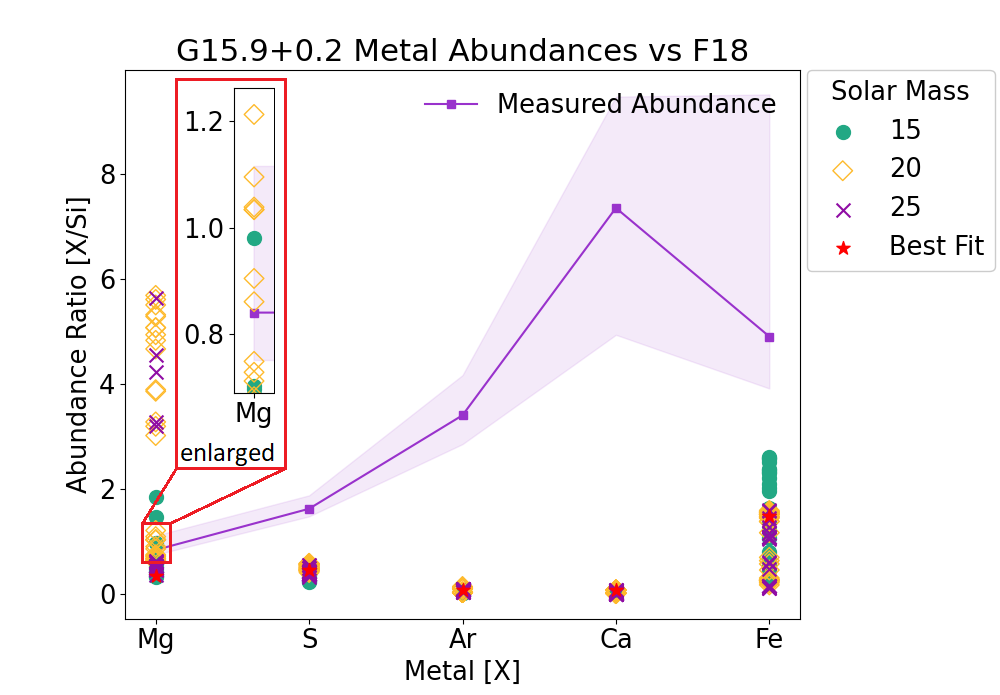}}
    		\subfloat[(f) J19 Model]{\includegraphics[angle=0,width=0.5\textwidth]{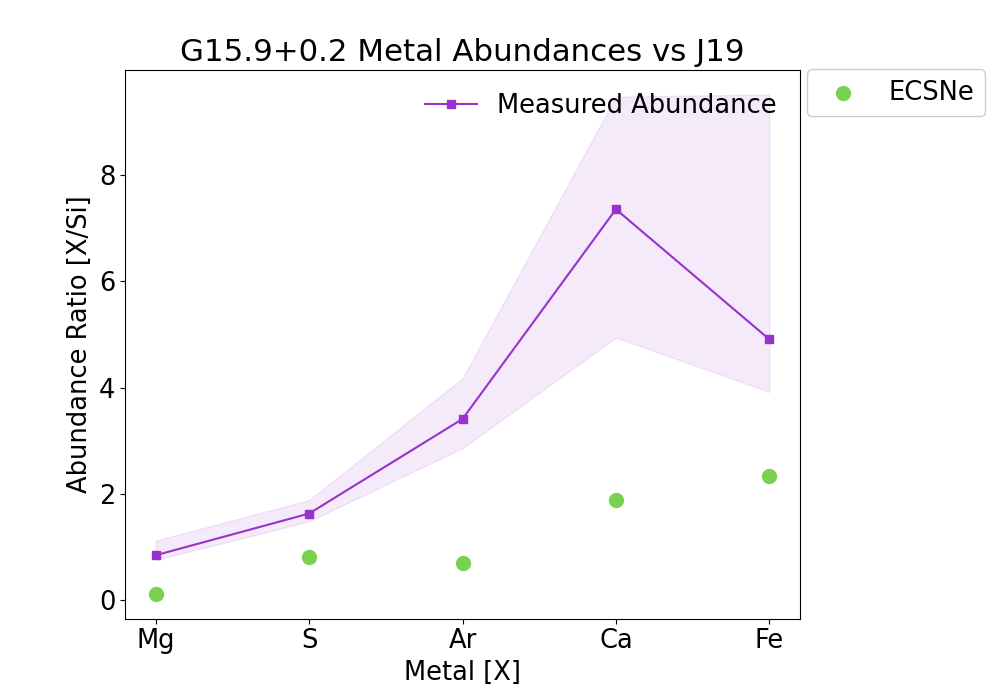}}
    
    		\caption{ G15.9+0.2: Best-fit abundances for Mg, S, Ar, Ca, and Fe relative to Si relative to solar values for the solar abundances from \protect\cite{Wilms}. The core-collapse nucleosynthesis models with predicted relative abundances [X/Si]/[X/Si]$_\odot$ are over-plotted for the models M03, N06, WW95, S16, and F18 with different masses labeled and in units of M$_\odot$. The F18 model has a red star indicating the best fit for the 20~M$_\odot$ model. The J19 model has only a singular data point for each model and represents the usual mass range for an electron-capture supernova (8--10~M$_\odot$). The M03 model has two scenarios where A refers to explosion energy $E_{51}\gtrsim10$ and B refers to explosion energy $E_{51}\approx1$. The N06 model has explosion energy $E_{51}$ for a majority of plots except for the last four plot points which refer to a multiplier of the canonical explosion energy $E_{51}$. }
    		\label{fig:g15Mass}	
		\end{center}
	\end{figure*}

    In Figure~\ref{fig:g15Mass}, We compare our abundance yields to the previously mentioned models in order to estimate the progenitor mass. From the progenitor models shown in Figure~\ref{fig:g15Mass}, there is only good agreement for the Mg value across all models. No model fits all of the data well. The best fit from all models comes from the S16 model, with the 21.6~M$_\odot$ which fits Mg best and is the closest fit for S. No model fits S, Ar, Ca, or Fe values. The WW95 model best fit is that of 18~M$_\odot$ with the 15~M$_\odot$ the next best fit which also trends as the closest fit for the other abundances. The N06 model has a best fit for Mg at 18~M$_\odot$. The F18 model has a best fit from the 20~M$_\odot$ model although the fit can be improved if we considered a mix of two explosion energies. These models add confidence to our S16 best fit. The M03 model only fits for the 40~M$_\odot$ model, but is inconsistent with the other models. We also calculated the ejecta yields for each element and summarize them in Table~\ref{tbl:massYields}. We compare the mass yields to the S16 model yields and found no model predicts all the ejecta mass. Additionally, if we look at individual elements, the progenitor mass estimates increase as we go to heavier elements (Mg to Fe) such that the range of estimates is 12--25.5~M$_\odot$. This could indicate a highly variable filling factor for each element. Since no model fits the data well, we consider a range of values such that the progenitor mass estimate is 18--22~M$_\odot$.
    
    There have been two previous mass estimates by \cite{G15} and \cite{G15XMM}. \cite{G15} using \textit{Chandra} data found enhanced ejecta from the full SNR fit for the elements Mg, Si, S, Ar, Ca, and Fe and compared to the progenitor model from \cite{N97} using the solar abundance values from \cite{anders} at a distance of 10~kpc. They calculated the ejecta mass for each element and found that the progenitor mass $\leq20$~M$_\odot$ which is in agreement with our assessment. \cite{G15XMM} is an \textit{XMM-Newton} study that examined 4 regions and the integrated spectrum spanning the whole SNR with enhanced ejecta for the elements Mg, Si, S, Ar, Ca, and Fe. They compared their abundance yields to the progenitor model \cite{N06} and used the solar abundances values from \cite{Wilms}. Similar to our results, Mg was the only element that fits the models whereas the abundances for S, Ar, Ca, and Fe were much higher than predicted by the progenitor models. \cite{G15XMM} estimated a progenitor mass of 25~M$_\odot$, which is not in agreement with our mass estimate nor the mass estimates from \cite{G15}. We note that both of these studies  reported results from a single progenitor model.

    Finally, we note that this remnant suffers from strong absorption.  Using a single thermal component to fit a broad range of elements (from Mg to Fe) might be an over-simplification. Therefore, we can not rule out a softer x-ray emitting component that might contain some of the lighter, not observed, elements (like O and Ne).  If true, this would then underestimate the abundance of the lighter elements if these elements are from a two-temperature gas. 
    
	\begin{figure*}
		\begin{center}
    		\subfloat[(a) M03 Model]{\includegraphics[angle=0,width=0.5\textwidth,scale=0.5]{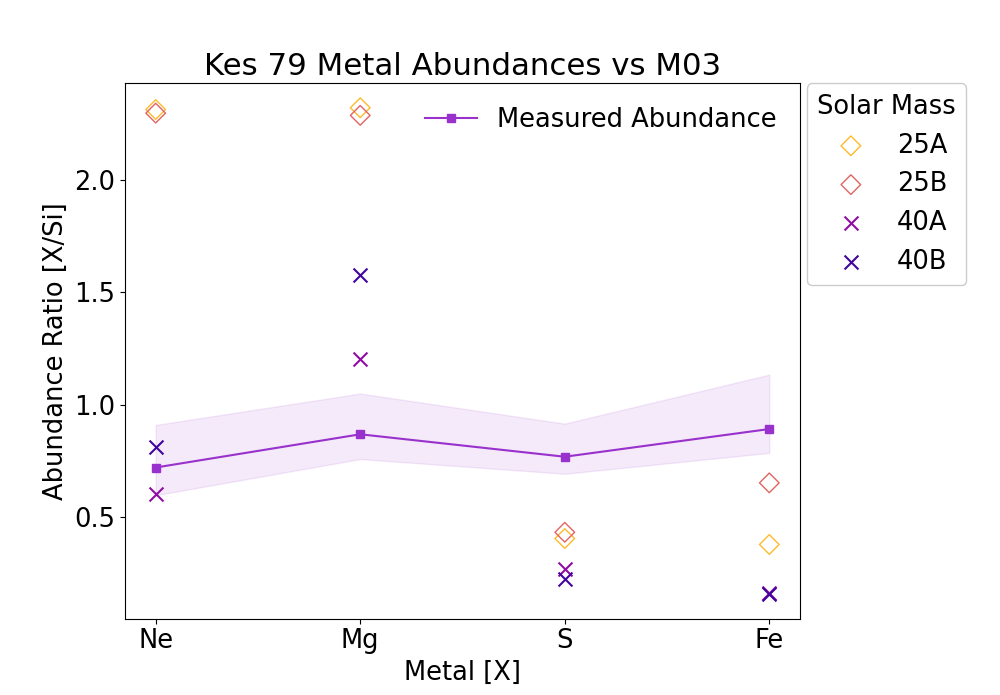}}
    		\subfloat[(b) N06 Model]{\includegraphics[angle=0,width=0.5\textwidth]{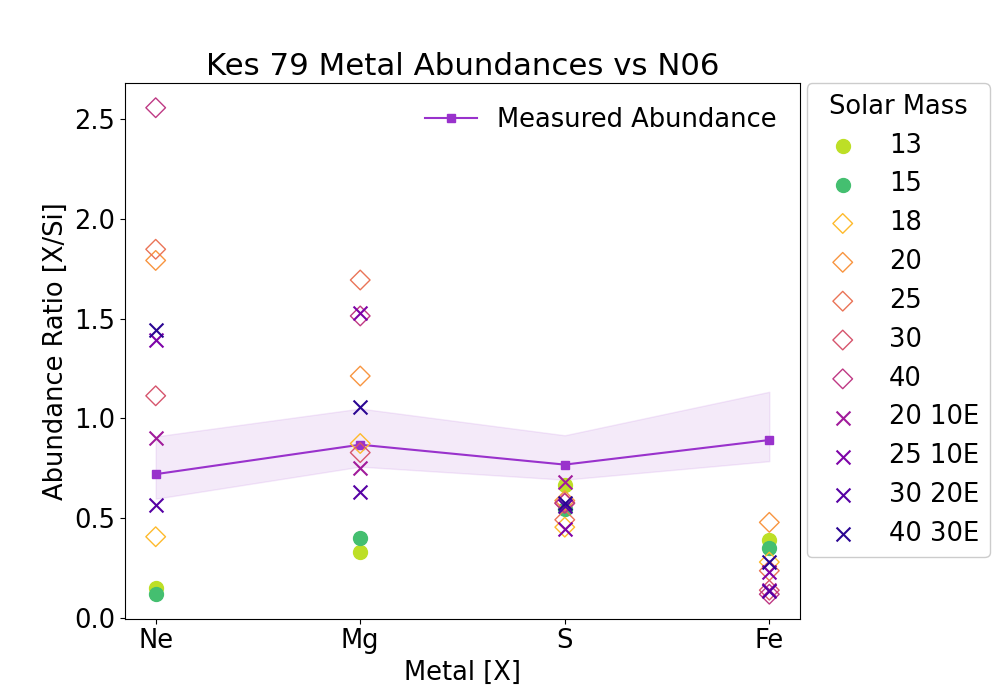}} \\
    		\subfloat[(c) WW95 Model]{\includegraphics[angle=0,width=0.5\textwidth]{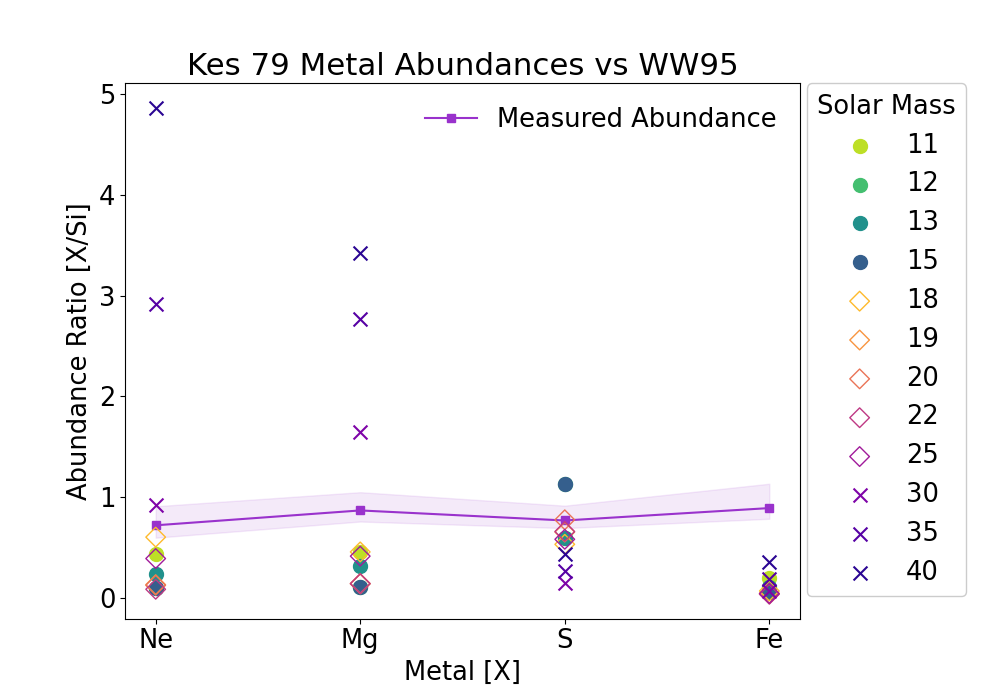}}
    		\subfloat[(d) S16 Model]{\includegraphics[angle=0,width=0.5\textwidth]{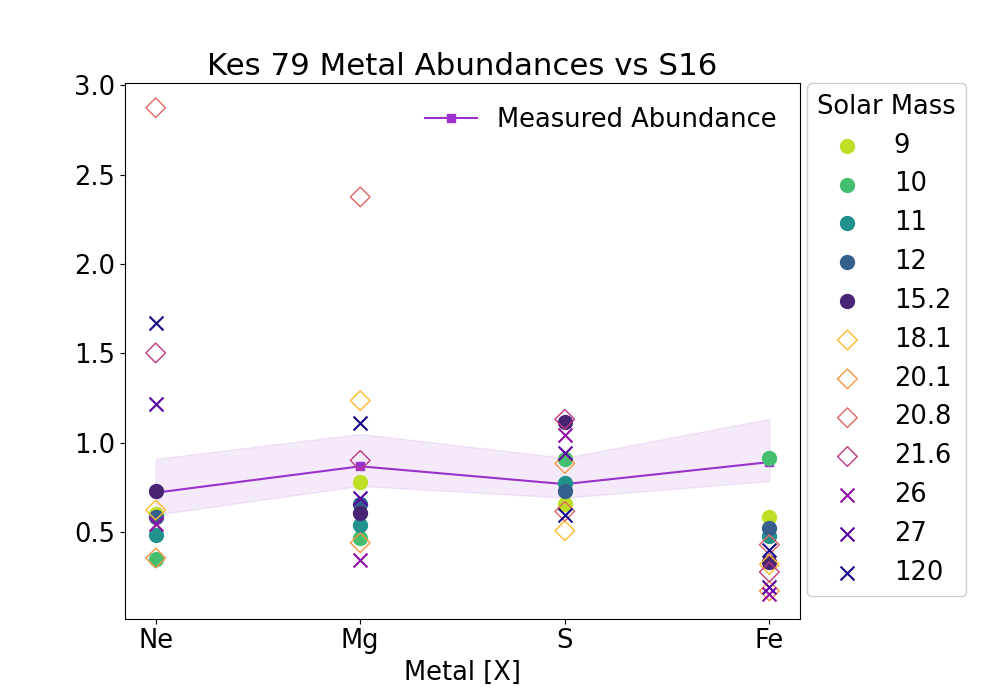}} \\
    		\subfloat[(e) F18 Model]{\includegraphics[angle=0,width=0.5\textwidth]{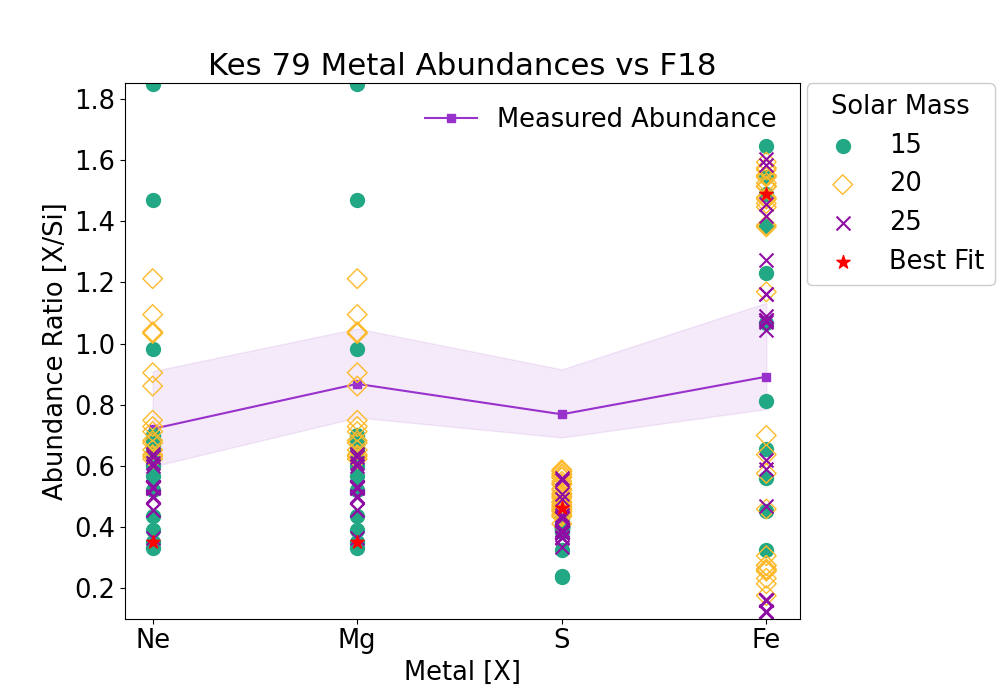}}
    		\subfloat[(f) J19 Model]{\includegraphics[angle=0,width=0.5\textwidth]{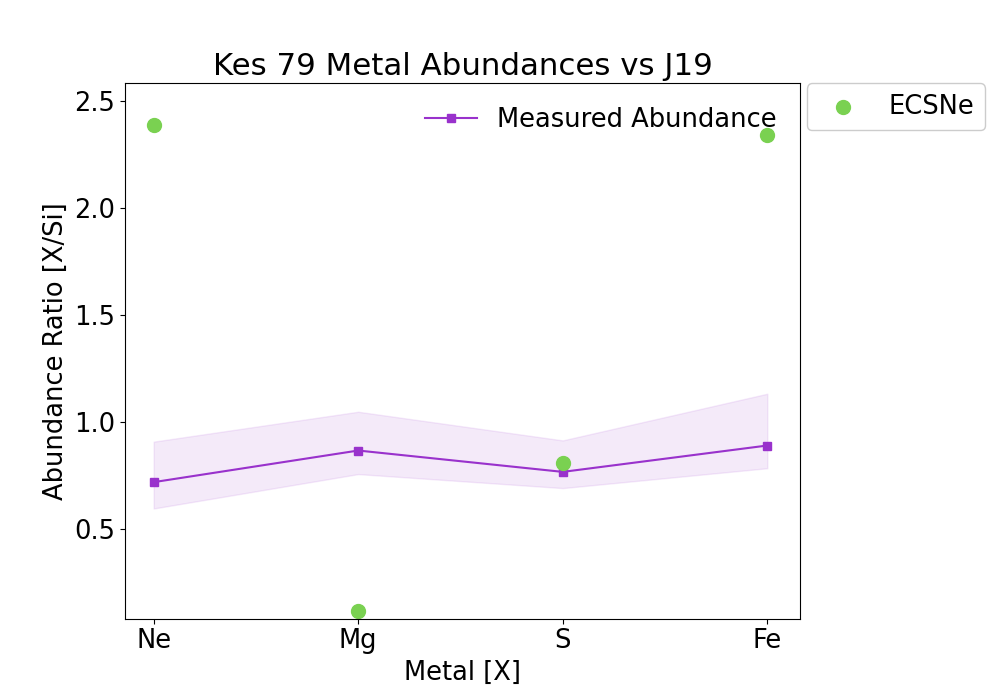}}
    
    		\caption{ Kes~79: Best-fit abundances for Ne, Mg, S, and Fe relative to Si relative to solar values from the solar abundances from \protect\cite{Wilms}. The core-collapse nucleosynthesis models with predicted relative abundances [X/Si]/[X/Si]$_\odot$ are over-plotted for the models M03, N06, WW95, S16, and F18 with different masses labeled and in units of M$_\odot$. The F18 model has a red star indicating the best fit for the 15~M$_\odot$ model. The J19 model has only a singular data point for each model and represents the usual mass range for an ECSN (8--10~M$_\odot$). The M03 model has two scenarios where A refers to explosion energy $E_{51}\gtrsim10$ and B refers to explosion energy $E_{51}\approx1$. The N06 model has explosion energy $E_{51}$ for a majority of plots except for the last four plot points which refer to a multiplier of the canonical explosion energy $E_{51}$. }   
    		\label{fig:kes79Mass}	
		\end{center}
	\end{figure*}     
    \begin{figure*}
		\begin{center}
    		\subfloat[(a) M03 Model]{\includegraphics[angle=0,width=0.5\textwidth,scale=0.5]{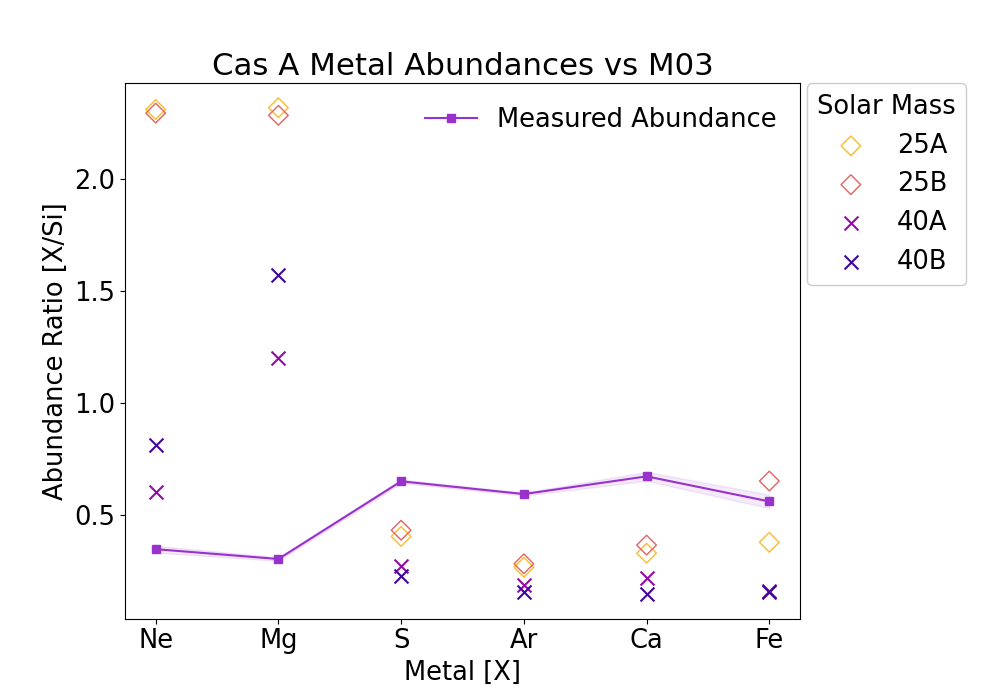}}
    		\subfloat[(b) N06 Model]{\includegraphics[angle=0,width=0.5\textwidth]{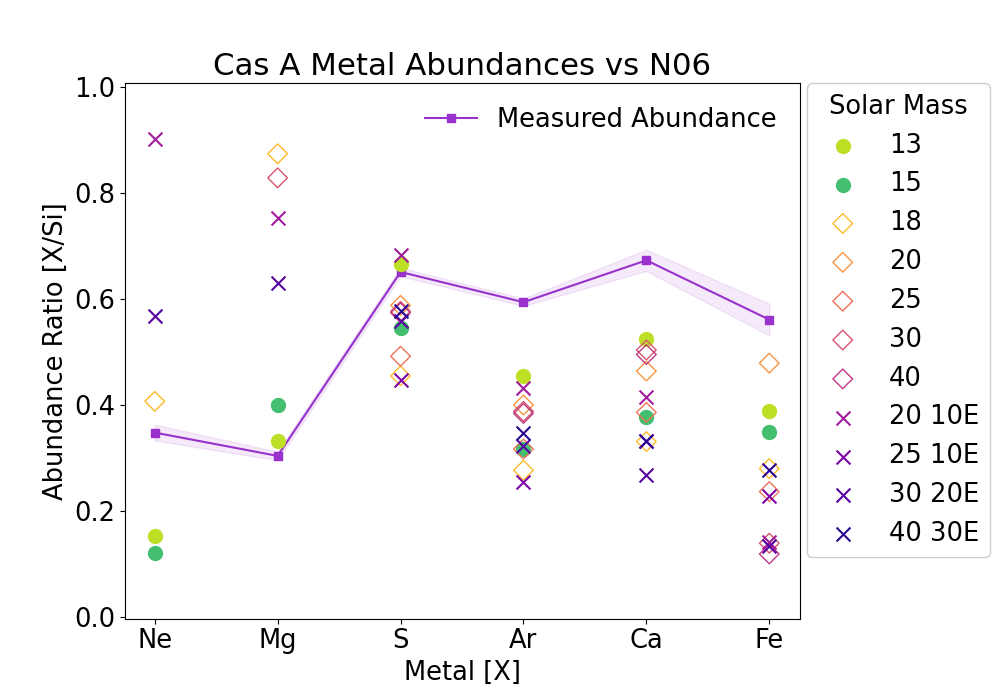}} \\
    		\subfloat[(c) WW95 Model]{\includegraphics[angle=0,width=0.5\textwidth]{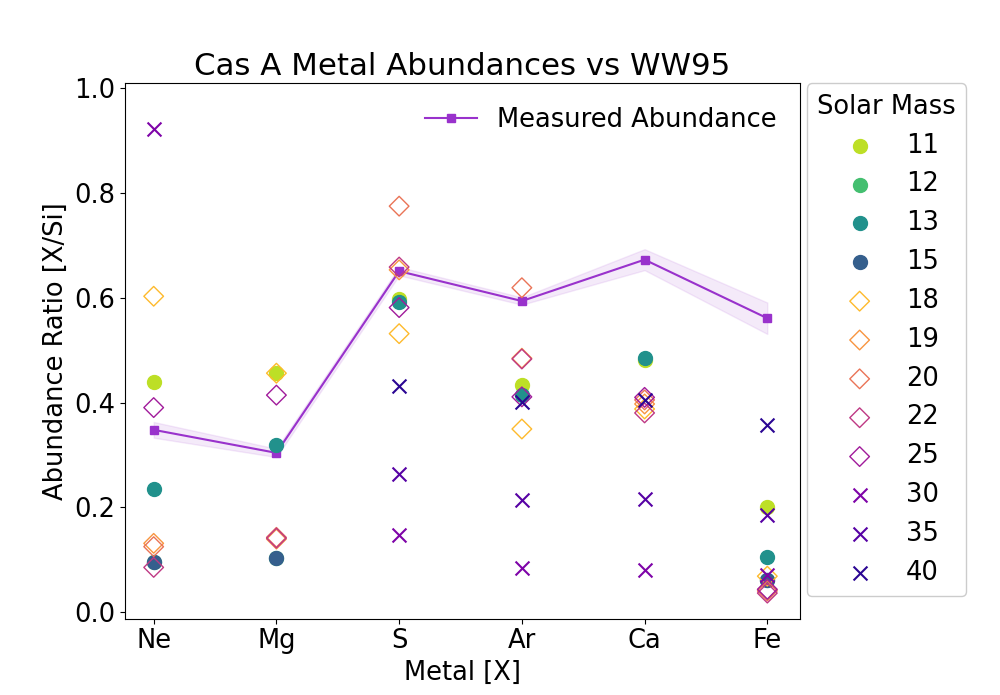}}
    		\subfloat[(d) S16 Model]{\includegraphics[angle=0,width=0.5\textwidth]{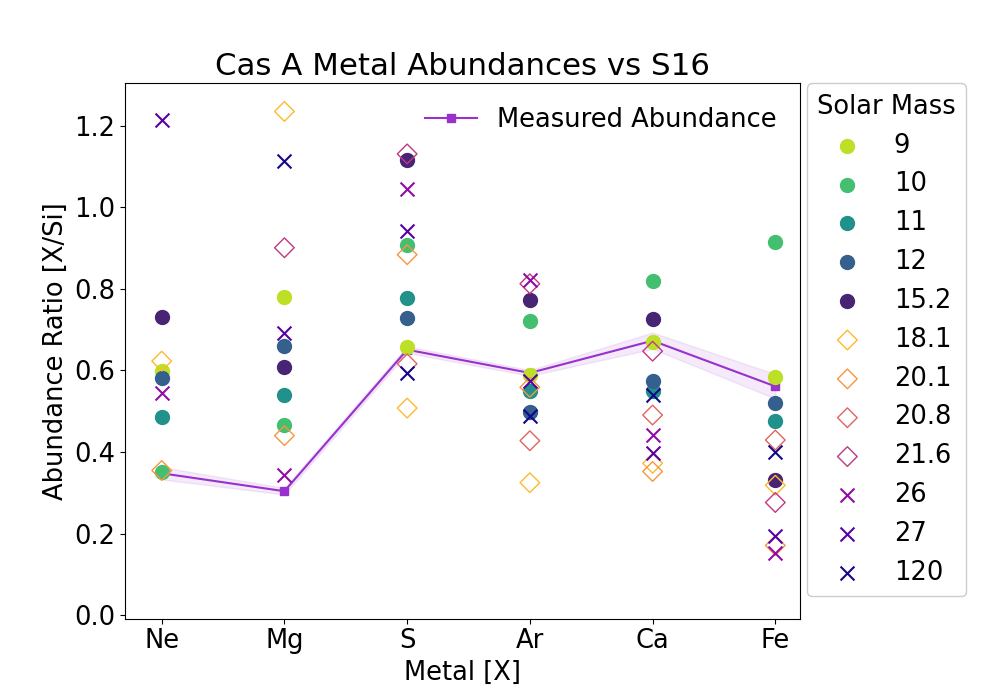}} \\
    		\subfloat[(e) F18 Model]{\includegraphics[angle=0,width=0.5\textwidth]{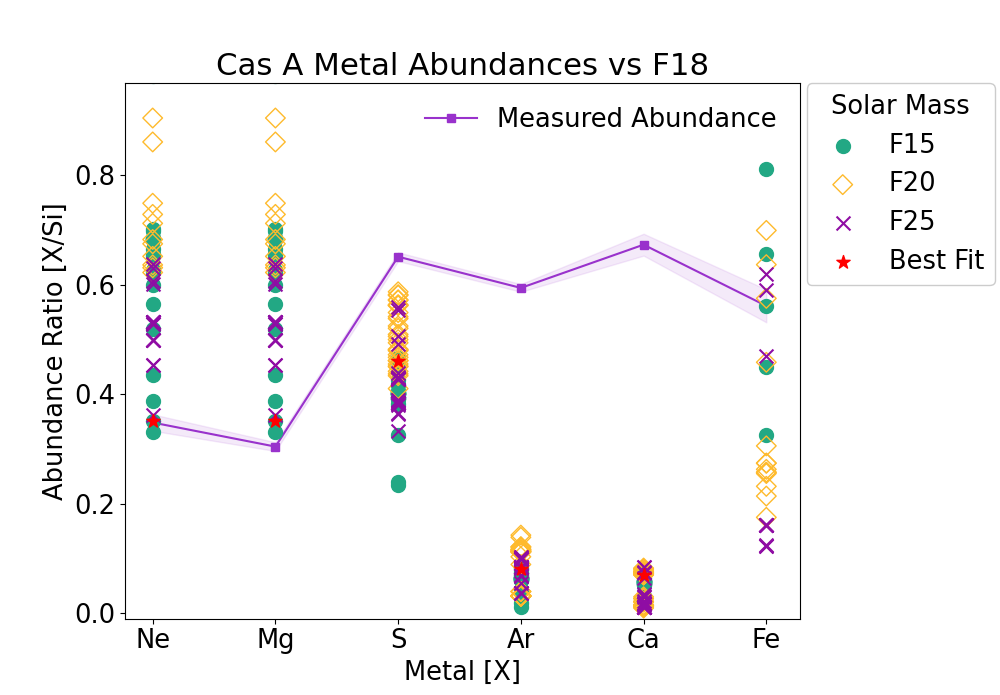}}
    		\subfloat[(f) J19 Model]{\includegraphics[angle=0,width=0.5\textwidth]{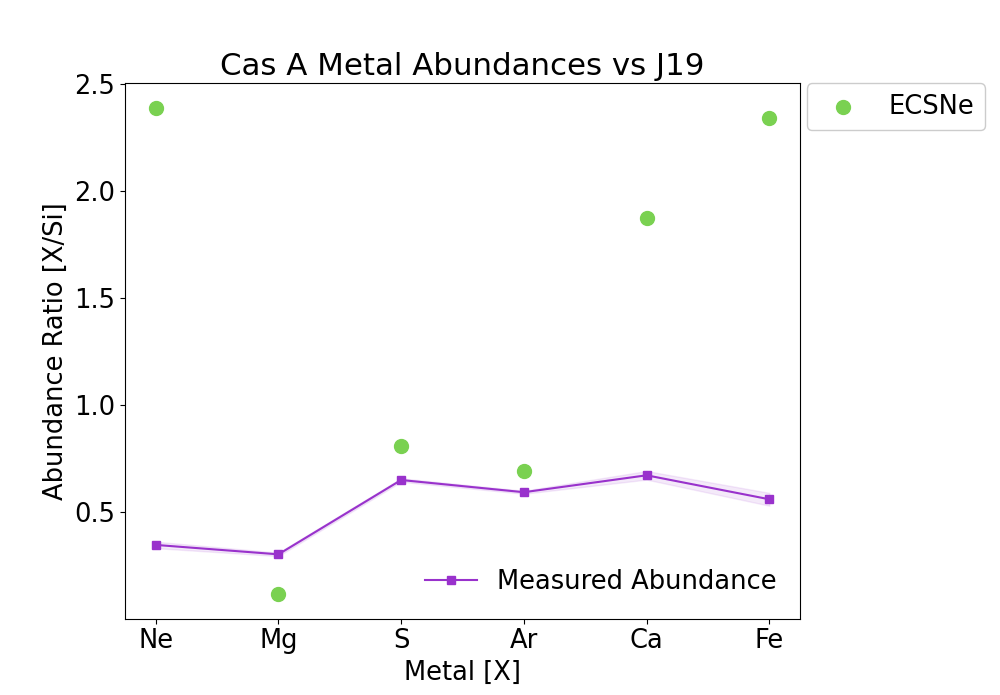}}
    		
    		\caption{ Cas~A: Best-fit abundances for Ne, Mg, S, Ar, Ca, and Fe relative to Si relative to solar values from the solar abundances from \protect\cite{Wilms}. The core-collapse nucleosynthesis models with predicted relative abundances [X/Si]/[X/Si]$_\odot$ are over-plotted for the models M03, N06, WW95, S16, and F18 with different masses labeled and in units of M$_\odot$. The red star indicates the best fit for a single explosion energy, 15~M$_\odot$ model. The J19 model has only a singular data point for each model and represents the usual mass range for an ECSN (8--10~M$_\odot$). The error range in the plots are due to standard error ($95\%$ confidence) to the mean (purple square). The M03 model has two scenarios where A refers to explosion energy $E_{51}\gtrsim10$ and B refers to explosion energy $E_{51}\approx1$. The N06 model has explosion energy $E_{51}$ for a majority of plots except for the last four plot points which refer to a multiplier of the canonical explosion energy $E_{51}$. }
    		\label{fig:casaMass}	
		\end{center}
	\end{figure*} 
 
\subsubsection{Kes~79}
    
    We compare our abundance yields to the previously mentioned models in Figure~\ref{fig:kes79Mass} in order to estimate the progenitor mass. From the models in Figure~\ref{fig:kes79Mass} we find the best fit from S16, with a 9~M$_\odot$ adequately fitting Ne, Mg, and S, but underpredicting the Fe abundance. The 10~M$_\odot$ model fits well for S and Fe, but not Ne, and Mg. We also found good fits from the N06 model for the 20~M$_\odot$ model which matches well for Ne and Mg. The WW95 and M06 models did not fit well for more than 1 element. We also examined the ejecta mass and compared the yields to the S16 model. We find that the S16 model yields were in good agreement for all elemental yields (except for Ar; mass estimates too small), with a progenitor mass range of 11.75--12.25~M$_\odot$. The WW95 model also showed trends to lower mass ejecta in the progenitor mass range of 11--13~M$_\odot$ for Ne, Mg, Si, S, but did not match well with Fe. N06 does not provide progenitor masses lower than 13~M$_\odot$ but trends towards our calculated values the lower the progenitor mass becomes. For the F18 models, a 15~M$_\odot$ progenitor fits well for Ne and we get a slight improvement to the fit when considering a mix of two different explosion energy models. The J19 model only fits well for S. If we examine the ejecta mass yields (see Table~\ref{tbl:massYields}) and compare to the S16 model, we get a range of progenitor masses also on the low mass end in the range 9--13~M$_\odot$. In summary, looking at all the models, this indicates a progenitor mass on the lower mass values, and we estimate it to be in the 9--13~M$_\odot$ range.

    We now compare our estimates to previous studies. \cite{kes79XMM} used two different methods to estimate the progenitor mass. First, from the size of the molecular cavity which found a lower limit $\geq14\pm2$~M$_\odot$, and secondly, from comparing the nucleosynthesis yields from 19 regions spanning the entire SNR for the elements Ne, Mg, Si, S, and Ar compared to the solar abundances from \cite{anders}. The abundance yields were compared to the progenitor models \cite{W95} and 2 different models from \cite{N06}. None of the models were well fit for each element and so estimated the mass based on the 20~M$_\odot$ model from \cite{N06}. Their final estimate was given as 15--20~M$_\odot$, which is higher than our mass range estimate. \cite{kes79Suzaku} examined \textit{Suzaku} data for 2 different regions and a full SNR fit. They compared the abundance yields for Ne, Mg, Al, Si, S, Ar, and Fe to the progenitor model \cite{W95}. None of the models were well fit by the data, but the closest match was a progenitor mass estimate of 30--40~M$_\odot$. None of these papers are in agreement, although the \cite{kes79XMM} value is closest to our estimates. 

\subsubsection{Cas~A}

    We compare our abundance yields to the previously mentioned models in Figure~\ref{fig:casaMass} in order to estimate the progenitor mass. Here we only include those regions with $\chi^2_\nu$<1.5 that were not labeled CSM as described previously. The best fit comes from the S16, 9~M$_\odot$ model which fits the data well for S, Ar, Ca, and Fe but not Ne or Mg, the lighter elements. However, the S16, 10~M$_\odot$ fits well for Ne but no explosion model fits Mg. The other models only fit for a single element. The F18, 15~M$_\odot$ model provides the its best fit for Ne and nearest data point for Mg. Another 15~M$_\odot$ model with a different explosion energy from the best fit also fits for Fe, which indicates a mixing of explosion energy would improve the best fit. Our best estimate for the progenitor mass from the abundance yields comes from the S16 model with a mass of 9--10~M$_\odot$. Next we consider the progenitor mass by comparing the observed mass yields (see Table~\ref{tbl:massYields}) to the progenitor mass yields of the S16 model. The best estimate from the observed yields is 10.25--14.1~M$_\odot$ where the lighter elements tend to be associated with the minimum value of the mass range estimate and the heavier elements with the maximum except for Fe which is best fit in the 12--13~M$_\odot$ range. Given that the mass yields act as an upper limit depending on the filling factor, we conclude a progenitor mass of 9--14~M$_\odot$. 
    
    There have been many previous progenitor mass estimates for Cas~A, with most estimates in the range of 15--25~M$_\odot$\footnote{http://snrcat.physics.umanitoba.ca/SNRrecord.php?id=G111.7m02.1} \citep{SNRCat} which is the assumed mass range associated with Cas~A's progenitor Type IIb (e.g., \cite{2006ApJ...640..891Y}). This is clearly not in agreement with our results, however \cite{CCOFeStudy} re-examined the progenitor mass of all CCSNRs by comparing the abundance ratio Fe/Si, which is sensitive to the carbon-oxygen core mass and therefore the progenitor mass. They found, using the \cite{HwangCasA} data, that the recalculated progenitor mass is actually $<15$~M$_\odot$. Additionally, we compared our Fe/Si ratio of 0.9 to their calculated value of $1.0\pm{0.1}$ (converting for different solar abundance tables) which would find the same conclusion of a progenitor mass is $<15$~M$_\odot$. We are in agreement with these results.

\subsubsection{Puppis~A}

\begin{figure*}
		\begin{center}
    		\subfloat[(a) M03 Model]{\includegraphics[angle=0,width=0.5\textwidth,scale=0.5]{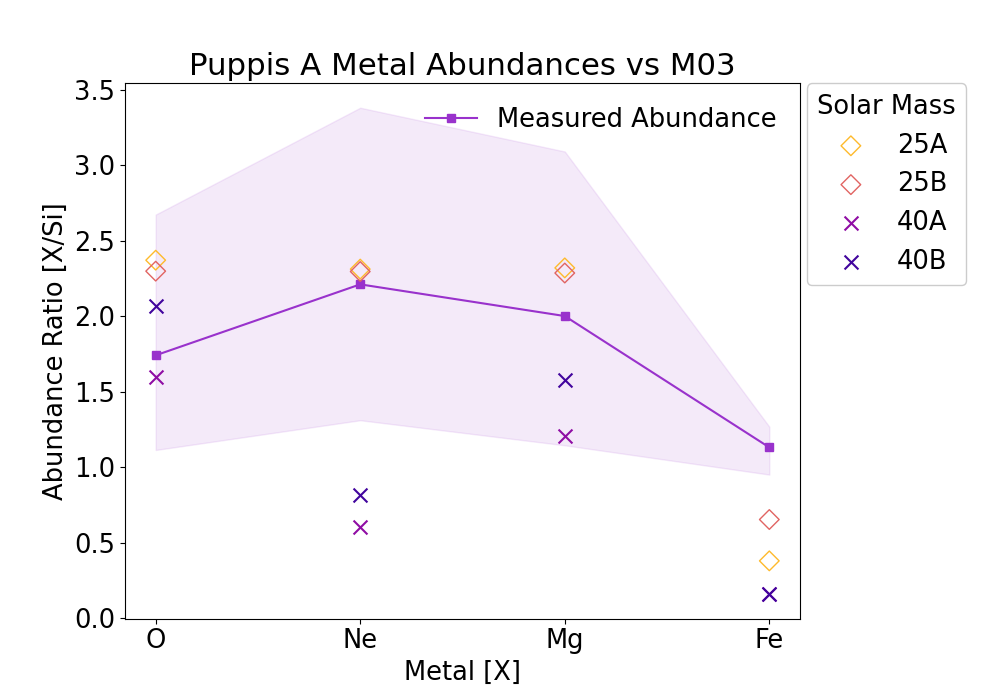}}
    		\subfloat[(b) N06 Model]{\includegraphics[angle=0,width=0.5\textwidth]{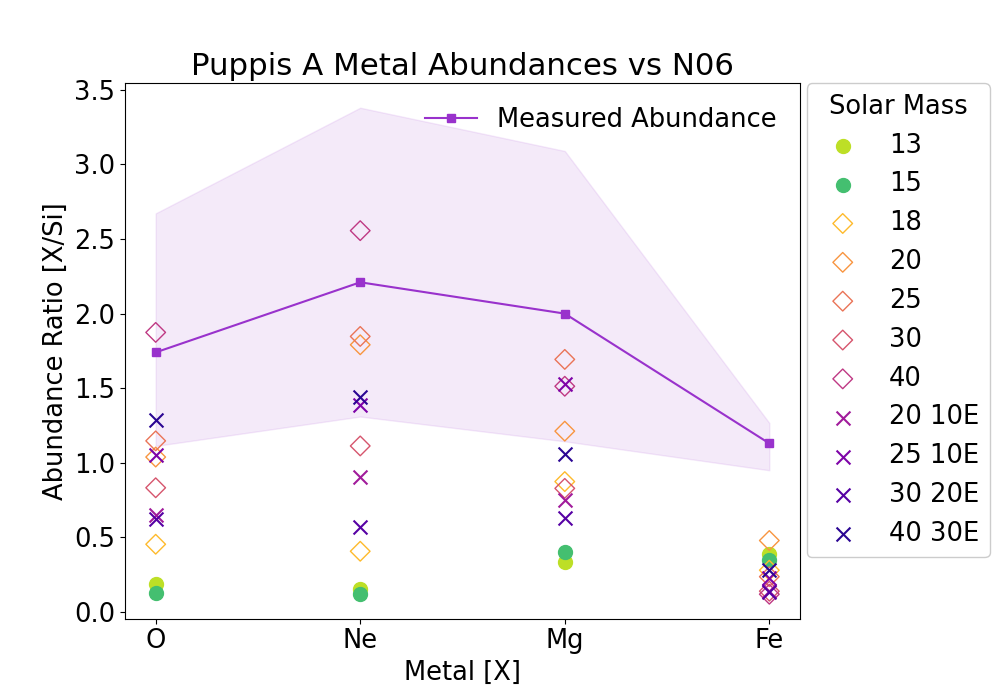}} \\
    		\subfloat[(c) WW95 Model]{\includegraphics[angle=0,width=0.5\textwidth]{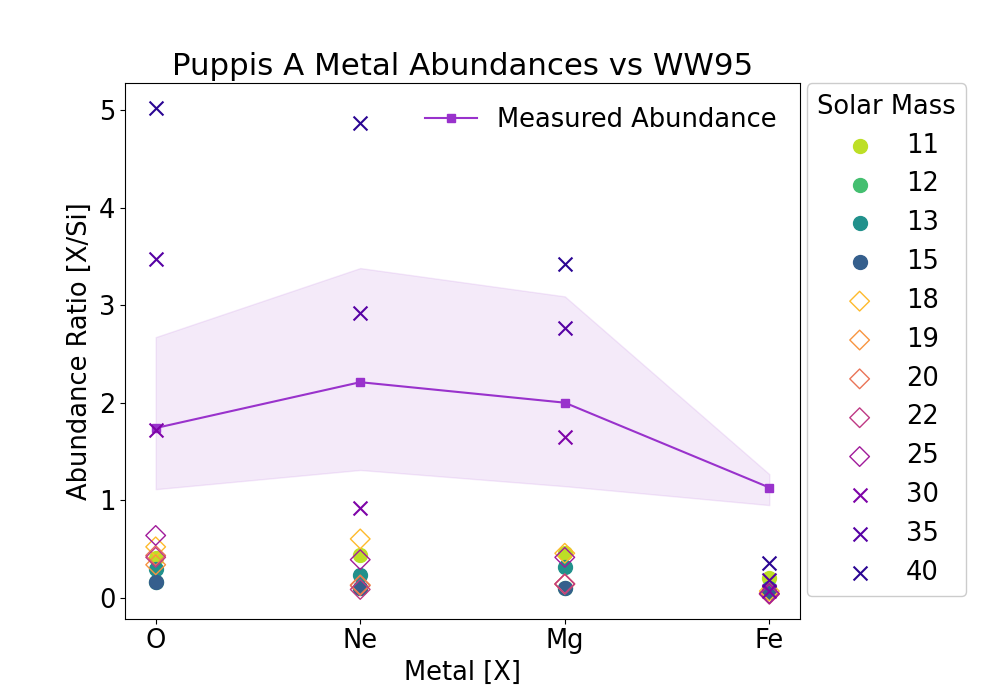}}
    		\subfloat[(d) S16 Model]{\includegraphics[angle=0,width=0.5\textwidth]{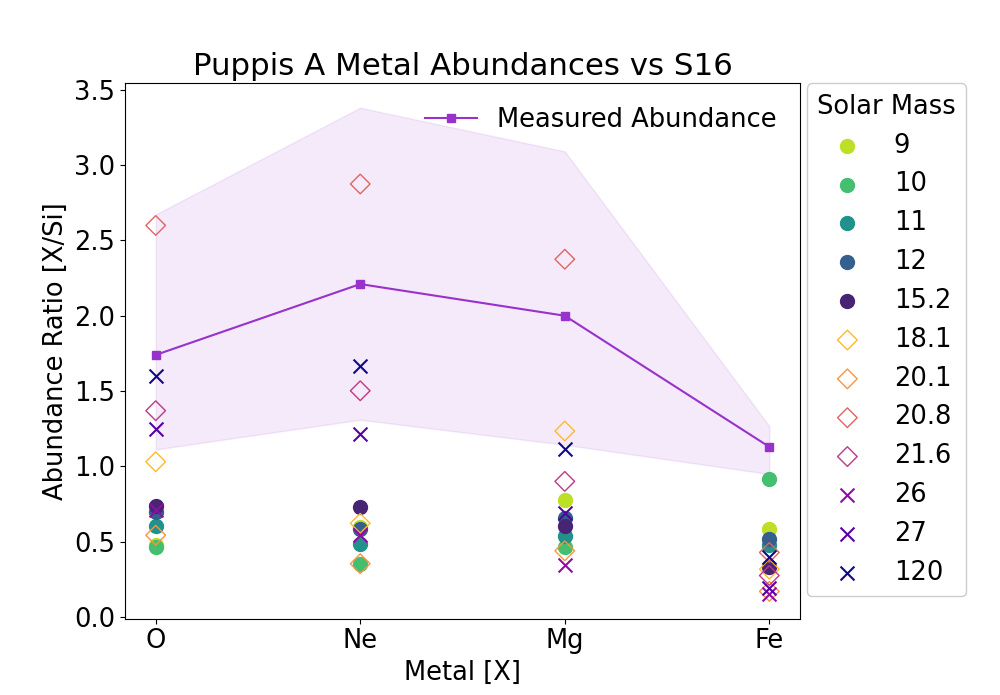}} \\
    		\subfloat[(e) F18 Model]{\includegraphics[angle=0,width=0.5\textwidth]{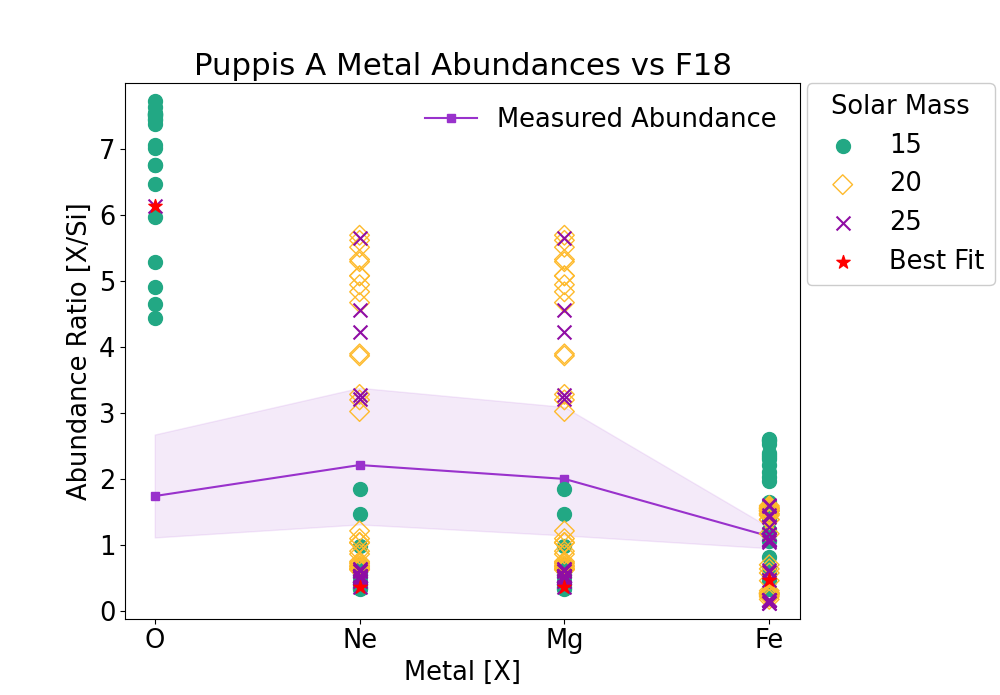}}
    		\subfloat[(f) J19 Model]{\includegraphics[angle=0,width=0.5\textwidth]{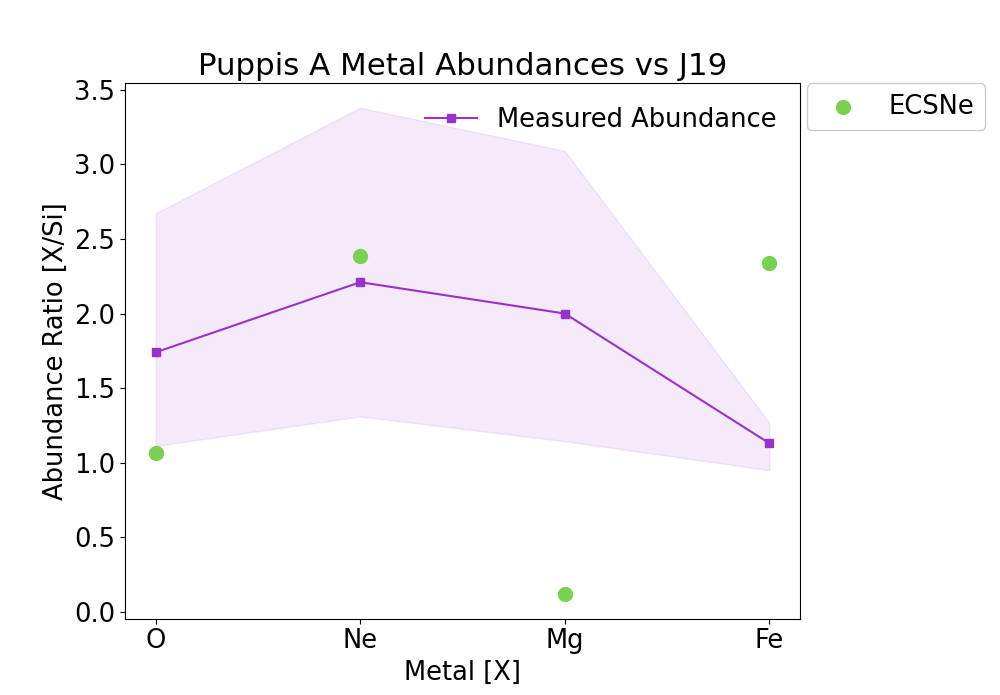}}
    		
    		\caption{ Puppis~A: Best-fit abundances for O, Ne, Mg, and Fe relative to Si relative to solar values from the solar abundances from \protect\cite{Wilms}. The core-collapse nucleosynthesis models with predicted relative abundances [X/Si]/[X/Si]$_\odot$ are over-plotted for the models M03, N06, WW95, S16, and F18 with different masses labeled and in units of M$_\odot$. The F18 model has a red star indicating the best fit for the 25~M$_\odot$ model. The J19 model has only a singular data point for each model and represents the usual mass range for an ECSN (8--10~M$_\odot$). The M03 model has two scenarios where A refers to explosion energy $E_{51}\gtrsim10$ and B refers to explosion energy $E_{51}\approx1$. The N06 model has explosion energy $E_{51}$ for a majority of plots except for the last four plot points which refer to a multiplier of the canonical explosion energy $E_{51}$.  }
    		\label{fig:pupaMass}	
		\end{center}
	\end{figure*}

    We compare our abundance yields to the previously mentioned models in Figure~\ref{fig:pupaMass} in order to estimate the progenitor mass. The best fit model comes from the N06 25~M$_\odot$ model where O, Ne, and Mg are in agreement. Additionally, the 40~M$_\odot$ model also fits well for O, Ne, and Mg and the 25~M$_\odot$ with $10^{52}$~erg explosion energy fits well for O and Ne. The S16, 21.6~M$_\odot$ and 120~M$_\odot$ model shows agreement for O and Ne and from the S16, 20.8~M$_\odot$ model for Ne and Mg. The F18 model finds a best fit from the 25~M$_\odot$ single explosion energy model which fits well for Fe and near the values for Ne and Mg. These results give a wide range of values, and so we consider the agreement between the three models and conclude a progenitor mass of 20--25~M$_\odot$. The ejecta mass in this case was not calculated as the enhanced ejecta could only be found in small knots not detected in our full SNR analysis.     
    
    The progenitor mass has been estimated in previous studies using the aforementioned \textit{Chandra} filaments and knots found in the East by \cite{puppAEastKnots}, and also found agreement with the ``Omega-knot'' as found by \cite{puppAOmegaKnot}. The data were compared to the progenitor model from \cite{R02} for the elements O, Ne, and Mg using the solar abundances from \cite{anders} and found an upper limit of $\leq25$~M$_\odot$. \cite{puppASuzaku} looked the SNR using \textit{Suzaku} data and found enhanced ejecta about the northern knot for O, Ne, Mg, Si, and Fe compared to the solar abundances of \cite{anders}. The enhanced abundances were compared to many progenitor models (\cite{W95,T96,R02,L03}), but found no perfect agreement with the observed mass ratios. The best agreement between observed mass ratios and predicted yields were for progenitor masses of 15--25~M$_\odot$ which agree with our results.

\subsubsection{G349.8+0.2}
    
\begin{figure*}
		\begin{center}
    		\subfloat[(a) M03 Model]{\includegraphics[angle=0,width=0.5\textwidth,scale=0.5]{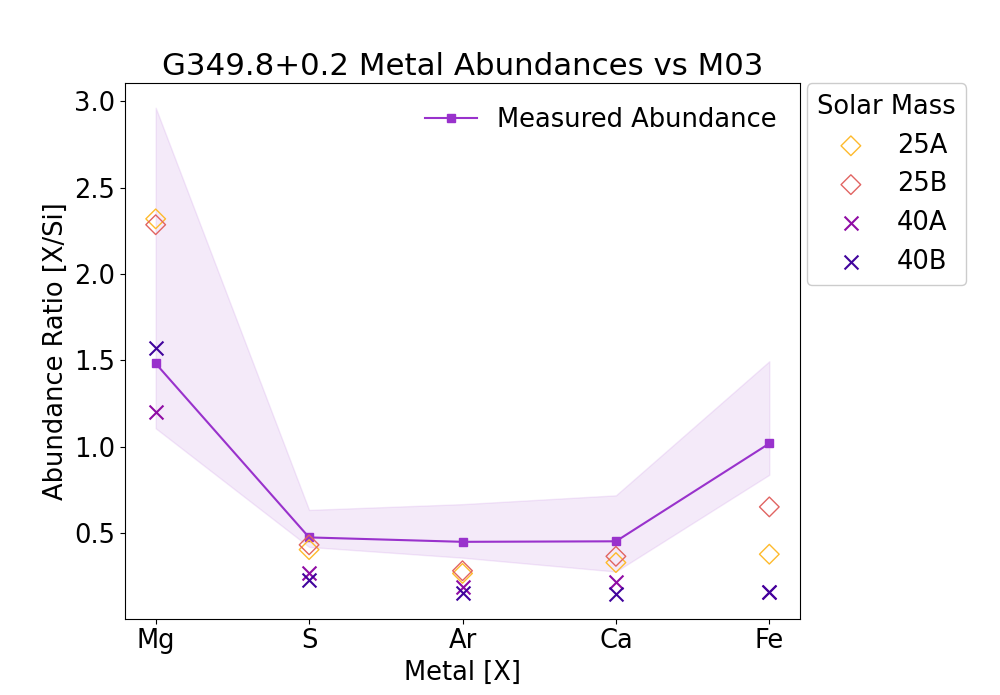}}
    		\subfloat[(b) N06 Model]{\includegraphics[angle=0,width=0.5\textwidth]{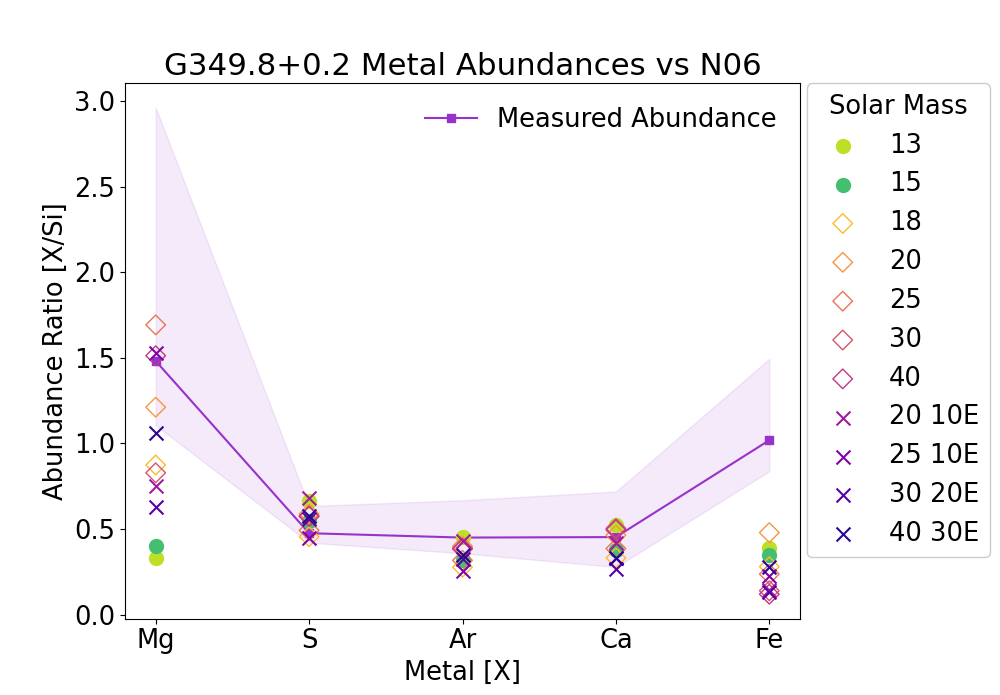}} \\
    		\subfloat[(c) WW95 Model]{\includegraphics[angle=0,width=0.5\textwidth]{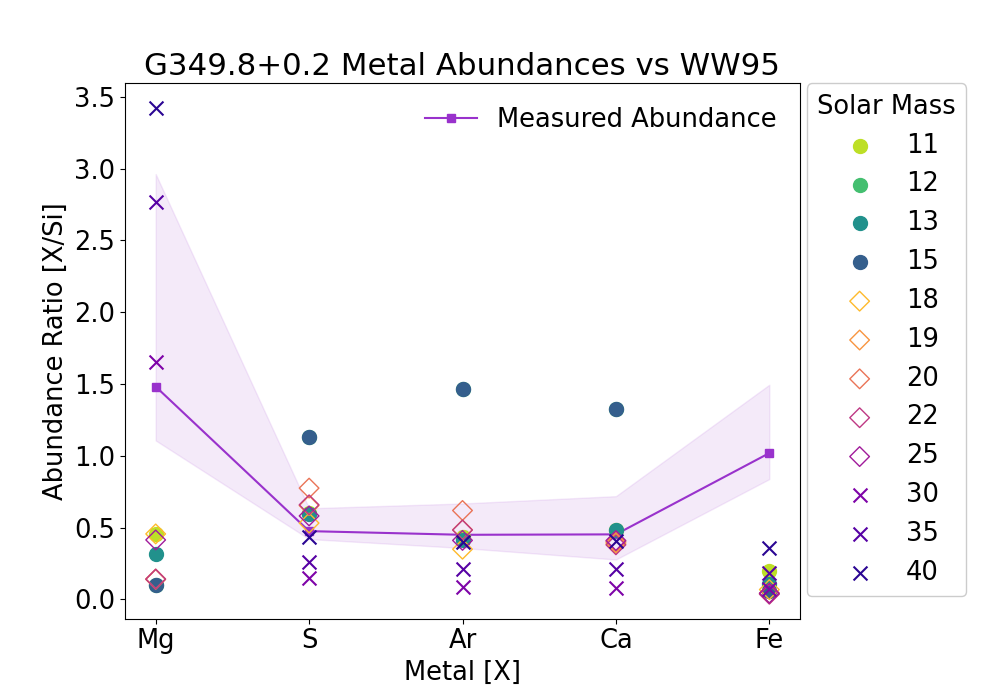}}
    		\subfloat[(d) S16 Model]{\includegraphics[angle=0,width=0.5\textwidth]{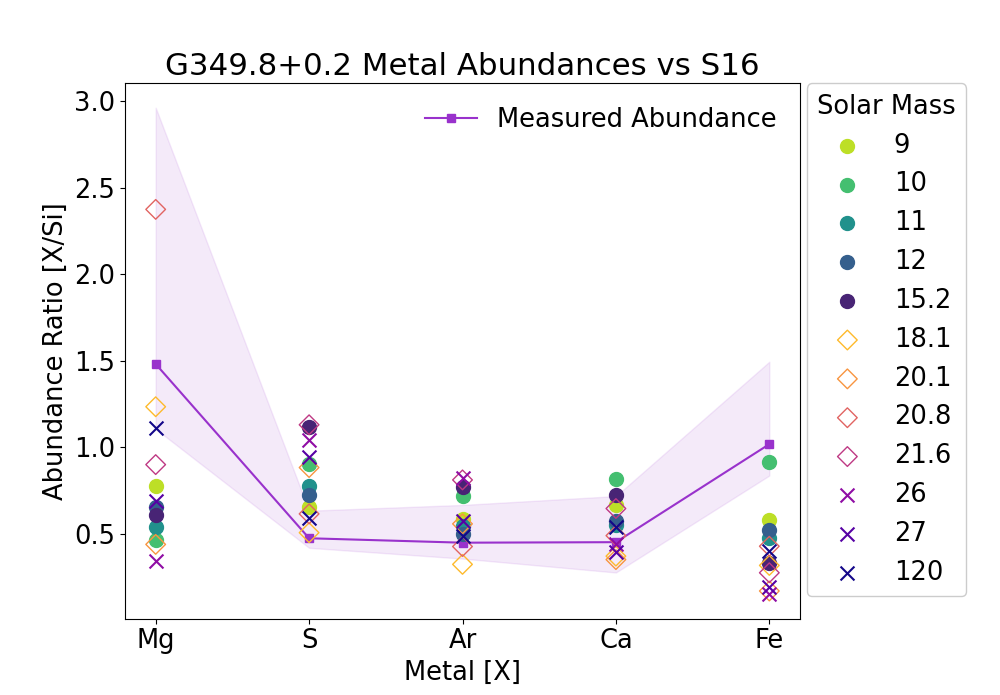}} \\
    		\subfloat[(e) F18 Model]{\includegraphics[angle=0,width=0.5\textwidth]{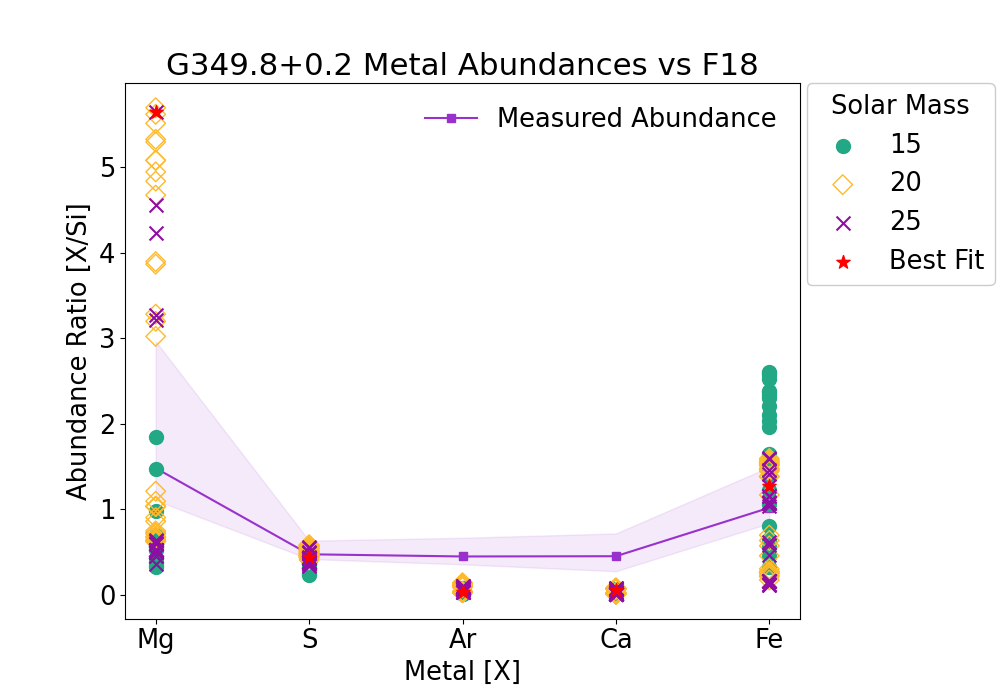}}
    		\subfloat[(f) J19 Model]{\includegraphics[angle=0,width=0.5\textwidth]{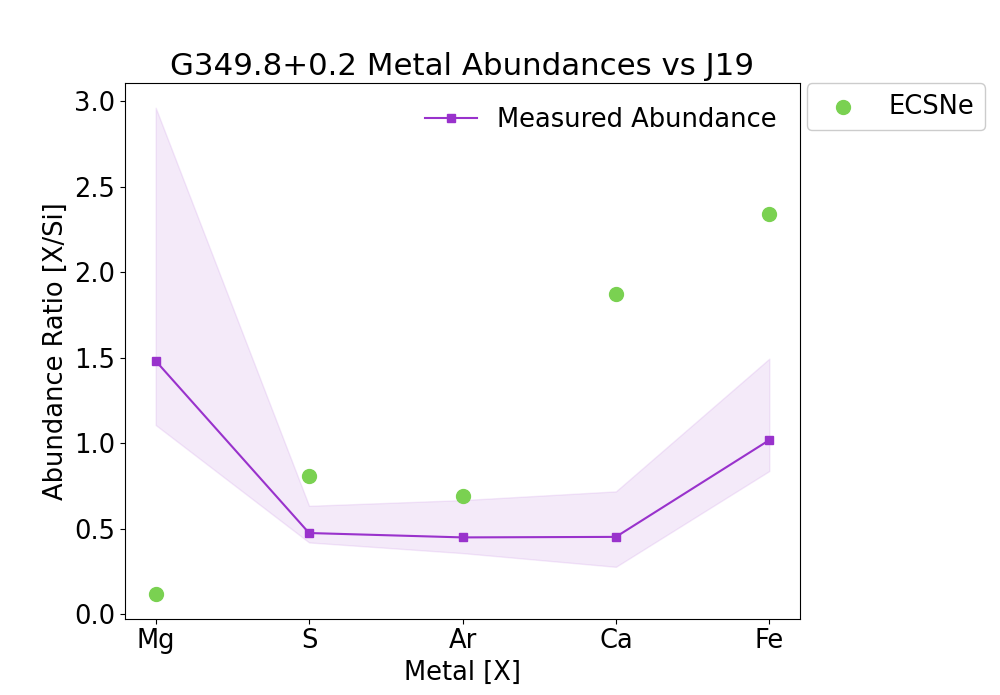}}
    
    		\caption{G349.8+0.2: Best-fit abundances for Mg, Ar, Ca, and Fe relative to Si relative to solar values from the solar abundances from \protect\cite{Wilms}. The core-collapse nucleosynthesis models with predicted relative abundances [X/Si]/[X/Si]$_\odot$ are over-plotted for the models M03, N06, WW95, S16, and F18 with different masses labeled and in units of M$_\odot$. The F18 model has a red star indicating the best fit for the 20~M$_\odot$ model. The J19 model has only a singular data point for each model and represents the usual mass range for an ECSN (8--10~M$_\odot$). The M03 model has two scenarios where A refers to explosion energy $E_{51}\gtrsim10$ and B refers to explosion energy $E_{51}\approx1$. The N06 model has explosion energy $E_{51}$ for a majority of plots except for the last four plot points which refer to a multiplier of the canonical explosion energy $E_{51}$. } 
    		\label{fig:g349Mass}	
		\end{center}
	\end{figure*}

    We compare our abundance yields to the previously mentioned models in Figure~\ref{fig:g349Mass} in order to estimate the progenitor mass. From the plots in Figure~\ref{fig:g349Mass} the best fit is from S16, where Mg, S, Ar and Ca are well fit by the 18.1--21.1~M$_\odot$. No model fits all abundances. The M06 model fits 25~M$_\odot$ progenitor for Mg, S, and Ca. The N06 and WW95 model have no multiple abundance values well fit across multiple elements. The F18, 20~M$_\odot$ model fits well for S and Fe. Looking at all the models, we find a best estimate for the progenitor mass of 18--25~M$_\odot$. The ejecta mass estimates (see Table~\ref{tbl:massYields}) indicate a lower mass estimate of 10--13~M$_\odot$ from the S16 model.
    
    A previous mass estimate finds a progenitor of mass 13--15~M$_\odot$ by calculating the ejecta mass for Si and S \citep{G349}. There is another mass estimate by \cite{distG349&G350} using \textit{Suzaku} data. The full SNR was fit with a 4-component model and found to have enhanced ejecta for Mg and Ni and subsolar-to-solar values for Al, Si, S, Ar, Ca, and Fe compared to solar abundances from \cite{anders}. The observed abundance ratios were compared to \cite{W95} and were found to be best fit by a progenitor mass of 35--40~M$_\odot$. None of these previous mass estimates agree with our results although the \cite{G349} estimates a low progenitor mass from the ejecta yields similar to our estimate from the yields.

\subsubsection{G350.1$-$0.3}
    
\begin{figure*}
		\begin{center}
    		\subfloat[(a) M03 Model]{\includegraphics[angle=0,width=0.5\textwidth,scale=0.5]{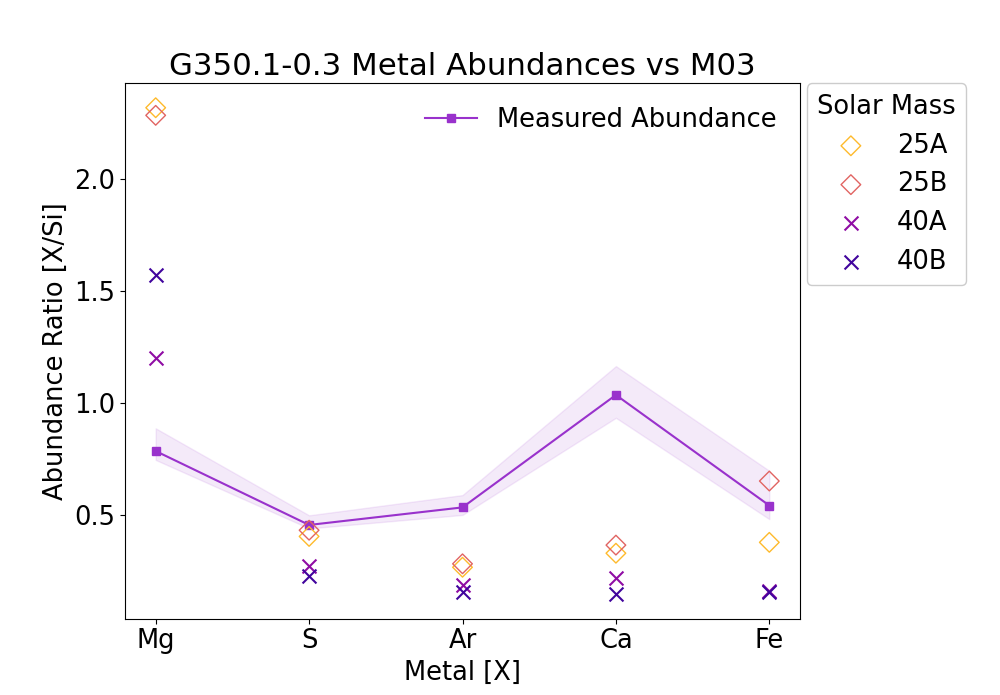}}
    		\subfloat[(b) N06 Model]{\includegraphics[angle=0,width=0.5\textwidth]{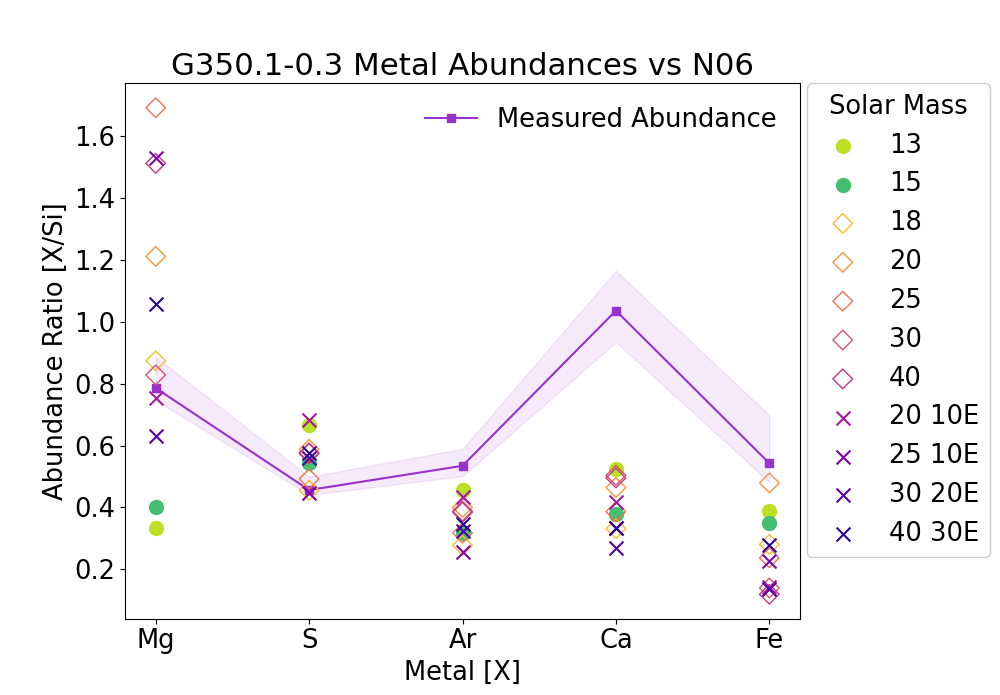}} \\
    		\subfloat[(c) WW95 Model]{\includegraphics[angle=0,width=0.5\textwidth]{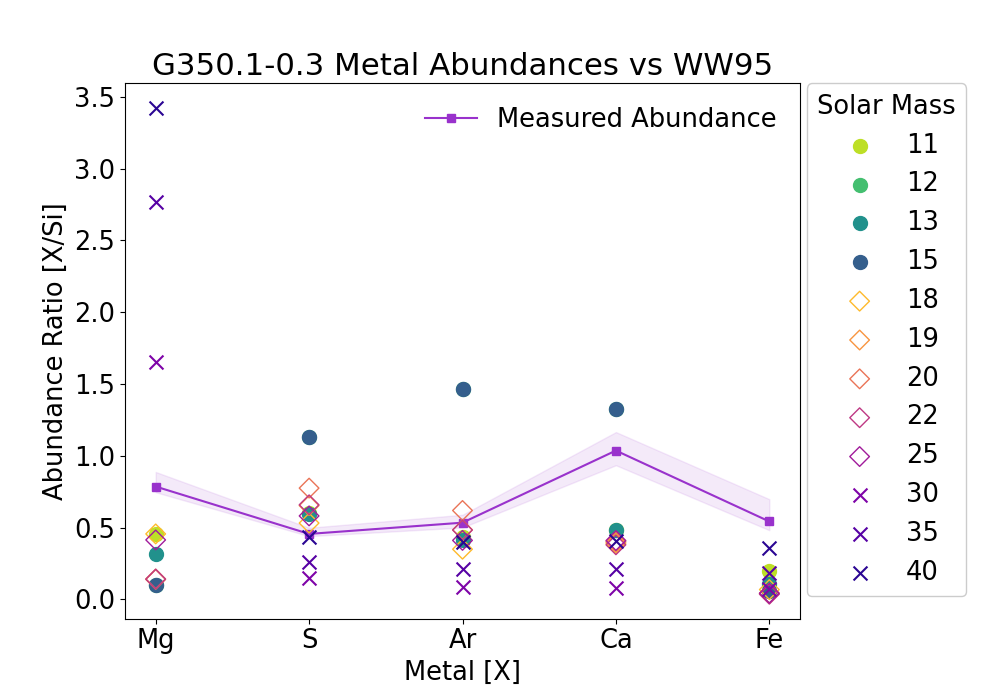}}
    		\subfloat[(d) S16 Model]{\includegraphics[angle=0,width=0.5\textwidth]{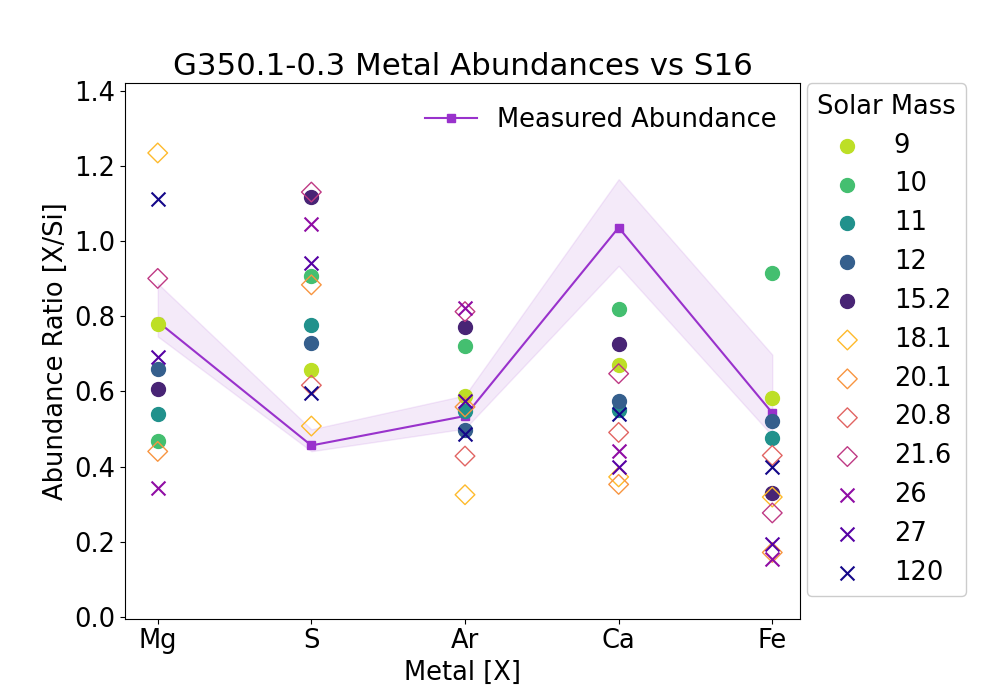}} \\
    		\subfloat[(e) F18 Model]{\includegraphics[angle=0,width=0.5\textwidth]{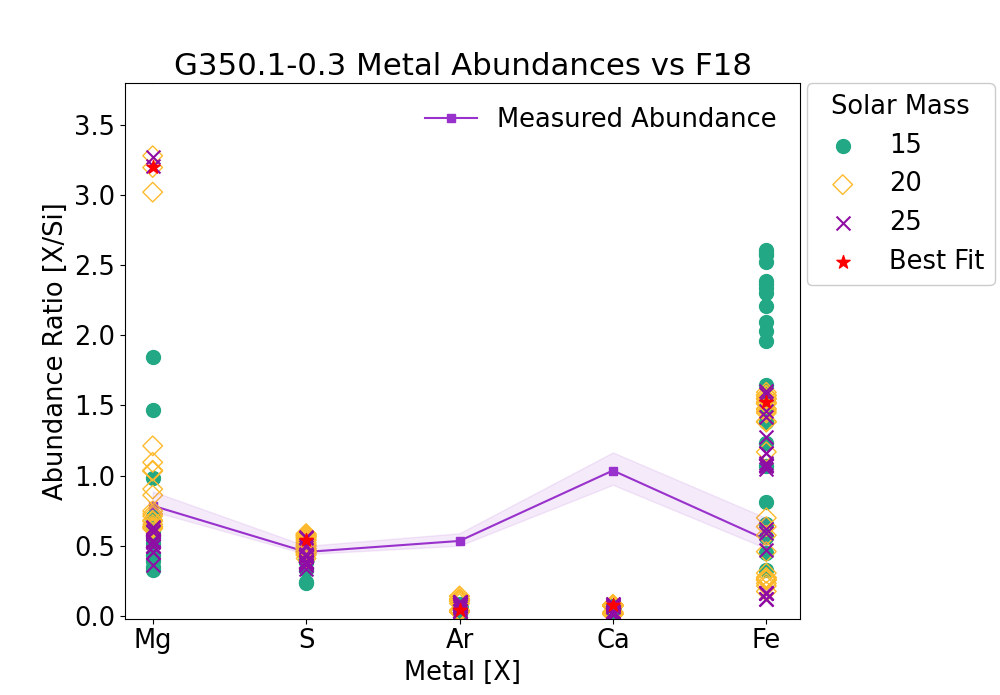}}
    		\subfloat[(f) J19 Model]{\includegraphics[angle=0,width=0.5\textwidth]{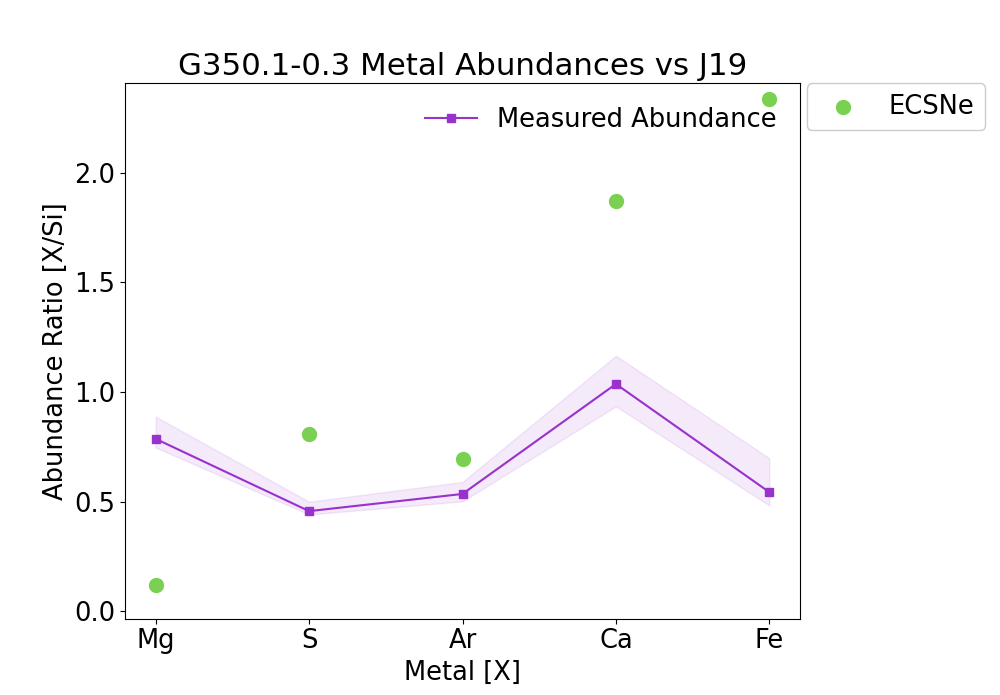}}
    
    		\caption{G350.1--0.3: Best-fit abundances for Mg, Ar, Ca, and Fe relative to Si relative to solar values from the solar abundances from \protect\cite{Wilms}. The core-collapse nucleosynthesis models with predicted relative abundances [X/Si]/[X/Si]$_\odot$ are over-plotted for the models M03, N06, WW95, S16, and F18 with different masses labeled and in units of M$_\odot$. The F18 model has a red star indicating the best fit for the 20~M$_\odot$ model. The J19 model has only a singular data point for each model and represents the usual mass range for an ECSN (8--10~M$_\odot$). The M03 model has two scenarios where A refers to explosion energy $E_{51}\gtrsim10$ and B refers to explosion energy $E_{51}\approx1$. The N06 model has explosion energy $E_{51}$ for a majority of plots except for the last four plot points which refer to a multiplier of the canonical explosion energy $E_{51}$. } 
    		\label{fig:g350Mass}	
		\end{center}
	\end{figure*}

    We compare our abundance yields to the previously mentioned models in Figure~\ref{fig:g350Mass} in order to estimate the progenitor mass. The best fit from Figure~\ref{fig:g350Mass} is from the S16, 9~M$_\odot$ model which fits for Mg, Ar, and Fe. The S16, 12~M$_\odot$ model also fits for Ar and Fe. The M03, 25~M$_\odot$B model fits for the S and Fe abundances, the N06, 18~M$_\odot$ model which fits for Mg and S, and the F18, 20~M$_\odot$ model fits for S and Fe with an improvement to the fit if we consider 2 explosion energies.  We then have 2 distinct estimates for the progenitor mass as 9--12~M$_\odot$ from the S16 model and 18--25~M$_\odot$ from the M03, N06, and F18 models. If we consider the mass yield estimates (see Table~\ref{tbl:massYields}) in comparison to the S16 model, we get an upper limit of 11--12~M$_\odot$ which is consistent with our estimates from the abundances. Therefore, we conclude a a progenitor mass of 9--12~M$_\odot$. 
    
    There has been one previous mass estimate by \cite{distG349&G350} using \textit{Suzaku} data. The full SNR was fit with a 4-component model and found enhanced ejecta for Mg, Al, Si, S, Ar, Ca, Fe, and Ni compared to the solar abundances from \cite{anders}. The observed abundance ratios were compared to \cite{W95} and found the data were best fit by a progenitor mass of 15--25~M$_\odot$, which did not fit well for Mg, Al, S, fe or Ni. They also estimated a total ejecta mass of 13~$f_h^{1/2}$D$_{9}^{5/2}$~M$_\odot$ which is comparable to our total ejecta mass of 17~$f_h^{1/2}$D$_{9}^{5/2}$~M$_\odot$. However, our progenitor mass estimate is not in agreement. 
  

\subsection{Abundance Ratios in Comparisons to Fe Core Mass}

\begin{figure*}
    	\includegraphics[angle=0,width=0.8\textwidth]{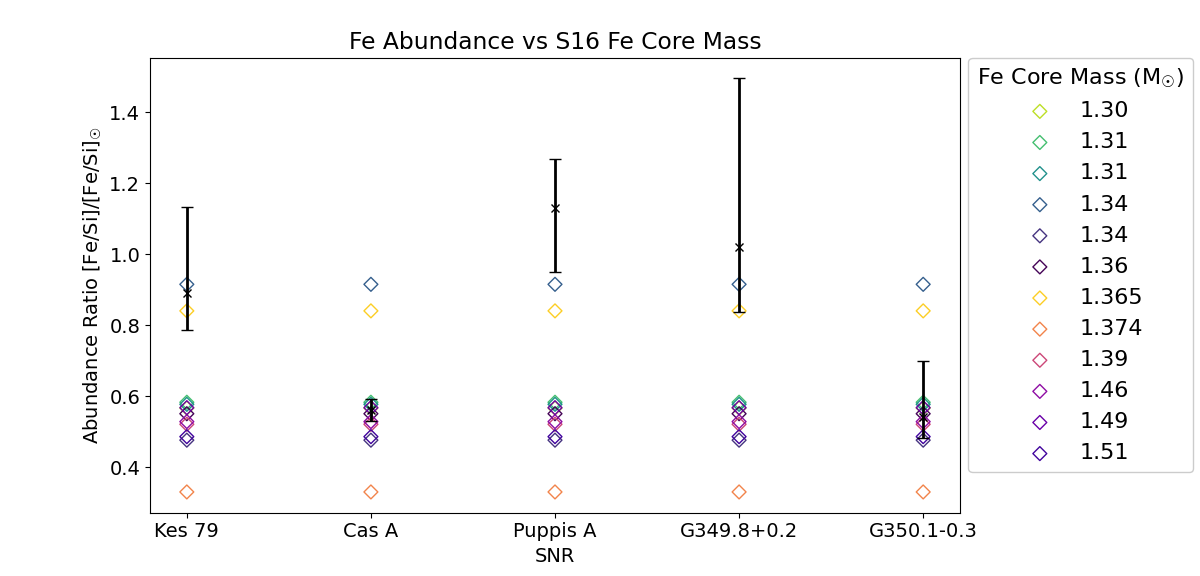}
    	\caption{Plotted is the abundance ratio for Fe/Si relative to the solar abundances from \protect\cite{Wilms} for each SNR in the sample (except for G15.9+0.2 which did not fit any model), versus the Fe core masses from the S16 model. The observed abundance ratios with errorbars are shown in black. }
    	\label{fig:ironCore}	
    \end{figure*}

   In the previous subsections, we followed the traditional method for calculating the progenitor mass by directly relating our observed abundance yields to the nucleosynthesis yields from the explosion products of a particular progenitor mass. However, the nucleosynthesis is highly dependent on the entropy in the core~\citep{2014frap.confE...4F}.  The iron core mass is set by the entropy, but the mapping from progenitor mass to iron core mass can be highly variable due to even small variations in the progenitor mass (see \cite{sukhbold2018}). As a result, we subsequently consider comparing our observed yields of Fe/Si, which is highly dependent on the iron core mass, to the final iron core mass from the models used in this paper. Among the suite of models we had considered, only three models list the iron core mass: WW95, F18, and S16. However, the WW95 model does not fit any of the Fe/Si abundances obtained from our observations, and the F18 model is capable of fitting all variations of the iron core masses depending on the explosion energy chosen. Since most of our mass estimates come from the comparison to the S16 model, we compare next our observed Fe/Si abundance ratios to the yields from the iron core mass from the S16 model. This is shown in Figure~\ref{fig:ironCore}. 

    We detail the results for each SNR in comparison to the S16 model here. For Kes~79, we find the Fe/Si abundance ratio fits well to an Fe core mass of 1.34--1.37~M$_\odot$, which correlates to a progenitor mass of 10--10.25~M$_\odot$ and matches well with our previous mass estimate of 9--13~M$_\odot$. For Cas~A, we find the Fe/Si abundance ratio fits well to an Fe core mass of 1.36--1.51~M$_\odot$, which correlates to a progenitor mass estimate of 12--14.9~M$_\odot$ and is a similar range to our previous mass estimate of 9--14~M$_\odot$. No model fits for Puppis~A although the closest is an iron core mass of 1.34~~M$_\odot$ which correlates to a progenitor mass of 10~M$_\odot$, and is therefore not in agreement with our estimate of 20--25~M$_\odot$. For G349.8+0.2, we find the Fe/Si abundance ratio fits well to an Fe core mass of 1.34--1.37~M$_\odot$, which correlates to a progenitor mass of 10--10.25~M$_\odot$ and does not match well with our previous mass estimate of 18--25~M$_\odot$. For G350.1--0.3, we find the Fe/Si abundance ratio fits well to an Fe core mass of 1.41--1.56~M$_\odot$, which correlates to a progenitor mass of 13.4--14.9~M$_\odot$ and is close to our previous mass estimate of 9--12~M$_\odot$. 

 \section{Discussion}
	
\subsection{Using Yields to Probe Core-Collapse Supernovae}

    \begin{figure*}
        \includegraphics[angle=0,width=0.6\textwidth]{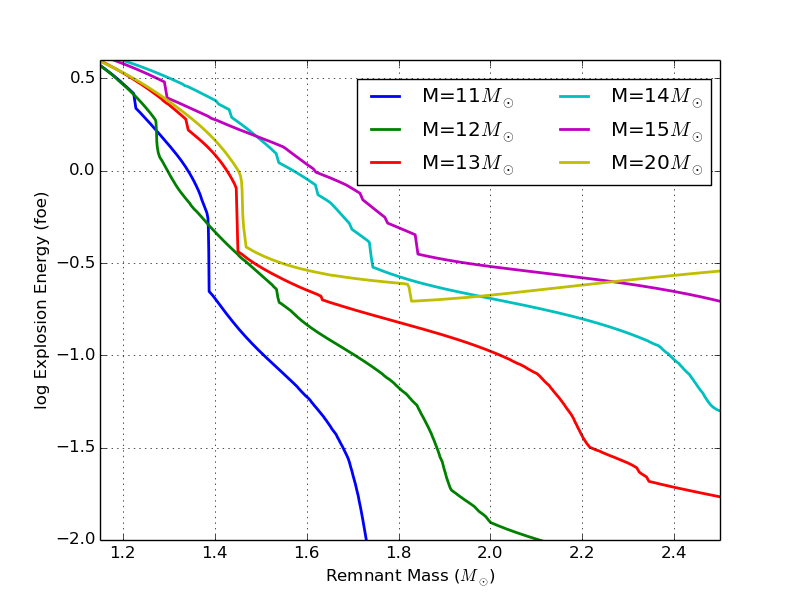}
        \caption{Explosion energy (in units of foe=10$^{51}$~ergs) versus remnant baryonic mass (excluding fallback) for a set of stellar progenitors~\protect\citep{2002RvMP...74.1015W}.  In the convective paradigm, the explosion energy is set by the energy in the convective region required to overcome the ram pressure of the infalling star. Once the explosion starts, further energy can be siphoned from the continuing downflows, but it is unlikely this will more than double the total explosion energy~\protect\citep{2012ApJ...749...91F}. The ram pressure of the infalling material evolves with time and, hence, does the explosion energy (typically, this energy decreases with time). In addition, the remnant mass at explosion increases with time as material continues to accrete onto the proto neutron star~\protect\cite[for a review, see][]{2006NewAR..50..492F}. This plot does 
        not include the increase in remnant mass from fallback.}
        \label{fig:enevm}	
    \end{figure*} 	

    In $\S$4, we focused on comparing the total yields of different explosion models.  One explanation for the difficulties in matching remnant observations with models could be that the remnant observations do not probe the total supernova yield (see $\S$5.3 for some caveats in the analysis).  Alternatively, because most suites of yields tie a single explosion energy based on an parameterized model, or a broad set of energies with a few different progenitors.  It could also be that we simply do not have the right combination of progenitor structure and supernova energy in the current set of models.  At this time, the explosion model behind core-collapse supernovae is not well enough understood to predict an exact energy for a given progenitor.  Under the convective paradigm, the explosion energy can be estimated (roughly to a factor of 2 accuracy) as the amount of energy in the convective region needed to overcome the ram pressure of the infalling star~\citep{2006NewAR..50..492F,2012ApJ...749...91F}.   For a given progenitor, the remnant mass and corresponding explosion energy (if the star does indeed explode), depends on progenitor structure and the time of explosion.  Figure~\ref{fig:enevm} shows the predicted explosion energy versus compact remnant mass (baryonic) at the time of explosion (excluding fallback).  Depending on the timescale of the explosion, the energy expected from a given progenitor can vary considerably.  Until we fully understand this explosion mechanism, the recipes in 1-dimensional explosion models provide a single possible instantiation of an explosion.

    However, even though the current grid of models do not cover the full parameter space of possible solutions, we can study these models and their yields and yield distributions.  In this section, we study the production of the yields typically measured in supernova remnants and their dependencies on the supernova explosion energy and its progenitor.  With this more detailed analysis, we then compare the specific yields and yield ratios from models to the observations.

\subsection{Trends in the Yields versus Enclosed Mass}
\label{sec:trends}

	\begin{figure*}
		\begin{center}
    		\subfloat[(a) 0.34\,foe]{\includegraphics[angle=0,width=0.45\textwidth,scale=0.5]{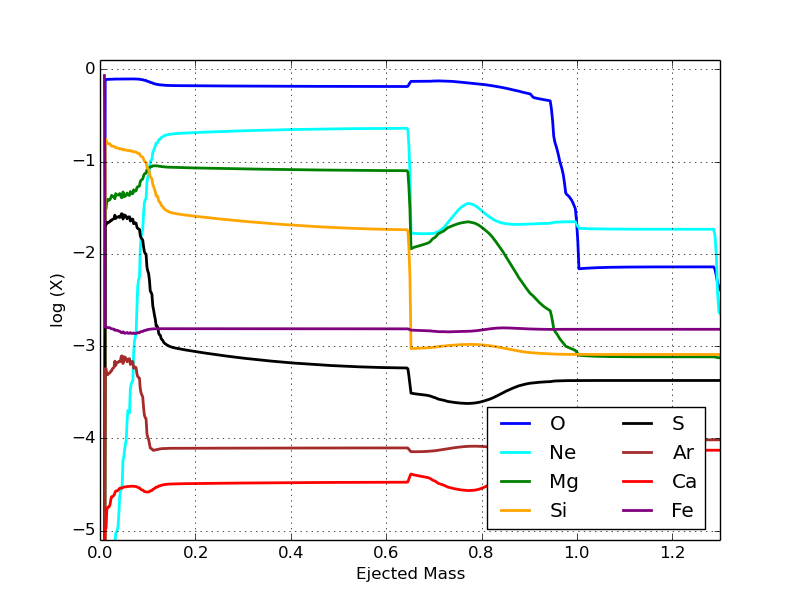}}
    		\subfloat[(b) 0.82\,foe]{\includegraphics[angle=0,width=0.45\textwidth]{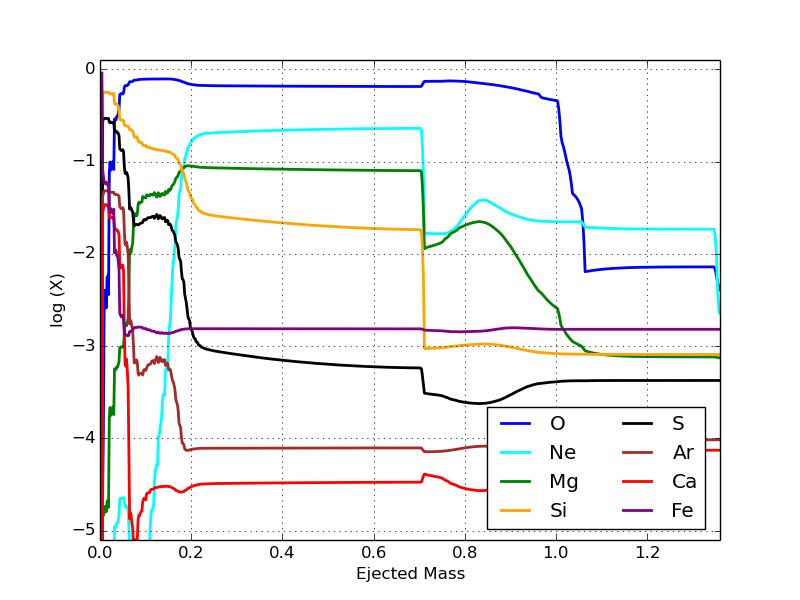}} \\
    		\subfloat[(c) 1.86\,foe]{\includegraphics[angle=0,width=0.45\textwidth]{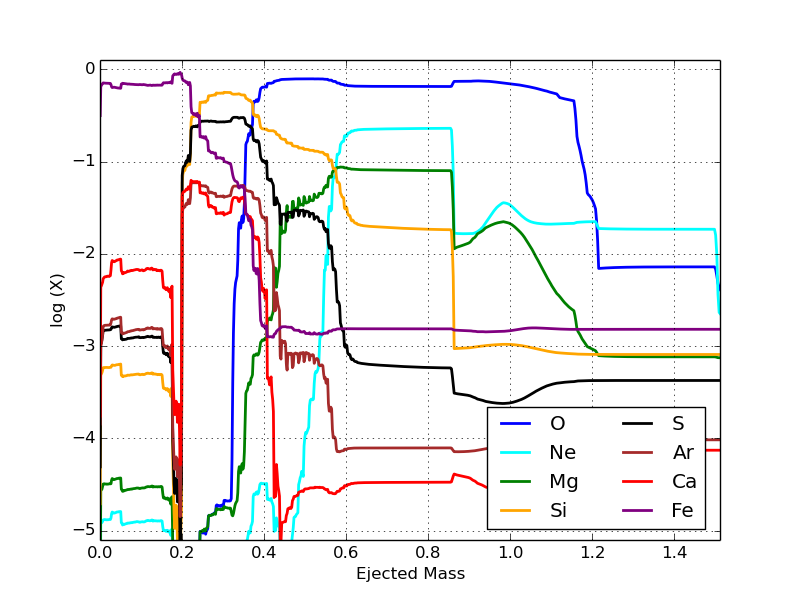}}
    		\subfloat[(d) 10.4\,foe]{\includegraphics[angle=0,width=0.45\textwidth]{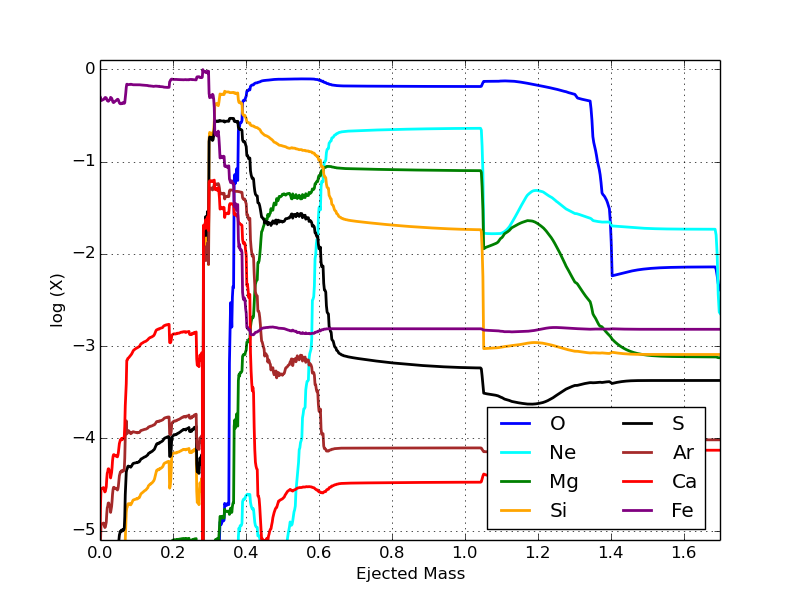}}
    
    		\caption{Yields versus enclosed ejecta mass for a 15\,M$_\odot$ with 4 different explosion energies:  0.34, 0.82, 1.86, and 10.4$\times 10^{51}\,{\rm erg}$ (1 foe=10$^{51}$~ergs). }
    		\label{fig:15msun}	
		\end{center}
	\end{figure*} 

	\begin{figure*}
		\begin{center}
    		\subfloat[(a) 0.65\,foe]{\includegraphics[angle=0,width=0.45\textwidth,scale=0.5]{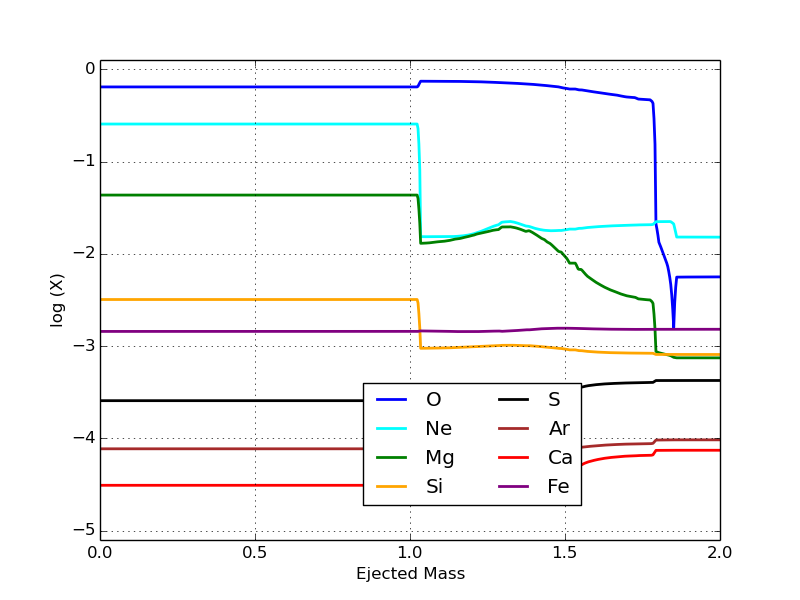}}
    		\subfloat[(b) 1.39\,foe]{\includegraphics[angle=0,width=0.45\textwidth]{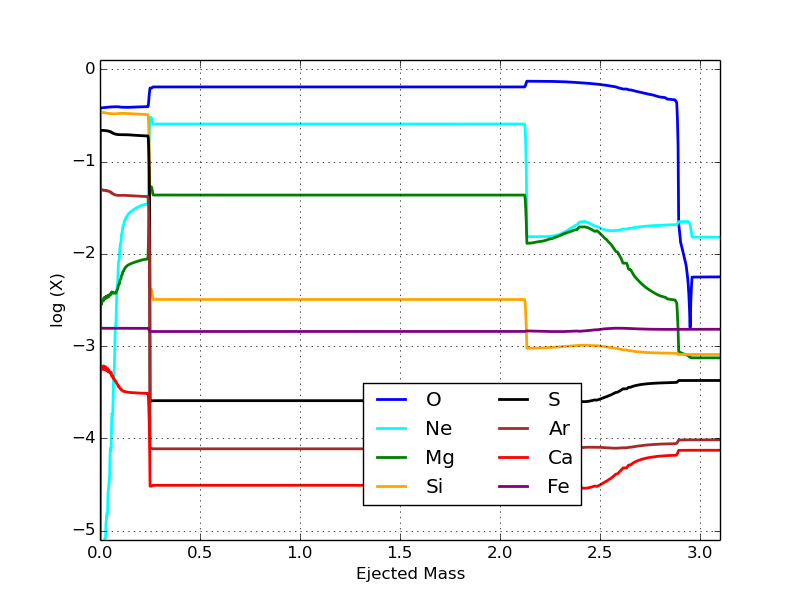}} \\
    		\subfloat[(c) 2.43\,foe]{\includegraphics[angle=0,width=0.45\textwidth]{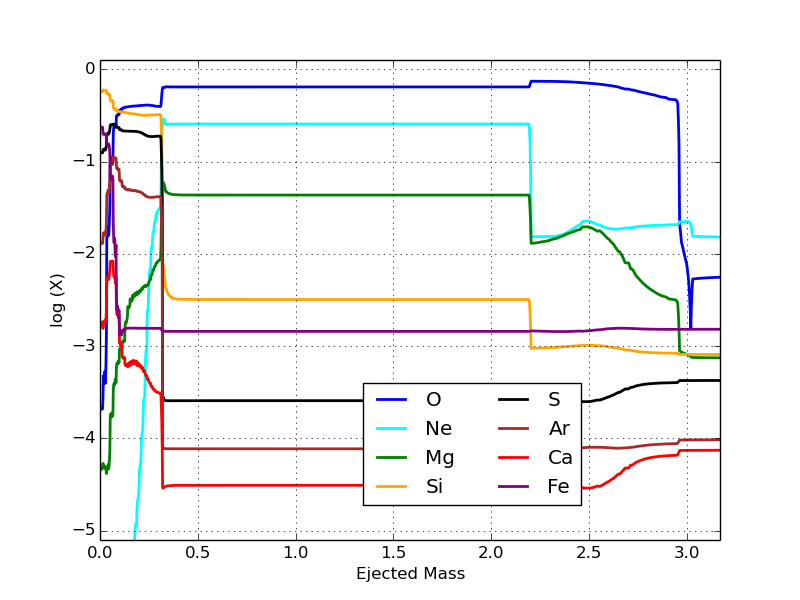}}
    		\subfloat[(d) 8.86\,foe]{\includegraphics[angle=0,width=0.45\textwidth]{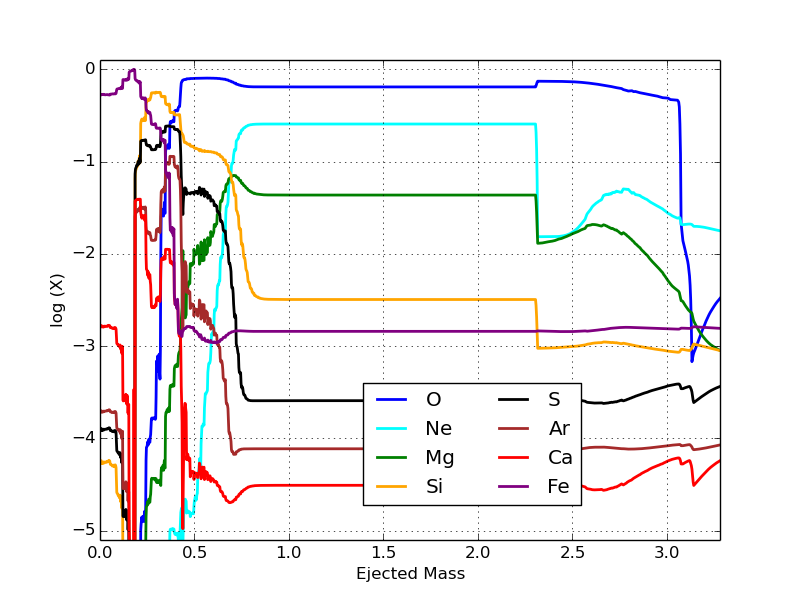}}
    
    		\caption{Yields versus enclosed ejecta mass for a 20\,M$_\odot$ with 4 different explosion energies:  0.65, 1.39, 2.43, and 8.86$\times 10^{51}\,{\rm erg}$.}
    		\label{fig:20msun}	
		\end{center}
	\end{figure*} 
	
    Figures~\ref{fig:15msun} and \ref{fig:20msun} show the abundance fraction of key supernova-remnant measured elements as a function of enclosed ejecta mass for a 15\,M$_\odot$ and 20\,M$_\odot$ progenitors each with 4 different explosion energies.  There are a few basic trends in these models.  First and foremost, the weaker explosions for both progenitors tend to have little or no iron ejecta.  The iron that is ejected is not synthesized in the explosion, but in the progenitor prior to collapse.  When iron is produced in the explosion, it is produced in the innermost layers of the ejecta (for weaker explosions, this innermost material falls back onto the compact remnant).  The Si, S, Ar, and Ca produced in the explosion are synthesized in the layer just above the iron layer, typically from explosion oxygen burning.  O, Mg, Ne are all produced in the progenitor and, for the most part, only destroyed in the explosion.

    The effect of the explosion energy is not just its affect on fallback.  To understand the energy dependence of the yields, it is important to understand nuclear burning.  Nuclear reactions require both high densities (to reduce the timescale for a reaction) and high energies (so the reactants can overcome energy barriers to cause burning).  In the equilibrium conditions in supernova explosions, the particle energy distribution fits a Maxwellian described by a single temperature and most numerical burning in astrophysical applications uses this equilibrium temperature.  The density of a strong shock ($\rho_{\rm shock}$in a radiation-pressure dominated gas is~\cite{}:
    \begin{equation}
        \rho_{\rm shock}=4 \rho_{\rm star}
    \end{equation}
    where $\rho_{\rm star}$ is the stellar density.  The temperature of the shocked material ($T_{\rm shock}$) is~\cite{}:
    \begin{equation}
        T_{\rm shock}=350\,keV (\rho_{\rm star}/10^6{\rm \, g \, cm^{-3}})^{0.25} (v_{\rm shock}/10^9 {\rm cm s^{-1}})^{0.5} 
    \end{equation}
    where $v_{\rm shock}$ is the shock velocity.  The stronger the shock, the higher the temperature and the stronger the reactions.  The supernova shock moves outward, heating material, causing silicon to burn to iron and oxygen to burn to Si, S, Ar, and Ca.  As the explosion energy, and hence shock velocity, increases, the temperature increases, extending the region where this burning can occur.  But as this burning front moves outward to lower-density regions, no amount of heating can drive burning.  For example, compare the iron production from the 1.86 and 10.4$\times 10^{51} {\rm erg}$ explosions of our 15\,M$_\odot$ progenitor.  This limit is even more important in the production of Si, S, Ar, and Ca.  The yield of these elements decreases with increasing energy as more iron is produced but the increased temperatures are unable to drive much more burning of oxygen.  Explosion energy can affect the nucleosynthetic yields, but it has its limits for the primary elements observed in supernova remnants.    

    Finally, notice that the relative abundances of different elements in each layer remain fairly constant.  This places strong constraints on our current remnant measurements that we discuss next.

\subsection{Yield Ratios:  Comparing Simulations to Observations}

    As we determined in $\S$4, none of our fitted models matched the total yields predicted from the suite of nucleosynthesis models available in the literature and used in this study. We first elaborate on the caveats in our presented analysis that may account for some of the discrepancy from an observational point of view: 
    1) While we have accounted for the ejecta component in inferring the metal abundances,  the one-component or two-component models likely underestimate the complexity of the ejecta in some of the selected regions, and may require multiple model components in order to more accurately represent the data. 
    2) It is possible that not all ejecta are shock-heated or even visible in the X-ray band; this is especially the case for the young SNR Cas~A (where Fe would not be fully shocked), and for the heavily absorbed remnants (like G15.9+0.2) where high extinction can hamper the detection of the lighter elements such as Oxygen and Neon. 
    3) The CCD spectral resolution of $Chandra$ and \textit{XMM-Newton} might introduce some degeneracy in the parameters which can prevent an accurate measurement of the chemical properties of the plasma. 
    4) We are averaging out the ejecta yields obtained from our spatially resolved spectroscopic study, and it's possible that some select regions are a better representation of the expected nucleosynthesis yields given the asymmetries expected in explosions. 
    5) There are systematic errors for plasma models (in our case, ATOMDB) and the calibration of the telescopes. 

    As mentioned above, since our remnants observations are limited to material that is shock-heated, it is possible that we are observing only specific portions of the ejecta. To test this, we compare select regions from the supernova yields in Figures~\ref{fig:15msun} and \ref{fig:20msun}.   These models are from the suite of explosion models studied in ~\cite{fryer18,andrews20}.  These 1-dimensional explosions included a prescription to mimic convection assuming that the convection quickly raises the entropy within the convective region. These models produce a range of explosion energies by including a convective timescale using entropy, not energy, as the energy-relevant quantity~\citep{fryer18}. We show the results from the 15 and 20\,$M_\odot$ progenitors in these studies, selecting a few models in that study with a range of convective growth times that correspond to different explosion energies.
 
    First and foremost, it is important to be aware of the different regions in the ejecta.  If elements are made in different regions, the relative abundances of these elements are very sensitive to this potential bias.  For example, if our observations are dominated by material in the innermost ejecta, we'd expect a very high iron to silicon ratio (anywhere from 100-1000).  It unlikely that we ever observe just this material without some mixing with the outer layers, but depending upon this mixing and this potential selection bias, the ratio can be easily high enough to explain the existing Fe/Si ratios in the data.  However, it is harder to explain abundance ratios of elements produced in the same region.

    One region probed by our remnants is the O, Mg, Ne region.  As we have shown, for the most part, O, Mg, Ne are primarily destroyed in the supernova explosion as the shock burns this layer to produce the silicon burning layer.  The ratio of these elements to Si can be used to probe the strength of the explosion.  If the explosion is weak, the Fe synthesized will fall back onto the compact remnant.  The shock will be too weak to burn O, Mg, and Ne to Si, S, Ca, and Ar.  So O, Mg and Ne will be strong and a remnant with high O/Si ratio is a clear signature of a low-energy explosion.  In contrast, a low O/Si ratio signals a high-velocity shock.  Within the oxygen layer, the O, Mg, Ne ratios probe stellar evolution and stellar mixing.  For our models, the O:Ne ratio is roughly 2-3 and the O:Mg ratio is 5-10.  In most of our models, Mg is the only element in this range that is observed.  But for Puppis A, O, Mg and Ne are all observed.  In most of the regions of Puppis A observed, O:Mg:Ne$\sim$1.  For Puppis A, the O:Mg ratio is difficult to explain with our progenitor stars and or their nuclear burning.

    Another region probed by supernova remnants is the silicon production layer.  In our models, most of the Si, Ar, Ca, and S in the explosion is produced in the same layer.  Since they are synthesized in the same ejecta region, any part of the remnant rich in one of these elements should be rich in the others.  Further, the abundance ratios of these elements should be set by these regions.  For example, for all of our models, in the production region for these elements, the ratio of Si:S is 1.2-2, the ratio of S:Ar or S:Ca is 3-10, the Ar:Ca ratio is 1-2.   These ratios persist for different explosion energies and progenitors.  These ratios tend to persist in the total yields from the WW95 models.  One of the peculiar aspects of our remnants is that the high Ar and Ca abundances (with respect to S or Si).  With our current progenitors and explosion energies, it is difficult to explain such observations.

\section{Conclusions}

    We have presented the first systematic X-ray study of CCO-hosting SNRs with the goal to address their supernova progenitors and explosion properties. Our sample consists of SNRs that show evidence of shock-heated ejecta: G15.9+0.2, Kes~79, Cas~A, Puppis~A, G349.7+0.2, and G350.1--0.3. In order to estimate the SN progenitor's mass, we fit the ejecta component of these remnants to infer their metal abundances, then compare the fitted elemental abundances from the spectral modelling to the nucleosynthesis yields from 5 of the most used stellar explosion models. We have also considered a recent electron capture supernova model. These models are labelled and referenced as follows: WW95 \citep{W95}, M03 \citep{M03}, N06 \citep{N06}; S16 \citep{S16}, F18 \citep{fryer18}, and J19 \citep{J19}. We find that none of the observed abundances fully match the explosion model yields for any of these commonly used models. In addition, we also considered a comparison between the Fe/Si abundance of each SNR in the sample to the iron core masses of the S16 model rather than the progenitor mass, which would avoid the non-linear mapping between progenitor mass and Fe core mass that can occur in these explosion models.

    That said, we present the best estimates for the progenitor mass for each SNR selected in our study, with the caveat that the final result is dependent on a few elements only, rather than all of them. Additionally, in order to aid our progenitor estimates, we look at the elemental mass yields, dependent on some filling factor, to compare to the progenitor model yields. We summarize our results in Table~\ref{tbl:prog}. Our list includes RCW~103, studied previously \citep{Mine} using the same models presented here. This SNR was studied separately given its unusual CCO which displayed a magnetar-like behaviour and is characterized by a very long 6.7-hour periodicity. In addition to the SN progenitor mass, we summarize the SN explosion energy determined from SNR dynamics. It is worth noting that except for Cas~A, this energy is smaller than the canonical 10$^{51}$~ergs. 

    \begin{table}
    \centering
        \begin{threeparttable}
            \caption{Summary of Explosion Energies and Progenitor Mass Estimates}
            \label{tbl:prog}
            \begin{tabular}{ccc}
                \hline
                \hline
                \footnotesize
                SNR & \begin{tabular}[c]{@{}@{}c} Explosion Energy \T\\ $(10^{50}~f_X^{-1/2}$~D$_{X}^{5/2})$~erg \B \end{tabular} & \begin{tabular}[c]{@{}@{}c} Progenitor \\ Mass (M$_\odot$) \end{tabular} \\
                \hline
                G15.9+0.2 &  $1.29^{+0.04}_{-0.06}$ & 18--22 \T\B \\
                Kes~79 & $6.3\pm{0.5}$ & 9--13 \B \\
                Cas~A & 15 & 9--14.1 \B \\
                Puppis~A & $3.0^{+1.5}_{-0.3}$ & 20--25 \B \\
                G349.8+0.2 & $1.5_{-0.5}^{+0.4}$ & 18--25 \B \\
                G350.1-0.3 & $1.1^{+0.2}_{-0.1}$ & 9--12 \B \\
                RCW~103 & 0.37 & 12--13 \B \\
                
                \hline
        
            \end{tabular}
            \begin{tablenotes}
                \linespread{1}
                \scriptsize
                \item Note - The subscript $X$ is a substitute variable that refers to the specific SNR's distance $D$ and filling factor $f$.
            \end{tablenotes}
        \end{threeparttable}
    \end{table}

    Our systematic study brings out the need to improve both on the observational data and the nucleosynthesis models used in the literature by the X-ray community. From the observational side, given the caveats discussed in this work, such a study will benefit from (a) future high-resolution X-ray spectroscopic studies to be acquired with the soon-to-be-launched (in 2023) XRISM mission \citep{XRISM} and with ATHENA \citep{ATHENA}, and (b) sensitive high-resolution imaging instruments such as the proposed AXIS Probe-Class mission that builds on the legacy of $Chandra$ and will be 10--50 times more sensitive than $Chandra$ to extended sources \citep{2019BAAS...51g.107M}.

    In old supernova remnants, the mass swept up in the explosion can become greater than the mass of the star itself.  The metals in this swept-up mass can contribute significantly to the total abundances, in some cases, dominating the observations.  We note that the abundance effects of this swept-up material are not included in most analyses, including those in this study.

    On the theoretical side, our study demonstrates weaknesses in current grids of spherically symmetric supernova explosions and their progenitors.  Although it is known that multi-dimensional models are needed to understand the structure of pre-collapse stellar progenitors~\citep{2009ApJ...690.1715A,2015ApJ...809...30A,2015ApJ...808L..21C}, grids of stellar models including these effects are beyond current computational abilities.  These models set the structure of the star and the relative sizes of different shell layers which, in turn, determine many of the observed abundance ratios.  In addition, only a handful of 3-dimensional explosions have been followed to late times~\citep{2003ApJ...594..390H,2005ApJ...635..487H, 
    Ellinger2013,2017ApJ...842...13W,2020ApJ...895...82V, 2021A&A...645A..66O} and most of the current results focus on the  $^{56}$Ni and $^{44}$Ti yields.  These asymmetries have been shown to alter the final nucleosynthetic yields of the explosion. 

    A systematic study of asymmetric explosions effects, combined with considering the time evolution of abundances (or shocked ejecta) in the considered models, is required to better understand the range of possible yields from a given progenitor.  These uncertainties are active areas of research and are beyond the scope of the current study; we defer a more complete study of their effect on the remnant yields to future work.

\section*{Acknowledgements}

    This research made use of NASA's Astrophysics Data System, HEASARC maintained at NASA's Goddard Space Flight Center, and the High-Energy Catalogue of Supernova Remnants (SNRcat)\footnote{\url{http://snrcat.physics.umanitoba.ca}} maintained at the University of Manitoba. We acknowledge the support of the Natural Sciences and Engineering Research Council of Canada (NSERC) through the Canada Research Chairs and the NSERC Discovery Grants programs, and the Canadian Space Agency (S.S.H.), the University of Manitoba's GETS program (C. B.), and the National Nuclear Security Administration of U.S.\ Department of Energy (Contract No.\ 89233218CNA000001) (C.L.F). We thank the referee for valuable comments that helped improve the clarity of the paper.

\section*{Data Availability}

    The data underlying this article are available in the \textit{Chandra} Data Archive, at https://cxc.harvard.edu/cda/, and the \textit{XMM-Newton} Science Archive, at https://www.cosmos.esa.int/web/xmm-newton/xsa/.




\bibliographystyle{mnras}
\bibliography{myref}



\appendix

\section{Spectral Analysis}

    Below, we present an in-depth discussion on the spectral analysis for each SNR in our sample. 

\subsection{G15.9+0.2}

    \begin{table*}
	    \caption{\textbf{G15.9+0.2:} Spectral properties of regions fit with the a VPSHOCK model with Mg, Si, S, Ar, Ca, and Fe (Ni tethered to Fe) allowed to vary. Abundances are in solar units from the abundance tables of \protect\cite{Wilms}. If an abundance value is missing, it was frozen at solar. For Region~0, we also added a Gaussian component to fit the Fe~K line in order to improve the fit. Data was grouped with a minimum of 20 counts per bins. }
			
		\label{tbl:G15RegionData} 
		\begin{tabular}{cccccccccccc}
		\hline
		Region & N$_\text{H}$ & kT & Mg & Si & S & Ar & Ca & Fe (Ni) & n$_\text{e}$t & $\chi^2_\nu$ (DOF) \T\\  		  
		& $\times 10^{22}$~cm$^{-2}$ & keV & & & & & & & $\times 10^{10}$~cm$^{-3}$~s &  & \B \\ 
		\hline
        0 & $5.5_{-0.2}^{+0.1}$ & $0.96_{-0.04}^{+0.07}$ & $0.8_{-0.1}^{+0.2}$ & $1.8_{-0.2}^{+0.2}$ & $3.8_{-0.4}^{+0.5}$ & $8.5_{-1.7}^{+1.1}$ & $15.7_{-5.6}^{+4.4}$ & ... & $3.0_{-0.4}^{+0.5}$ & 1.27 (363) \T\B \\
        1 & $6.2_{-0.5}^{+1.1}$ & $1.17_{-0.16}^{+0.09}$ & $1.9_{-0.5}^{+2.9}$ & $2.8_{-0.5}^{+2.2}$ & $2.7_{-0.2}^{+1.3}$ & $3.3_{-0.7}^{+1.7}$ & ... & $3.9_{-2.0}^{+13.0}$ & $7.1_{-0.9}^{+2.9}$ & 0.99 (395) \B \\
        2 & $6.7_{-0.2}^{+0.2}$ & $0.95_{-0.04}^{+0.03}$ & $7.3_{-1.1}^{+2.0}$ & $3.8_{-0.4}^{+0.9}$ & $4.1_{-0.3}^{+0.8}$ & $7.1_{-1.7}^{+1.8}$ & $10.6_{-6.4}^{+5.4}$ & $13.7_{-2.5}^{+9.2}$ & $5.9_{-0.9}^{+0.9}$ & 1.41 (403) \B \\
        3 & $4.3_{-0.2}^{+0.3}$ & $0.91_{-0.09}^{+0.07}$ & $0.7_{-0.1}^{+0.2}$ & $1.5_{-0.2}^{+0.1}$ & $2.4_{-0.6}^{+0.7}$ & $6.5_{-3.4}^{+4.2}$ & ... & ... & $4.2_{-0.9}^{+2.1}$ & 0.98 (361) \B \\
        4 & $4.8_{-0.2}^{+0.2}$ & $0.95_{-0.05}^{+0.05}$ & $1.2_{-0.2}^{+0.2}$ & $1.9_{-0.2}^{+0.2}$ & $4.4_{-0.7}^{+0.9}$ & $5.1_{-2.2}^{+3.3}$ & ... & ... & $2.6_{-0.4}^{+0.5}$ & 1.10 (392) \B \\
        \hline
		\end{tabular}
	\end{table*}

	\begin{table*}
	    \caption{\textbf{G15.9+0.2:} Spectral properties for the full SNR fit with a VPSHOCK+PSHOCK model with Mg, Si, S, Ar, and Ca allowed to vary for the hard (H) VPSHOCK component and the soft (S) PSHOCK abundances were frozen at solar. Fe was frozen to solar as it did not affect the fit. Data was grouped with a minimum of 20 counts per bins. }
			
		\label{tbl:G15FullSNRRegionData} 
		\begin{tabular}{cccccccccccc}
		\hline
		N$_\text{H}$ & kT$_\text{H}$ & Mg & Si & S & Ar & Ca & n$_\text{e}$t$_\text{H}$ & kT$_\text{S}$ & n$_\text{e}$t$_\text{S}$ & $\chi^2_\nu$ (DOF) \T \\  		  
		$\times 10^{22}$~cm$^{-2}$ & keV & & & & & & $\times 10^{10}$~cm$^{-3}$~s & keV & $\times 10^{9}$~cm$^{-3}$~s & \B \\ 
		\hline
		$3.46_{-0.16}^{+0.07}$ & $2.2_{-0.2}^{+0.3}$ & $1.7_{-0.3}^{+0.6}$ & $8.2_{-0.8}^{+2.0}$ & $13.0_{-1.6}^{+2.9}$ & $17.0_{-3.6}^{+5.8}$ & $10.6_{-4.8}^{+7.1}$ & $3.18_{-0.05}^{+0.12}$ & $1.27_{-0.06}^{+0.01}$ & $2.7_{-0.7}^{+0.2}$ & 1.30 (542) \T\B \\	
		\hline
		\end{tabular}
	\end{table*}	
        
    G15.9+0.2 has four data sets, where obsID 5530, 6288, and 6289 were taken within a few days of each other but are significantly shorter than obsID 16766. Therefore, spectra from obsID 5530, 6288, and 6289 were extracted separately then combined using the CIAO task \textit{combine\_spectra}, and then simultaneously fit with the spectra from obsID 16766. We limited the energy to the range 0.3--7.0~keV band. The spectra were best fit with a single VPSHOCK model, with the exception of region 0 which required a secondary Gaussian component to be fit for the Fe~K line. The hydrogen column density, temperature, ionization timescale, and normalization were first freed and then fit, leaving the abundances set to solar. Then, we fit and free the abundances one at time starting with Mg, Si, S, Ar, Ca, then Fe (with Ni tethered to Fe). In some instances, where freeing one of the aforementioned element did not improve the fit, it was frozen at solar. Since all the regions were well fit by a one-component model, we also fit the whole SNR with a two-component VPSHOCK+PSHOCK model in order to separate the forward shocked ISM from the reverse shocked ejecta.
        
    The spectral fits for the individual regions can be found in Table~\ref{tbl:G15RegionData} and the full SNR fit in Table~\ref{tbl:G15FullSNRRegionData}. Each region was well fit ($\chi^2_\nu$<1.4) by a single component VPSHOCK associated with the ejecta. We find that the column density is quite high across the SNR with a range of 4.3--6.7$\times 10^{22}$~cm$^{-2}$, where the lowest column density is associated with the diffuse northern portion of the SNR (Region~3) and the highest associated with the southern shell (Region~2). The temperatures range from 0.91--1.17~keV, the lowest temperature is again associated with the diffuse northern region (Region 3) and the hottest temperature us associated with the southern shell (Region~1). The ionization timescale ranges from 3.0--7.1$\times10^{10}$~cm$^{-3}$~s and is relatively consistent across the remnant, while also indicating the SNR is still in NEI. For each region selected, we found above-solar abundances for Mg, Si, S, Ar and in some cases Ca and Fe. Adding a secondary component in search of the blast wave component did not improve the fits. We also examined the global spectrum which was best fit by a two-component model. The column density is less than the region values with $3.46^{+0.07}_{-0.16}\times10^{22}$~cm$^{-2}$, with higher temperature for the shocked ejecta component of $2.2^{+0.3}_{0.2}$~keV and overall higher abundances. The ionization timescale is similar to that found in the regions with $3.18^{+0.12}_{-0.05}\times10^{10}$~cm$^{-3}$~s. The blast wave component has a lower temperature of $0.27^{+0.01}_{-0.06}$ and a low ionization timescale of $2.7^{+0.2}_{-0.7}\times10^9$~cm$^{-3}$~s. 
    
    We can compare our results to the study by \cite{G15}, which is based on the \textit{Chandra} obsID 16766 and which used solar abundances from \cite{anders} for 7 regions. We find good agreement for our temperature range and ionization timescale. Our hydrogen column density is much higher but is expected with the different solar abundances tables. Additionally, \cite{G15} found both solar abundance regions and enhanced abundance regions for Mg, Si, S, Ar,  Ca, and Fe, whereas we did not detect a solar (blast wave) component and had significantly higher abundance values. However, the larger ellipse region of the bright shell seems to indicate large abundances, although poorly constrained, but were more consistent with our results. \cite{G15XMM} performed an \textit{XMM-Newton} study for 4 regions using the same solar abundance tables as this study. While their column density is slightly lower than our range, and their temperature slightly higher, they do find enhanced ejecta in all regions, which is more in agreement with our results for Mg, Si, S, Ar, and Ca.  
    
\subsection{Kes~79}

	\begin{table*}
	    \caption{\textbf{Kes~79:} Spectral properties of regions fit with the either a single component VNEI or a two-component VNEI+APEC where the VNEI abundances for Mg, Si, S, Ar, and Fe (with nickel tethered to Fe) were allowed to vary and the APEC abundances were frozen at solar. Abundances are in solar units from the abundance tables of \protect\cite{Wilms}. If an abundance value is missing, it was frozen at solar. If a temperature is missing for the APEC component, the region was fit with a single VNEI model. Data was grouped with a minimum of 20 counts per bins, except for Region~9, the faint region in the north, which was grouped with a minimum of 40 counts per bins.  }
			
		\label{tbl:Kes79RegionData} 
        \resizebox{\textwidth}{!}{
		\begin{tabular}{ccccccccccccc}
		\hline
		& & VNEI & & &  & & & & & APEC & \T \\
		Region & N$_\text{H}$ & kT & Ne & Mg & Si & S & Ar & Fe (Ni) & n$_\text{e}$t & kT & $\chi^2_\nu$ (DOF) \\  		  
		& $\times 10^{22}$~cm$^{-2}$ & keV & & & & & & & $\times 10^{10}$~cm$^{-3}$~s & keV & \B \\ 
		\hline
        0 & $2.4_{-0.2}^{+0.3}$ & $0.92_{-0.08}^{+0.08}$ & $0.4_{-0.2}^{+0.2}$ & $1.0_{-0.2}^{+0.3}$ & $0.9_{-0.2}^{+0.3}$ & $0.8_{-0.2}^{+0.2}$ & $1.4_{-0.6}^{+0.8}$ & $0.8_{-0.2}^{+0.4}$ & $6.3_{-1.4}^{+2.4}$ & ... & 1.00 (194) \T\B \\
        1 & $2.3_{-0.2}^{+0.1}$ & $1.1_{-0.1}^{+0.3}$ & $2.3_{-1.1}^{+2.1}$ & $1.8_{-0.6}^{+2.0}$ & $2.9_{-0.5}^{+1.9}$ & $2.2_{-0.5}^{+1.7}$ & ... & ... & $5.7_{-1.8}^{+3.5}$ & $0.22_{-0.02}^{+0.02}$ & 1.23 (179) \B \\
        2 & $2.0_{-0.1}^{+0.2}$ & $0.82_{-0.06}^{+0.0.02}$ & $1.6_{-0.7}^{+0.5}$ & $2.2_{-0.3}^{+0.3}$ & $1.9_{-0.1}^{+0.6}$ & $1.6_{-0.4}^{+0.5}$ & $3.9_{-1.8}^{+5.2}$ & $1.8_{-0.4}^{+0.0.9}$ & $8.2_{-0.5}^{+1.3}$ & ... & 1.22 (173) \B \\
        3 & $1.82_{-0.08}^{+0.17}$ & $0.81_{-0.09}^{+0.06}$ & $0.9_{-0.4}^{+0.4}$ & $1.4_{-0.2}^{+0.4}$ & $1.5_{-0.2}^{+0.3}$ & $1.2_{-0.3}^{+0.5}$ & ... & $1.3_{-0.3}^{+1.0}$ & $7.3_{-1.7}^{+5.1}$ & ... & 1.30 (184) \B \\
        4 & $2.6_{-0.2}^{+0.2}$ & $0.89_{-0.10}^{+0.12}$ & $1.7_{-0.8}^{+0.7}$ & $1.1_{-0.4}^{+0.8}$ & $1.5_{-0.3}^{+0.6}$ & $1.2_{-0.3}^{+0.5}$ & ... & $0.5_{-0.4}^{+1.2}$ & $6.134_{-0.041}^{+0.002}$ & $0.20_{-0.02}^{+0.02}$ & 1.26 (178) \B \\
        5 & $2.19_{-0.09}^{+0.22}$ & $0.74_{-0.07}^{+0.11}$ & $1.1_{-0.3}^{+0.9}$ & $1.7_{-0.2}^{+0.4}$ & $1.45_{-0.09}^{+0.60}$ & $1.3_{-0.2}^{+0.5}$ & ... & $1.3_{-0.3}^{+1.1}$ & $10.6_{-4.3}^{+7.7}$ & ... & 1.23 (228) \B \\
        6 & $1.8_{-0.1}^{+0.2}$ & $0.88_{-0.15}^{+0.13}$ & $0.8_{-0.7}^{+0.9}$ & $2.9_{-0.3}^{+2.0}$ & $3.1_{-0.2}^{+2.0}$ & $1.2_{-0.4}^{+1.0}$ & ... & $3.5_{-0.7}^{+2.9}$ & $6.8_{-1.6}^{+7.8}$ & ... & 1.19 (191) \B \\
        7 & $2.7_{-0.1}^{+0.1}$ & $0.86_{-0.10}^{+0.09}$ & $1.1_{-0.6}^{+0.8}$ & $1.0_{-0.4}^{+0.2}$ & $1.6_{-0.2}^{+0.7}$ & $1.4_{-0.4}^{+0.7}$ & ... & $0.9_{-0.6}^{+0.2}$ & $3.9_{-1.3}^{+3.9}$ & $0.16_{-0.01}^{+0.02}$ & 1.27 (177) \B \\
        8 & $2.9_{-0.2}^{+0.1}$ & $0.91_{-0.09}^{+0.10}$ & $1.9_{-0.9}^{+1.7}$ & $1.1_{-0.4}^{+1.0}$ & $2.6_{-0.7}^{+1.0}$ & $1.8_{-0.4}^{+0.7}$ & ... & $2.0_{-0.5}^{+0.9}$ & $5.0_{-1.5}^{+2.1}$ & $0.22_{-0.02}^{+0.04}$ & 0.95 (176) \B \\
        9 & $2.2_{-0.2}^{+0.4}$ & $0.70_{-0.10}^{+0.07}$ & $0.7_{-0.4}^{+1.2}$ & $1.1_{-0.2}^{+0.6}$ & $1.1_{-0.2}^{+0.6}$ & $0.30_{-0.26}^{+0.32}$ & ... & $1.2_{-0.5}^{+1.3}$ & $8.7_{-2.7}^{+13.5}$ & ... & 1.32 (121) \B \\
        10 & $2.67_{-0.08}^{+0.28}$ & $0.90_{-0.08}^{+0.05}$ & $1.8_{-1.0}^{+2.0}$ & $2.4_{-0.3}^{+0.8}$ & $2.4_{-0.2}^{+0.5}$ & $2.0_{-0.4}^{+0.6}$ & ... & $2.4_{-0.5}^{+1.8}$ & $4.7_{-0.8}^{+1.2}$ & ... & 1.42 (234) \B \\
        11 & $2.1_{-0.1}^{+0.2}$ & $0.75_{-0.08}^{+0.08}$ & $0.79_{-0.49}^{+0.04}$ & $1.4_{-0.2}^{+0.7}$ & $1.28_{-0.08}^{+0.59}$ & $1.1_{-0.3}^{+0.7}$ & ... & $1.7_{-0.5}^{+1.4}$ & $7.1_{-2.2}^{+3.4}$ & ... & 1.02 (190) \B \\	
		\hline
		\end{tabular}}
	\end{table*}    
    The three data sets (1982, 17453, and 17659) for kes~79 were extracted and simultaneously fit in the energy range 0.3--7.0~keV. Each region was fit with a one-component VNEI. The hydrogen column density, temperature, ionization timescale, and normalization were first freed and then fit. Then the abundances for Ne, Mg, Si, S, and Fe (Ni is tethered to Fe) were freed and fit one at a time. We also attempted to fit a secondary VNEI component to account for the forward shock for some of the regions, but the ionization timescale tended to the maximum allowed value (5$\times10^{13}$~cm$^{-3}$~s) and so was fit with a secondary APEC component instead, which improved the statistics of the fit.
    
    The spectral fits can be found in Table~\ref{tbl:Kes79RegionData}. Each region was well fit ($\chi^2_\nu$<1.4) by either a single component VNEI or a two component VNEI+APEC. The single component VNEI could either be associated with enhanced ejecta (Region 2, 3, 5, 6, 10, 11, 12) or the blast wave component (Reg 0, 9). In the case of the two component VNEI+APEC, the VNEI was best associated with enhanced ejecta and the APEC component was best associated with the blast wave component. The column density ranged from 1.8--2.9$\times 10^{22}$~cm$^{-2}$, with the lowest values in the southern portion (region~3 and region~6) of the SNR and the highest value in the northeast portion of the bright, knotty emission (region~8). The ejecta is best associated with the hard component temperature of 0.75--1.1~keV, with most regions in the 0.8--0.9~keV range. The blast wave is best associated with the soft component temperature 0.16--0.22~keV, however we do see some harder single component fits with solar abundance, with temperatures of 0.92~keV and 0.70~keV, which indicates that there may be a range of temperatures that can be associated with the blast wave component. The enhanced ejecta regions have a range of ionization timescales of 8.2--11$\times10^{10}$~cm$^{-3}$~s indicating that it is still in NEI. The soft component tended towards CIE in the two component fits, whereas the single component regions associated with the blast wave had values of 6.3$\times10^{10}$~cm$^{-3}$~s and 8.7$\times10^{10}$~cm$^{-3}$~s indicating they are still in NEI. The abundances associated with the enhanced ejecta finds slightly supersolar values for Ne, Mg, Si, S, Ar, and Fe. The abundances for the blast wave components were subsolar to solar values. 
    
    There are three previous studies by which we can compare our results. The first is a \textit{Chandra} study by \cite{kes79Chandra}, which use the obsID~1987 and solar abundances from \cite{anders}. They performed a global fit of the SNR using a single VNEI (our data matched best with their SPEX fit), which found a column density of 1.67$\times 10^{22}$~cm$^{-2}$ and a temperature of 0.76~keV, slightly lower than our values. However, they found an ionization timescale of 9.8$\times10^{10}$~cm$^{-3}$~s, and slightly enhanced abundances for Ne, Mg, Si, S, and Fe, which matches our results well. They did not detect the soft component that we associated with the blast wave. Another study that looks at \textit{XMM-Newton} data was performed by \cite{kes79XMM} for 19 regions, using solar abundances from \cite{anders}, who found their data was best fit by a two component NEI+VNEI model. Their hydrogen column density was $\sim1.75\times 10^{22}$~cm$^{-2}$, which is slightly lower than our column density range. They found that the hard component, with temperatures ranging from 0.7--1.4~keV, was best associated with enhanced ejecta for Ne, Mg, Si, S, Ar, and Fe and ionization timescales $\sim6\times10^{10}$~cm$^{-3}$~s, which matches well with our results. Additionally, they found the soft component, with temperatures ranging from 0.18-0.22~keV, was best associated with the blast wave, with some regions in CIE or ionization timescales with large error ranges or no upper bounds, that also matches with our results. However, they did not find a range of temperatures associated with the blast wave components akin to our single component blast wave models. \cite{kes79Suzaku} examined \textit{Suzaku} data by which we can compare with their spatially resolved spectroscopy. They fit their data with a two component VNEI+APEC model, with slightly enhanced abundances for Ne, Mg, Al, Si, S, Ar, Ca, and Fe associated with the hard NEI component, with temperatures ranging from 0.82-1.0~keV, and an ionization timescale of $\sim6\times10^{10}$~cm$^{-3}$~s. The blast wave component was best associated with the soft APEC component with temperatures of 0.22~keV. We find good agreement with their results. 
    
\subsection{Cas~A}

	\begin{figure*}
		\begin{center}
		\subfloat[(a) N$_H$]{\includegraphics[angle=0,width=0.28\textwidth]{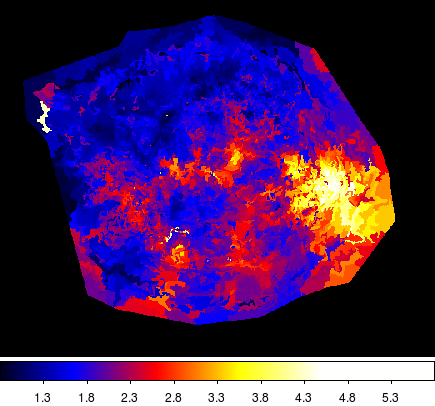}} \
		\subfloat[(b) kT]{\includegraphics[angle=0,width=0.28\textwidth]{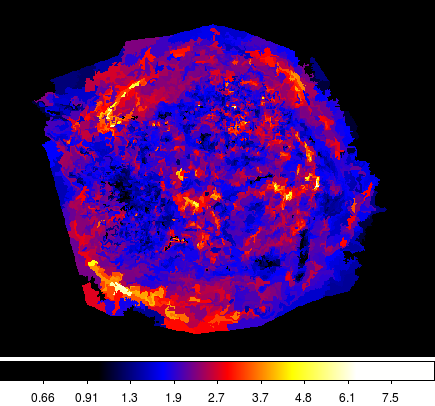}} \
		\subfloat[(c) n$_e$t]{\includegraphics[angle=0,width=0.28\textwidth]{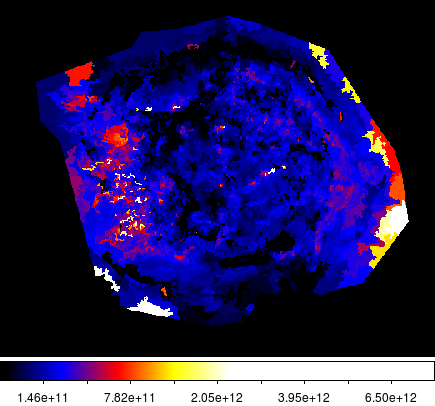}}
		\\
		\subfloat[(d) Ne]{\includegraphics[angle=0,width=0.28\textwidth]{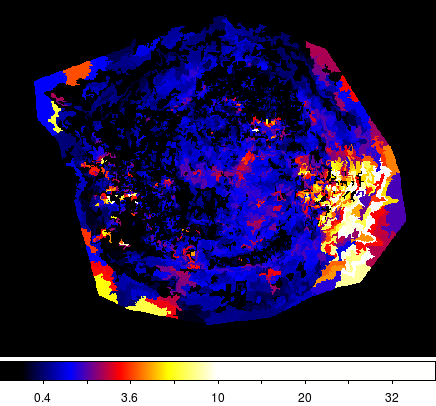}} \
		\subfloat[(e) Mg]{\includegraphics[angle=0,width=0.28\textwidth]{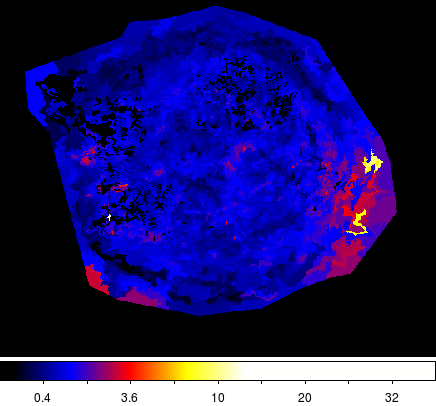}} \
		\subfloat[(f) Si]{\includegraphics[angle=0,width=0.28\textwidth]{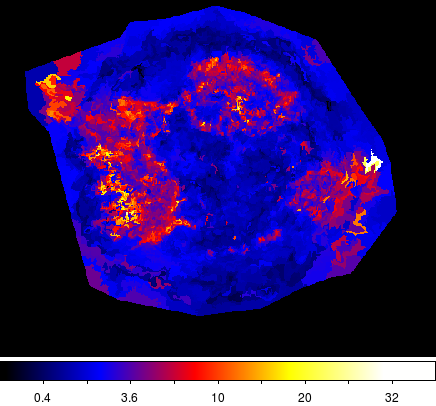}}
		\\
		\subfloat[(g) S]{\includegraphics[angle=0,width=0.28\textwidth]{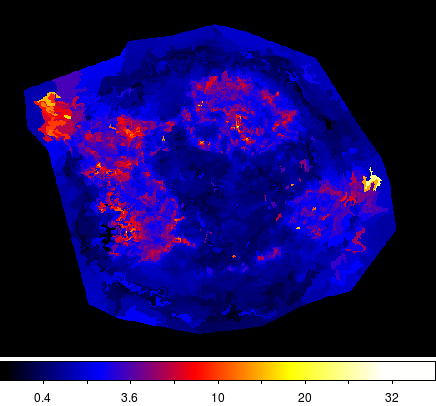}} \
		\subfloat[(h) Ar]{\includegraphics[angle=0,width=0.28\textwidth]{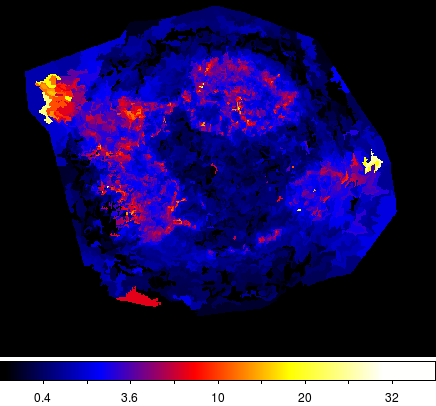}} \
		\subfloat[(i) Ca]{\includegraphics[angle=0,width=0.28\textwidth]{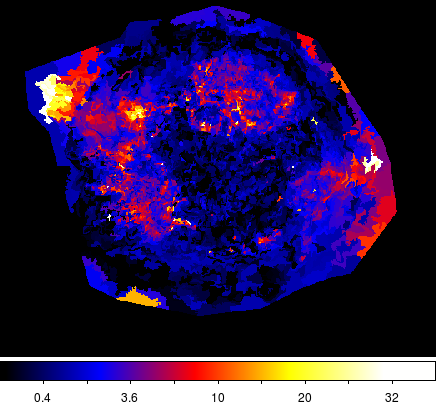}}	
		\\
		\subfloat[(j) Fe]{\includegraphics[angle=0,width=0.28\textwidth]{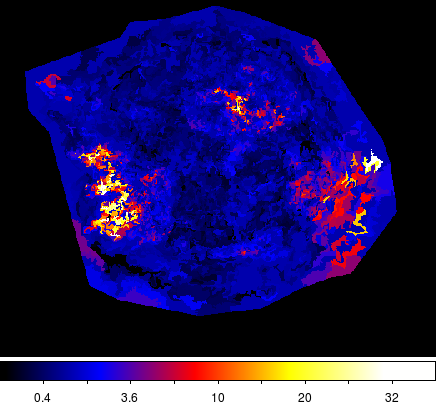}} \
		\subfloat[(k) $\chi^2_\nu$]{\includegraphics[angle=0,width=0.28\textwidth]{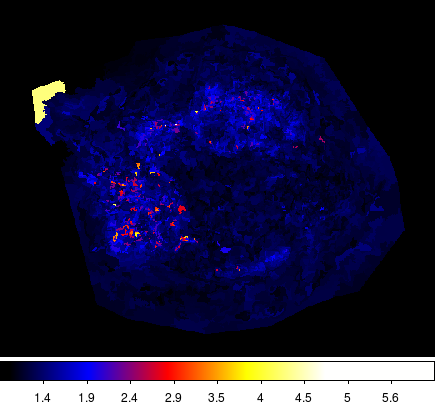}} \
		\subfloat[(k) Blast Wave]{\includegraphics[angle=0,width=0.28\textwidth]{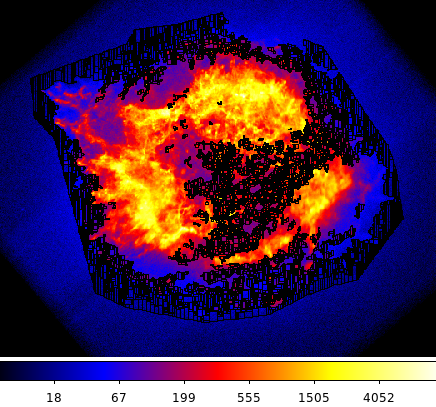}}

		\caption{ Region maps for the given parameter maps. All maps use a square-root scale except for the $\chi^2_\nu$ and N$_H$ map which use a linear scale. N$_H$ has units of cm$^{-2}$ and n$_e$t has units of cm$^{-3}$~s. The abundance maps are relative to the solar values of \protect\cite{Wilms}. The abundance and $n_et$ maps are truncated at the high end. (k) Shows the regions (black) that were designated as forward shocked regions. }
		\label{fig:casaRegMap}	
		\end{center}
	\end{figure*}   
    
    For Cas~A, we examined a total of 6251 regions from the million-second observation performed with \textit{Chandra} that was separated into nine different observations. The spectra were extracted and then simultaneously fit in the 0.3--10~keV range. Each region was fit with a one-component VPSHOCK model following the analysis done by \cite{casaHwang}. The hydrogen column density, temperature, ionization timescale, and normalization were first freed and then fit. The redshift was allowed to vary next, as the ejecta have significant bulk motion but the redshift was freed to achieve good fits rather than used as a reliable velocity measurement (see \cite{HwangCasA}). Finally, the abundances were fit and freed one at a time from Ne to Fe, where Ni was tethered to Fe. O is considered to be the primary source of the continuum and held at solar (see \cite{casaHwang}). In certain regions, the Fe~K line was poorly fit and so we added a secondary VPSHOCK model where only the temperature, ionization age, normalization, and Fe abundance were allowed to vary. In certain regions, the Fe~K line was poorly fit and so we added a secondary VPSHOCK model where only the temperature, ionization age, normalization, and Fe abundance were allowed to vary. This did not significantly improve the fits and so we only used the single VPSHOCK fits. Next we attempted to model the blast wave component of the emission with a single VPSHOCK as described in \cite{casaHwang} where He is 3 times the solar value relative to H, and N about 15 times the solar value. If the reduced chi-squared was less than 1.5, it was considered the forward shocked CSM component. We also attempted to model the prominent non-thermal component present in Cas~A with a secondary power law component (CSM+PL) which we subsequently considered part of the forward shock. In this way, we can separate the blast wave component from the shock-heated ejecta. In total, we found 1166 forward shocked regions, both CSM and CSM+PL. Because of the large number of regions, we did not perform error analysis on the individual model fits. 
    
    The results of the fits are reported as regions maps in Figure~\ref{fig:casaRegMap} for the single VPSHOCK fits and the location of the blast wave. Most regions of the single VPSHOCK fits were well fit ($\chi^2_\nu$<2.0) with only a few regions that were not, which are mostly found in the knots found in the inner ring of emission and one region that would be consider background near the breakout on the northeast. The median goodness of fit for all regions is $\chi^2_\nu=1.2$. The hydrogen column density is higher on the west, and falls off towards the east with a range from 0.8--5.8$\times 10^{22}$~cm$^{-2}$ and a median value 1.8$\times 10^{22}$~cm$^{-2}$. The temperatures range from 0.57--9.1~keV, but a majority fall within the 0.57--4.0~keV and a median of 2.0~keV. The larger temperature regions are found in the outer rim of the shell and in some of the bright knots in the center. The ionization timescale is mostly below $10^{11}$~cm$^{-3}$~s which indicates the majority of the remnant is in NEI, and the highest timescales fall on the background regions around the edge which were not completely masked out. The abundance maps fall into 2 similar mappings. The Si map has the highest emission in the northeast blowout, the inner rim emission and the enhancements in the west. S, Ar, Ca, and Fe follow similar arrangements, with an exception to Fe which does not have as much emission in the northeast breakout. Ne is mostly subsolar to solar (with a median of 0.5 and a mean of 1.3), with large enhancement in the west. Mg follows a similar mapping as to Ne with a median 0.64 and a mean of 0.78. We can draw a direct comparison to \cite{casaHwang} who used the same data and whose work we followed in our analysis. We note the differences in analysis: the solar abundances were from \cite{anders} whereas we used \cite{Wilms}; the spectra were summed together before fitting as opposed to our work here where the data were simultaneously fit over the 9 data sets; the enhanced ejecta models tethered Ni to Fe, Ca to Ar, and sometimes Ar to S; and a secondary component was added to some of the enhanced ejecta fits in order to improve statistics (secondary VPSHOCK to improve Fe~K). We see similar spatial trends in the maps for each variable, although the specific values can be different. For example our N$_H$ maps have similar trends of higher values in the west and lower values in the east, but tend to have overall higher values than that of \cite{casaHwang} (as expected with the \cite{Wilms} abundance model). We also found similar regions selected as predominantly blast wave components and pure Fe emission in the inner ring of emission. We can also compare our elemental mass yields. For the lighter elements Mg and Ne, we find larger mass estimates which is likely due to the different solar abundance tables. We find comparable Si and S mass yields but under estimate the yields from Ar. For the Fe mass, we find similar mass estimates to that of the single VPSHOCK fits of \cite{casaHwang}, however they found an additional purely Fe component in their models which we did not find significantly improved our fit. This gave them an additional Fe mass which we cannot account for in our model. 
    
    \subsection{Puppis~A}
	\begin{table*}
	    \caption{\textbf{Puppis~A:} Spectral properties of regions with enhanced ejecta, fit with a one-component VPSHOCK model where abundances for O, Ne, Mg, Si, and Fe (with nickel tethered to Fe) were allowed to vary. Abundances are in solar units from the abundance tables of \protect\cite{Wilms}. Data was grouped with a minimum of 20 counts per bins except for the ``omega knot'', which was grouped with a minimum of 40 counts per bins. The Chandra knots were fit together, but with the normalization constants allowed to vary between the knots. Additionally, we tested with variable temperatures between the 2 knots but there was no appreciable difference between the fits.  }
			
		\label{tbl:PupARegionData} 
		\begin{tabular}{cccccccccc}
		\hline
		Region & N$_\text{H}$ & kT & O & Ne & Mg & Si & Fe (Ni) & n$_\text{e}$t & $\chi^2_\nu$ (DOF) \T \\  		  
		& $\times 10^{22}$~cm$^{-2}$ & keV & & & & & & $\times 10^{11}$~cm$^{-3}$~s & \B \\ 
		\hline
        Chandra knots & 0.48$_{-0.05}^{+0.06}$ & 0.55$_{-0.04}^{+0.03}$ & 1.7$_{-0.4}^{+-0.5}$ & 2.5$_{-0.3}^{+0.5}$ & 2.4$_{-0.4}^{+0.5}$ & 0.6$_{-0.5}^{+0.6}$ & 0.3$_{-0.1}^{+0.1}$ & 0.57$_{-0.12}^{+0.12}$ & 1.25 (197) \T\B \\
        Omega Knots & 0.49$_{-0.04}^{+0.03}$ & 0.56$_{-0.04}^{+0.09}$ & 9.2$_{-1.5}^{+7.4}$ & 6.9$_{-1.3}^{+5.5}$ & 6.3$_{-1.4}^{+4.6}$ & 4.9$_{-1.1}^{+4.4}$ & 0.2$_{-0.1}^{+0.3}$ & 1.8$_{-0.5}^{+0.5}$ & 1.23 (1063) \B \\
        North & 0.34$_{-0.04}^{+0.03}$ & 0.55$_{-0.01}^{+0.02}$ & 1.1$_{-0.2}^{+0.2}$ & 1.7$_{-0.1}^{+0.3}$ & 1.4$_{-0.1}^{+0.3}$ & 0.7$_{-0.1}^{+0.2}$ & 0.70$_{-0.09}^{+0.07}$ & 2.2$_{-0.10}^{+0.03}$  & 1.08 (576) \B \\
        East & 0.35$_{-0.06}^{+0.07}$ & 0.54$_{-0.04}^{+0.06}$ & 1.3$_{-0.2}^{+0.4}$ & 1.5$_{-0.2}^{+0.2}$ & 1.4$_{-0.3}^{+0.2}$ & 1.9$_{-0.6}^{+0.3}$ & 1.6$_{-0.3}^{+0.2}$ & 1.0$_{-0.2}^{+0.3}$ & 1.04 (608) \B \\	
		\hline
		\end{tabular}
	\end{table*}	
    
    The 56 regions as selected by the \textit{contbin} algorithm were first fit with a VPSHOCK model. In some instances in the soft x-rays, there were significant differences between the two MOS cameras, and so the spectra were restricted to 0.5--5.0~keV range. This difference may be due to the contamination on the MOS--1 chip \citet{XMMMos1Contam}. In general, most spectra were fit in the 0.2--8.0~keV range. The column density, the temperature, the ionization timescale, and the normalization constant were first fit while holding the abundances at solar. Following that, we freed and fit O, Mg, Si, S, Ca, Ar, and Fe (Ni tethered to Fe) one at a time. In some instances, a secondary PSHOCK (or APEC) model was added if statistically significant to account for a secondary component. None of the regions as selected \textit{contbin} showed evidence of enhanced ejecta and so manually selected knots were chosen as seen in Figure~\ref{fig:regions}. These knots were much smaller than the algorithm selected regions with a lower count rate. Some of the regions have been discovered in previous studies like the \textit{Chandra} knots and the Omega knot \citet{puppAEastKnots, puppAOmegaKnot}. Two western ejecta knots (see West Knot 1, Table~\ref{tbl:PupARegionData}) were simultaneously fit together to improve statistics, only allowing their normalization constants to vary between each other, and a secondary western knot (West Knot 2) was also discovered. \cite{puppAEastKnots} found evidence of an enhanced filament, however, the statistics were poor with respects to the abundances when allowing O to vary and therefore were not included in the study. Additionally, \cite{puppAOmegaKnot} found two knots in the central region, but only the northern most knot had high enough counts to allow O to vary (called Omega knot in this study). The manually selected regions were fit as follows. The column density, temperature, normalization constants, and ionization timescales were first fit, then abundances were freed and fit beginning with O, Mg, Fe (Ni tethered to Fe), and Si one at a time.
    
    The full 56 region fits are not presented in this paper, but we can summarize the results here. Each region was fit with a single VPSHOCK model, and a secondary PSHOCK or APEC was added if it improved the fit. In general, the spectra were reasonably fit with a $\chi^2_\nu\approx1.7$ for most regions. The northernmost section was the least well fit ($\chi^2_\nu>2.0$ for 3 cases) due to only having a single data in large window mode, which meant some cameras did not cover at least 80\% of the region's area and therefore was excluded from the extraction. We find a hydrogen column density with a range of 0.15--0.35$\times 10^{22}$~cm$^{-2}$, with an average value of 0.24$\times 10^{22}$~cm$^{-2}$. Temperatures range from 0.48--0.98~keV, with a median value of $\sim$0.65~keV. Ionization timescale indicates that the plasma is still in NEI with values $\sim$1.0$\times10^{11}$~cm$^{-3}$~s. The global abundances tend to subsolar to solar values for O, Ne, Mg, Si, S, and Fe. A full SNR analysis was performed by \cite{puppAFeatureTailored} which selected regions via surface brightness and by \cite{puppAEastKnots} which performed a radially resolved spectral analysis. These studies found values for N$_H$ that tended to be slightly higher at $\sim$0.3$\times 10^{22}$~cm$^{-2}$, but both found temperature ranges consistent with our results, as well as similar ionization timescales and subsolar abundances. 
    
    The spectral fits for the manually selected regions can be found in Table~\ref{tbl:PupARegionData}. These regions were small with low counts, and in some instances, had unconstrained abundances unless oxygen is frozen. In those instances, we removed them from the study. The \textit{Chandra} data found enhanced ejecta for O, Ne, and Mg, in the manually selected knots whereas the \textit{XMM-Newton} data found these regions to be only slightly enhanced, with only the West Knot 2 showing enhanced Fe. We find a higher N$_H$ for the \textit{Chandra} data about $\sim0.48\times 10^{22}$~cm$^{-2}$ whereas the \textit{XMM-Newton} regions are more inline with the density range from the \textit{contbin} selected regions with 0.34--0.35$\times 10^{22}$~cm$^{-2}$. Temperatures for the enhanced ejecta regions tend to be about 0.55~KeV except for 1 of the knots (Omega Knot S) which has a much lower temperature of 0.29~keV. The ionization timescales indicate that the plasma is still in NEI with values $\sim$1.0$\times10^{11}$~cm$^{-3}$~s, much like the \textit{contbin} selected regions.
    
\subsection{G349.7+0.2}

	\begin{table*}
	    \caption{\textbf{G349.7+0.2:} Spectral properties of regions fit with a one-component VPSHOCK model where abundances for  Mg, Si, S, Ar, Ca, and Fe (with nickel tethered to Fe) were allowed to vary. Abundances are in solar units from the abundance tables of \protect\cite{Wilms}. Data was grouped with a minimum of 20 counts per bins. If an abundance value is missing, it was frozen at solar. }
			
		\label{tbl:G349RegionData} 
		\begin{tabular}{ccccccccccc}
		\hline
		Region & N$_\text{H}$ & kT & Mg & Si & S & Ar & Ca & Fe (Ni) & n$_\text{e}$t & $\chi^2_\nu$ (DOF) \T \\  		  
		& $\times 10^{22}$~cm$^{-2}$ & keV & & & & & & & $\times 10^{12}$~cm$^{-3}$~s & \B \\ 
		\hline
        0 & $9.9_{-0.4}^{+0.8}$ & $1.19_{-0.12}^{+0.03}$ & $4.1_{-1.8}^{+7.3}$ & $2.4_{-0.4}^{+1.6}$ & $0.9_{-0.1}^{+0.5}$ & $0.4_{-0.3}^{+0.3}$ & ... & $2.4_{-0.7}^{+2.7}$ & $0.86_{-0.21}^{+1.36}$ & 1.03 (208) \T\B \\
        1 & $10.2_{-0.4}^{+0.6}$ & $0.96_{-0.05}^{+0.04}$ & $10.0_{-2.9}^{+5.1}$ & $3.2_{-0.6}^{+1.4}$ & $1.1_{-0.2}^{+0.3}$ & $0.7_{-0.4}^{+0.6}$ & $1.6_{-1.0}^{+1.2}$ & $3.1_{-1.5}^{+5.5}$ & $1.6_{-0.2}^{+0.2}$ & 0.90 (194) \B \\
        2 & $9.6_{-0.2}^{+1.1}$ & $1.09_{-0.12}^{+0.02}$ & $2.5_{-0.4}^{+6.3}$ & $1.7_{-0.2}^{+1.4}$ & $1.07_{-0.08}^{+0.48}$ & $1.0_{-0.4}^{+0.6}$ & $0.8_{-0.6}^{+0.6}$ & $1.3_{-0.5}^{+1.6}$ & $0.79_{-0.01}^{+13.5}$ & 1.06 (197) \B \\
        3 & $8.9_{-2.2}^{+1.3}$ & $1.36_{-0.18}^{+0.03}$ & $0.6_{-0.3}^{+2.0}$ & $1.6_{-0.2}^{+0.8}$ & $1.0_{-0.1}^{+0.2}$ & $1.1_{-0.3}^{+0.4}$ & $0.6_{-0.5}^{+0.5}$ & $1.4_{-0.5}^{+1.2}$ & $3.4_{-0.5}^{+2.1}$ & 1.18 (208) \B \\
        4 & $10.2_{-1.1}^{+1.3}$ & $1.21_{-0.13}^{+0.14}$ & $4.0_{-2.7}^{+12.6}$ & $2.2_{-0.5}^{+2.5}$ & $1.0_{-0.1}^{+0.7}$ & $1.1_{-0.1}^{+0.8}$ & $1.4_{-0.8}^{+0.8}$ & $2.4_{-1.0}^{+2.2}$ & $0.50_{-0.14}^{+0.56}$ & 0.92 (215) \B \\
        5 & $10.2_{-0.3}^{+0.7}$ & $1.23_{-0.12}^{+0.07}$ & ... & $2.2_{-0.4}^{+0.6}$ & $1.2_{-0.2}^{+0.2}$ & ... & ... & $3.0_{-1.1}^{+2.6}$ & $0.40_{-0.85}^{+0.20}$ & 1.06 (226) \B \\
		\hline
		\end{tabular}
	\end{table*}

	\begin{table*}
	    \caption{\textbf{G349.7+0.2:} Spectral properties for the full SNR fit with a VPSHOCK+PSHOCK model with Mg, Si, S, Ar, and Ca allowed to vary for the hard (H) VPSHOCK component and the soft (S) PSHOCK abundances were frozen at solar. Fe and Ca was frozen to solar as they did not affect the fit. Data was grouped with a minimum of 20 counts per bins. }
			
		\label{tbl:G349FullSNRRegionData} 
		\begin{tabular}{ccccccccccc}
		\hline
		N$_\text{H}$ & kT$_\text{H}$ & Mg & Si & S & Ar & n$_\text{e}$t$_\text{H}$ & kT$_\text{S}$ & n$_\text{e}$t$_\text{S}$ & $\chi^2_\nu$ (DOF) \T \\  		  
		$\times 10^{22}$~cm$^{-2}$ & keV & & & & & $\times 10^{12}$~cm$^{-3}$~s & keV & $\times 10^{10}$~cm$^{-3}$~s & \B \\ 
		\hline
	    $10.3_{-0.5}^{+0.3}$ & $1.34_{-0.11}^{+0.08}$ & $13.4_{-2.3}^{+2.1}$ & $4.0_{-0.2}^{+1.1}$ & $1.3_{-0.3}^{+0.2}$ & $2.3_{-0.6}^{+0.9}$ & $1.0_{-0.1}^{+0.5}$ & $0.86_{-0.04}^{+0.16}$ & $7.8_{-1.6}^{+1.5}$ & 0.96 (347) \T\B \\
		\hline
		\end{tabular}
	\end{table*}
 
    G349.7+0.2 has one data set (obsID 2785) with 6 regions that were fit over the energy range 0.8--7.0~keV with a single VPSHOCK model. The column density, temperature, normalization constant, and ionization timescale were first freed and fit. Then the abundance variables for Mg, Si, S, Ar, Ca, and Fe (Ni tethered to Fe). In some cases where freeing Mg, Ar, or Ca did not improve the fit, they were frozen at solar. In most cases where the presence of the Fe~K line was present, it was well accounted for by the single VPSHOCK model. Since our 6 regions did not find evidence of a blast wave component, we fit the whole SNR with a VPSHOCK+PSHOCK model in order to differentiate the shocked ejecta from the forward shock. 
    
    The spectral fits for the individual regions can be found in Table~\ref{tbl:G15RegionData} and the global fit in Table~\ref{tbl:G349FullSNRRegionData}. The column density is quite high with a range of 8.8--10.5$\times 10^{22}$~cm$^{-2}$, the temperatures range from 1.1--1.4~keV, and the ionization timescales range from 3.4--15.9$\times10^{11}$~cm$^{-3}$~s which suggests the SNR is approaching ionization equillibrium. The abundances for Mg, Si, S, Ar, Ca, and Fe are super solar except in some instances for S and Ar. We fit the full SNR with a VPSHOCK+PSHOCK where the VPSHOCK was best associated with shock heated ejecta, with a temperature of $1.34^{+0.3}_{-0.5}$~keV and an ionization timescale of $1.0^{+0.5}_{-0.1}\times10^{12}$~cm$^{-3}$~s similar to that found in the individual regions, and the PSHOCK was best associated with the forward shock, with solar abundances, a temperature of $0.86^{+0.16}_{0.04}$~keV and an ionization timescale of $7.8\times10^{+0.16}_{-1.6}$~cm$^{-3}$~s indicating the blast wave component is not yet in CIE.   
    
    We can compare to \cite{G349} which examined the same data set for 6 regions as well as the full SNR. Their 6 regions found a slightly smaller column density in the range of 6.5--7.8$\times 10^{22}$~cm$^{-2}$, however our temperature and ionization timescale are similar, as well as the slightly enhanced abundances for Si and S. As for the global fit, our column density is much higher, but our soft and hard temperatures are similar. However, \cite{G349} found a secondary blast-wave component in CIE whereas our blast wave component was still in NEI. \cite{distG349&G350} also performed a \textit{Suzaku} study for the whole SNR with a 4 component model CIE+NEI+Gaussian(Al~K)+Gaussian(Fe~K) which found enhanced ejecta for Mg and Ni. Our hydrogen column density is much higher than their fit value of 6.4$\pm{0.1}\times 10^{22}$~cm$^{-2}$ and they also found a forward shock component in CIE with a lower temperature of $0.60\pm{0.04}$~keV. The hard component associated with the shocked ejecta has a temperature of $1.24\pm{0.3}$~keV and a ionization timescale of $20\pm{3}\times10^{11}$~cm$^{-3}$~s which is similar to our global fit. 
    
\subsection{G350.1-0.3}
    
	\begin{table*}
	    \caption{\textbf{G350.1--0.3:} Spectral properties of regions fit with the either a single component VNEI or a two-component VNEI+APEC where the VNEI abundances for Mg, Si, S, Ar, and Fe (with nickel tethered to Fe) were allowed to vary and the APEC abundances were frozen at solar. Abundances are in solar units from the abundance tables of \protect\cite{Wilms}. If an abundance value is missing, it was frozen at solar. If a temperature is missing for the APEC component, the region was fit with a single VNEI model. Data was grouped with a minimum of 20 counts per bins.  }
			
		\label{tbl:G350RegionData} 
        \resizebox{\textwidth}{!}{
		\begin{tabular}{ccccccccccccc}
		\hline
		& & VNEI & & &  & & & & & APEC & \T \\
		Region & N$_\text{H}$ & kT &  Mg & Si & S & Ar & Ca & Fe (Ni) & n$_\text{e}$t & kT & $\chi^2_\nu$ (DOF) \\  		  
		& $\times 10^{22}$~cm$^{-2}$ & keV & & & & & & & $\times 10^{11}$~cm$^{-3}$~s & keV & \B \\ 
		\hline
        0 & $6.0_{-0.1}^{+0.2}$ & $2.16_{-0.26}^{+0.08}$ & $17.7_{-2.8}^{+5.7}$ & $21.8_{-1.4}^{+4.9}$ & $7.5_{-0.7}^{+1.2}$ & $6.1_{-1.0}^{+1.5}$ & $10.9_{-1.2}^{+2.2}$ & $6.5_{-0.7}^{+2.2}$ & $2.1_{-0.1}^{+0.4}$ & $0.37_{-0.03}^{+0.03}$ & 1.51 (824) \T\B \\
        1 & $6.0_{-0.1}^{+0.2}$ & $1.48_{-0.10}^{+0.08}$ & $11.7_{-1.8}^{+3.3}$ & $12.8_{-1.4}^{+2.1}$ & $4.6_{-0.5}^{+2.3}$ & $5.4_{-0.9}^{+1.6}$ & $12.6_{-1.9}^{+4.7}$ & $11.9_{-3.8}^{+5.9}$ & $0.94_{-0.06}^{+0.10}$ & $0.28_{-0.02}^{+0.01}$ & 1.06 (793) \B \\
        2 & $5.82_{-0.08}^{+0.21}$ & $1.49_{-0.13}^{+0.06}$ & $8.5_{-0.6}^{+2.3}$ & $12.6_{-0.6}^{+1.8}$ & $5.0_{-0.2}^{+0.5}$ & $6.0_{-0.4}^{+0.8}$ & $10.5_{-1.2}^{+2.5}$ & $8.6_{-1.7}^{+4.9}$ & $1.40_{-0.07}^{+0.25}$ &  $0.30_{-0.01}^{+0.02}$ & 1.30 (814) \B \\
        3 & $6.0_{-0.1}^{+0.2}$ & $1.18_{-0.07}^{+0.01}$ & $10.2_{-1.4}^{+2.2}$ & $8.0_{-0.8}^{+2.1}$ & $2.7_{-0.3}^{+0.5}$ & $3.2_{-0.6}^{+0.6}$ & $7.7_{-1.0}^{+2.3}$ & $7.7_{-1.6}^{+5.0}$ & $2.2_{-0.1}^{+0.5}$ & $0.31_{-0.02}^{+0.03}$ & 1.21 (796) \B \\
        4 & $5.95_{-0.07}^{+0.23}$ & $1.59_{-0.13}^{+0.04}$ & $12.4_{-1.3}^{+4.1}$ & $10.3_{-0.5}^{+1.6}$ & $2.9_{-0.2}^{+0.4}$ & $3.1_{-0.4}^{+0.6}$ & $5.3_{-0.7}^{+1.4}$ & $2.7_{-0.6}^{+1.3}$ & $4.6_{-0.2}^{+0.1}$ & $0.32_{-0.02}^{+0.03}$ & 1.27 (854) \B \\
        5 & $6.0_{-0.1}^{+0.2}$ & $1.27_{-0.06}^{+0.11}$ & $8.1_{-0.9}^{+2.8}$ & $16.0_{-1.6}^{+3.7}$ & $7.0_{-0.7}^{+1.2}$ & $6.4_{-1.2}^{+1.7}$ & $11.9_{-2.9}^{+3.6}$ & $12.8_{-3.3}^{+5.7}$ & $1.6_{-0.2}^{+0.2}$ & $0.287_{-0.008}^{+0.017}$ & 1.43 (779) \B \\
        6 & $4.45_{-0.05}^{+0.07}$ & $0.66_{-0.02}^{+0.01}$ & $0.7_{-0.1}^{+0.1}$ & $1.18_{-0.07}^{+0.08}$ & $0.777_{-0.007}^{+0.015}$ & $1.0_{-0.1}^{+0.2}$ & $1.1_{-1.0}^{+3.3}$ & $0.10_{-0.10}^{+0.07}$ & $6.907_{-0.0007}^{+0.0250}$ & ... & 1.14 (811) \B \\
        7 & $4.29_{-0.09}^{+0.19}$ & $0.96_{-0.02}^{+0.05}$ & $1.24_{-0.09}^{+0.05}$ & $2.55_{-0.07}^{+0.36}$ & $1.51_{-0.06}^{+0.16}$ & $1.7_{-0.3}^{+0.3}$ & $4.7_{-1.0}^{+0.8}$ & $<0.24$ & $2.7_{-0.3}^{+0.2}$ & ... & 1.39 (826) \B \\
        8 & $5.9_{-0.2}^{+0.3}$ & $1.4_{-0.1}^{+0.1}$ & $4.0_{-1.0}^{+1.5}$ & $8.8_{-0.9}^{+1.8}$ & $3.6_{-0.3}^{+0.7}$ & $4.5_{-0.7}^{+1.1}$ & $8.0_{-1.4}^{+2.5}$ & $6.0_{-2.4}^{+5.2}$ & $1.5_{-0.2}^{+0.3}$ & $0.35_{-0.03}^{+0.02}$ & 1.21 (824) \B \\
        9 & $5.0_{-0.2}^{+0.4}$ & $0.91_{-0.04}^{+0.05}$ & $1.1_{-0.2}^{+0.4}$ & $1.4_{-0.1}^{+0.2}$ & $1.10_{-0.08}^{+0.11}$ & $1.6_{-0.3}^{+0.3}$ & $2.1_{-1.1}^{+0.6}$ & $0.4_{-0.2}^{+0.7}$ & $3.31_{-0.04}^{+0.31}$ & ... & 1.15 (787) \B \\
        10 & $4.6_{-0.1}^{+0.2}$ & $0.87_{-0.05}^{+0.03}$ & $2.0_{-0.2}^{+0.3}$ & $3.0_{-0.2}^{+0.2}$ & $1.5_{-0.1}^{+0.1}$ & $1.7_{-0.2}^{+0.2}$ & $3.0_{-0.9}^{+1.1}$ & ... & $6.1_{-0.9}^{+1.7}$ & ... & 1.32 (739) \B \\
        11 & $5.4_{-0.3}^{+0.6}$ & $0.73_{-0.03}^{+0.03}$ & $1.7_{-0.4}^{+1.1}$ & $1.9_{-0.1}^{+0.4}$ & $1.11_{-0.07}^{+0.18}$ & $1.4_{-0.3}^{+0.4}$ & $3.1_{-1.1}^{+1.5}$ & $1.0_{-0.4}^{+1.5}$ & $4.1_{-0.4}^{+0.3}$ & ... & 1.43 (752) \B \\	
		\hline
		\end{tabular}}
	\end{table*}  
 
    G350.1$-$0.3 regions were fit over the 0.5--10~keV range using a single VPSHOCK model. The column density, temperature, normalization constant, and ionization timescale were first freed and fit. Then the abundance variables for Mg, Si, S, Ar, Ca, and Fe (Ni tethered to Fe). In some cases, a secondary APEC model was added in order to account for the blast wave component of the SNR whose abundances were held at solar. The regions were well fit with $\chi^2_\nu<1.51$. 
    
    The spectral fits can be found in Table~\ref{tbl:G350RegionData}. The column density has a range of 4.3--6.5$\times 10^{22}$~cm$^{-2}$ with the brighter eastern portion around $\sim6.0\times 10^{22}$~cm$^{-2}$ and the fainter western region around $\sim5.0\times 10^{22}$~cm$^{-2}$. Some regions were best fit with a two-component APEC+VPSHOCK model, where the CIE component was best associated with the blast wave and a temperature range of 0.29--0.37~keV, and the NEI component was best associated with the shock-heated ejecta for Mg, Si, S, Ar, Ca, and Fe with temperatures of 1.5--2.2~keV and ionization timescales with a range of 0.9--6.1$\times10^{11}$~cm$^{-3}$~s. Almost all the single component regions were best characterized with enhanced ejecta except for Region~6, which has solar to subsolar abundances and can be associated with the blast wave component. Region~6 has a hotter temperature of 0.66~keV than the temperatures associated with the CIE blast-wave component of the two-component fits and a ionization timescale of $6.9\times10^{11}$~cm$^{-3}$~s, which could indicate a range of temperatures associated with the forward shock.
    
    There were 3 studies that previously examined G350.1$-$0.3, all using the solar abundance tables from \cite{Wilms}. A previous \textit{Chandra} study by \cite{G350Chandra} examined the same data for 6 regions. These regions were fit with a single VNEI with abundances for Mg, Si, S, Ca, Ar, and Fe allowed to vary but found no two-component model fits. They have similar column density to our results as well as similar ionization timescale and abundance values for the shock-heated ejecta component. An \textit{XMM-Newton} study by \cite{g350_gaensler} fit the whole SNR with a two-component VPSHOCK+RAYMOND which found a shock heated ejecta component with temperature of 1.46~keV, an ionization timescale of $3.0\times10^{11}$~cm$^{-3}$~s, and a forward shock component with temperature 0.36~keV in CIE, which is in agreement with our results. They also found large enhancement for Ne, Mg, Si, S, Ca, and Fe (in some cases $>20$), which is much higher than our results. \cite{distG349&G350} performed a \textit{Suzaku} study for the whole SNR which found a 4 component model CIE+NEI+Gaussian(Al~K)+Gaussian(Fe~K) with enhanced ejecta for Mg, Al, Si, S, Ar, Ca, Fe, and Ni. The CIE component had solar abundances with a temperature of 0.48~keV and the NEI component had enhanced abundances with temperature of 1.51~keV and an ionization timescale of $3.5\times10^{11}$~cm$^{-3}$~s, which is in agreement with our results.

\section{Derived X-ray Properties}

    The tabulated results for the derived X-ray properties of individual regions for each SNR.

\begin{landscape}
    
    \begin{table}
	\begin{center}
		\caption{ Derived X-ray properties for G15.9+0.2. Subscript `h' refers to the hard component. We use a distance of 8.5~kpc and a radius of 155'' for the full SNR region. } 
		
		\label{tbl:G15derived1}
    	\begin{tabular}{c c c c c c c} 
			
            \hline
			Parameter & Full SNR & Region 0 & Region 1 & Region 2 & Region 3 & Region 4 \T \B \\
						
		    \hline	
            Emission Measure, EM$_h$ ($\times 10^{57}$~$f_h$~D$_{8.5}^{-2}$~cm$^{-3}$) & 3.3$_{-0.7}^{+0.5}$ & 17.6$_{-3.5}^{+2.6}$ & 9.48$_{-1.1}^{+2.1}$ & 12.1$_{-1.4}^{+1.5}$ & 7.81$_{-1.6}^{+2.4}$ & 11.6$_{-1.7}^{+2.1}$ \T\B  \\
			Electron Density, n$_e{_h}$ ($f_h^{-1/2}$~D$_{8.5}^{-1/2}$~cm$^{-3}$) & 0.35$_{-0.04}^{+0.02}$ & 3.12$_{-0.3}^{+0.2}$ & 2.0$_{-0.1}^{+0.2}$ & 1.9$\pm{0.1}$ &  1.0$\pm{0.1}$ & 1.49$\pm{0.1}$ \B \\
			Shock Age, t$_\text{shock}{_h}$ ($f_h^{1/2}$~D$_{8.5}^{1/2}$~yr) & 2870$\pm{90}$ & 310$\pm{50}$ & 1120$_{-150}^{+490}$  & 1000$\pm{140}$ & 1420$_{-330}^{+490}$ & 560$_{-90}^{+120}$ \B \\
			\hline
				
		\end{tabular} 
	\end{center}
    \end{table}

    \begin{table}
		\begin{center}
		\caption{ Derived X-ray properties for Kes~79. Subscripts `h' and `s' denote the hard and soft components, respectively. Regions 1, 4, 7, and 8 were fit with a soft APEC component and therefore do not have a shock wave age estimate. Region 0 and 9 are listed as soft components due to their association with the forward shock even though they have higher temperature fits. All other regions were single component VNEI fits. We use a distance of 7.1~kpc and a full SNR radius of 6'.} 
		\label{tbl:kes79derived}
    		\begin{tabular}{ccccccccccccc} 

				\hline
				Parameter & Reg 0 & Reg 1 & Reg 2 & Reg 3 & Reg 4 & Reg 5 & Reg 6 & Reg 7 & Reg 8 & Reg 9 & Reg 10 & Reg 11 \T \B \\
						
		        \hline
                Emission Measure, EM$_h$ & ... & 8.7$_{-1.0}^{+0.8}$ & 18.74$\pm{0.06}$ & 22.3$_{-0.9}^{+1.6}$ & 22.1$_{-1.9}^{+2.4}$ & 39.8$_{-3.0}^{+2.3}$ & 8.98$_{-0.7}^{+0.3}$ & 26.0$_{-1.3}^{+2.0}$ & 19.0$_{-1.6}^{+1.7}$ & ... & 29.4$_{-1.0}^{+1.9}$ & 23.9$_{-1.8}^{+1.1}$ \T \\
                ($\times 10^{56}$~$f_h$~D$_{7.1}^{-2}$~cm$^{-3}$) \\
				Electron Density, n$_e{_h}$ & ... & 0.77$_{-0.05}^{+0.04}$ & 1.03$\pm{0.02}$ & 0.93$_{-0.02}^{+0.03}$ & 0.88$_{-0.04}^{+0.05}$ & 0.85$_{-0.03}^{+0.02}$ & 0.33$_{-0.01}^{+0.06}$ & 0.77$_{-0.02}^{+0.03}$ & 0.75$\pm{0.03}$ & ... & 0.55$_{-0.01}^{+0.02}$ & 0.48$_{-0.02}^{+0.01}$  \\
				($f_h^{-1/2}$~D$_{7.1}^{-1/2}$~cm$^{-3}$) \\
				Shock Age, t$_\text{shock}{_h}$ & ... & 2360$_{-780}^{+1440}$ & 2520$_{-150}^{+430}$ & 2480$_{-590}^{+1740}$ & 2210$_{-90}^{+120}$ & 4000$_{-1600}^{+2900}$ & 6400$_{-1500}^{+7500}$ & 1610$_{-540}^{+1590}$ & 2100$_{-630}^{+890}$ & ... & 2730$_{-480}^{+710}$ & 4700$_{-1500}^{+2200}$ \\
				 ($f_h^{1/2}$~D$_{7.1}^{1/2}$~yr) \\
				Emission Measure, EM$_s$  & 0.032$\pm{0.002}$ & 0.44$_{-0.04}^{+0.08}$ & ... & ... & 1.20$\pm{0.2}$ & ... & ... & 2.31$\pm{0.2}$ & 0.78$\pm{0.10}$ & 0.027$\pm{-0.002} $ & ... & ...   \\
				($\times 10^{59}$~$f_s$~D$_{7.1}^{-2}$~cm$^{-3}$) \\
				Electron Density, n${_e}_s$  & 1.24$_{-0.03}^{+0.04}$ & 0.77$_{-0.04}^{+0.07}$ & ... & ... & 6.48$_{-0.5}^{+0.7}$ & ... & ... & 7.3$_{-0.3}^{+0.4}$ & 4.8$\pm{0.3}$ & 0.34$\pm{0.01}$ & ... & ...   \\
				($f_s^{-1/2}$~D$_{7.1}^{-1/2}$~cm$^{-3}$) \\
				Ambient Density, n${_0}_s$  & 0.258$_{-0.007}^{+0.008}$ & 0.161$_{-0.008}^{+0.070}$ & ... & ... & 1.4$\pm{0.1}$ & ... & ... & 1.52$_{-0.07}^{+0.08}$ & 1.00$\pm{0.06}$ & 0.070$_{-0.002}^{+0.003}$ & ... & ...  \\
				($f_s^{-1/2}$~D$_{7.1}^{-1/2}$~cm$^{-3}$) \\
				Shock Age, t$_\text{shock}{_s}$ & 1610$_{-370}^{+610}$ & ... & ... & ... & ... & ... & ... & ... & ... & 8200$_{-2500}^{+12700}$ & ... & ...  \\ 
				 ($f_s^{1/2}$~D$_{7.1}^{1/2}$~yr) \B \\
        		\hline
				
			\end{tabular}
		\end{center}
	\end{table}
 
    \begin{table}
		\begin{center}
		\caption{ Derived X-ray properties for Puppis~A. Subscript `h' refers to the hard component. } 
		
		\label{tbl:pupaderived1}
    		\begin{tabular}{ccccc} 

				\hline
				Parameter & Chandra Knots & Omega Knots & North Knot & East Knot \T \B \\
				
		        \hline
                Emission Measure, EM$_h$ ($\times 10^{56}$~$f_h$~D$_{1.3}^{-2}$~cm$^{-3}$) & $1.31\pm(0.04)$ & $1.27^{+0.01}_{-0.06}$ & $18.6^{+0.2}_{-0.3}$ & $1.09\pm{0.03}$ \T\B \\
				Electron Density, n$_e{_h}$ ($f_h^{-1/2}$~D$_{1.3}^{-1/2}$~cm$^{-3}$) & $3.00\pm{0.05}$ & $4.4^{+0.2}_{-0.1}$ & $3.64\pm{0.02}$ & $0.89\pm{0.01}$ \B \\
				Shock Age, t$_\text{shock}{_h}$ ($f_h^{1/2}$~D$_{1.3}^{1/2}$~yr) & $500\pm{60}$ & $1300\pm{200}$ & $1900\pm{200}$ & $3700\pm{400}$ \B \\
				\hline
				
			\end{tabular}
		\end{center}
	\end{table}
\end{landscape}

\begin{landscape}
    \begin{table}
		\begin{center}
		\caption{ Derived X-ray properties for G349.8+0.2. Subscript `h' refers to the hard component. } 
		
		\label{tbl:G349derived1}
            \resizebox{\textheight}{!}{
    		\begin{tabular}{c c c c c c c c} 

				\hline
				Parameter & Full SNR & Region 0 & Region 1 & Region 2 & Region 3 & Region 4 & Region 5 \T\B \\
				
		        \hline
                Emission Measure, EM$_h$ ($\times 10^{58}$~$f_h$~D$_{12}^{-2}$~cm$^{-3}$) & 8.7$_{-0.7}^{+0.8}$ & 3.5$_{-0.15}^{+0.07}$ & 4.10$_{-0.21}^{+0.09}$ & 4.24$_{-0.12}^{+0.18}$ & 2.93$_{-0.05}^{+0.26}$ & 3.57$_{-0.21}^{+0.14}$ & 3.33$_{-0.09}^{+0.16}$ \T\B \\
				Electron Density, n$_e{_h}$ ($f_h^{-1/2}$~D$_{12}^{-1/2}$~cm$^{-3}$) & 2.76$\pm{0.1}$ & 0.310$_{-0.006}^{+0.003}$ & 0.298$_{-0.008}^{+0.003}$ & 0.188$_{-0.003}^{+0.004}$ & 0.169$_{-0.001}^{+0.008}$ &  0.137$_{-0.004}^{+0.003}$ & 0.0886$_{-0.001}^{+0.002}$ \B \\
				Shock Age, t$_\text{shock}{_h}$ ($f_h^{1/2}$~D$_{12}^{1/2}$~kyr) & 11.5$_{-1.2}^{+5.7}$ & 88.3$_{-21.8}^{+139.4}$ & 170$\pm{19}$ & 134$_{-17}^{+23}$ & 64$_{-10}^{+41}$ & 116$_{-32}^{+130}$ & 141$_{-30}^{+70}$ \B \\

				\hline
				
			\end{tabular}}
		\end{center}
	\end{table}

    \begin{table}
		\begin{center}
		\caption{ Derived X-ray properties for G350.1--0.3. Subscripts `h' and `s' denote the hard and soft components, respectively. } 

		\label{tbl:G350derived1}
    		\begin{tabular}{ccccccccccccc} 

				\hline
				Parameter & Reg 0 & Reg 1 & Reg 2 & Reg 3 & Reg 4 & Reg 5 & Reg 6 & Reg 7 & Reg 8 & Reg 9 & Reg 10 & Reg 11 \T\B \\
				
		        \hline
                Emission Measure, EM$_h$ & 2.5$_{-0.3}^{+0.5}$ & 4.6$\pm{0.4}$ & 4.8$\pm{0.4}$ & 7.5$\pm{0.9}$ & 5.4$_{-0.6}^{+0.7}$ & 4.8$_{-0.8}^{+0.4}$ & ... & 20$_{-3}^{+1}$ & 5.6$_{-0.9}^{+0.7}$ & 28$_{-5}^{+4}$ & 19$_{-2}^{+3}$ & 42$_{-6}^{+7}$ \T \\
                ($\times 10^{57}$~$f_h$~D$_{4.5}^{-2}$~cm$^{-3}$) \\
				Electron Density, n$_e{_h}$  & 7.3$_{-0.5}^{+0.7}$ & 9.4$_{-0.4}^{+0.5}$ & 7.2$\pm{0.3}$ & 8.2$\pm{0.5}$ & 5.0$\pm{0.3}$ & 5.0$_{-0.4}^{+0.2}$ & ... & 2.24$_{-0.15}^{+0.08}$ & 2.8$\pm{0.2}$ & 3.9$_{-0.3}^{+0.6}$ & 3.8$_{-0.2}^{+0.3}$ & 5.1$_{-0.3}^{+0.4}$  \\
				($f_h^{-1/2}$~D$_{4.5}^{-1/2}$~cm$^{-3}$) \\
				Shock Age, t$_\text{shock}{_h}$ & 920$_{-90}^{+210}$ & 320$_{-0.20}^{+0.40}$ & 610$_{-40}^{+120}$ & 830$_{-70}^{+200}$ & 2900$_{-200}^{+600}$ & 1000$_{-200}^{+140}$ & ... & 3800$_{-500}^{+300}$ & 1700$_{-300}^{+400}$ & 2700$_{-200}^{+500}$ & 5200$_{-800}^{+1500}$ & 2500$_{-300}^{+280}$ \\
				 ($f_h^{1/2}$~D$_{4.5}^{1/2}$~yr) \\
				Emission Measure, EM$_s$  & 1.0$\pm{0.3}$ & 4.1$_{-0.9}^{+1.6}$ & 2.6$\pm{0.5}$ & 2.5$_{-0.7}^{+1.1}$ & 2.0$_{-0.3}^{+0.6}$ & 3.0$_{-0.7}^{+0.5}$ & 0.47$_{-0.04}^{+0.05}$ & ... & 1.8$_{-0.4}^{+1.1}$ & ... & ... & ...   \\
				($\times 10^{58}$~$f_s$~D$_{11.5}^{-2}$~cm$^{-3}$) \\
				Electron Density, n${_e}_s$  & 47$_{-6}^{+5}$ & 88$_{-9}^{+18}$ & 53$\pm{5}$ & 48$_{-7}^{+10}$ & 30$_{-3}^{+4}$ & 39$_{-5}^{+3}$ & 3.0$_{-0.1}^{+0.2}$ & ... & 16$_{-2}^{+5}$ & ... & ... & ...   \\
				($f_s^{-1/2}$~D$_{4.5}^{-1/2}$~cm$^{-3}$) \\
				Ambient Density, n${_0}_s$  & 10$\pm{0.1}$ & 18$_{-2}^{+4}$ & 11$\pm{1}$ & 10$_{-1}^{+2}$ & 6.3$_{-0.5}^{+0.9}$ & 8.1$_{-1.0}^{+0.7}$ & 0.62$_{-0.02}^{+0.03}$ & ... & 3.3$_{-0.3}^{+1.0}$ & ... & ... & ... \\
				($f_s^{-1/2}$~D$_{4.5}^{-1/2}$~cm$^{-3}$) \B \\				
				\hline
				
			\end{tabular}
		\end{center}
	\end{table}
\end{landscape}

\bsp	
\label{lastpage}
\end{document}